\documentclass[twocolumn]{aa} 

\usepackage{lscape}
\usepackage{natbib}
\usepackage{graphicx}
\usepackage{longtable}
\usepackage{xcolor}
\usepackage{txfonts}
\begin{document}

   \title{Hot subdwarf binaries from the MUCHFUSS project}

   \subtitle{Analysis of 12 new systems and a study of the short-period binary population
}

   \author{T.~Kupfer \inst{1}
     \and S.~Geier   \inst{2}
      \and U.~Heber   \inst{3}
         \and R.~H.~{\O}stensen  \inst{4}
     \and B.~N.~Barlow \inst{5}
     \and P.~F.~L.~Maxted \inst{6}
     \and C.~Heuser   \inst{3}
     \and V.~Schaffenroth \inst{3,7}
         \and B.~T.~G\"ansicke \inst{8}
          }
          \offprints{T.\,Kupfer,\\\email{t.kupfer@astro.ru.nl}}

   \institute{Department of Astrophysics/IMAPP, Radboud University Nijmegen, P.O. Box 9010, 6500 GL Nijmegen, The Netherlands
   \and
            European Southern Observatory, Karl-Schwarzschild-Str. 2, 85748 Garching, Germany
            \and
            Dr. Karl Remeis-Observatory \& ECAP, Astronomical Institute, Friedrich-Alexander University Erlangen-Nuremberg, Sternwartstr. 7, D 96049 Bamberg, Germany
            \and 
            Institute of Astronomy, KU Leuven, Celestijnenlaan 200D, B-3001 Heverlee, Belgium
            \and
            Department of Physics, High Point University, 833 Montlieu Avenue, High Point, NC 27262, USA
            \and
            Astrophysics Group, Keele University, Staffordshire, ST5 5BG, UK
            \and
            Institute for Astro- and Particle Physics, University of Innsbruck, Technikerstr. 25/8, 6020 Innsbruck, Austria
            \and 
            Department of Physics, University of Warwick, Coventry CV4 7AL, UK
                       }

   \date{Received ; accepted }

 
  \abstract{
 The project Massive Unseen Companions to Hot Faint Underluminous Stars from SDSS (MUCHFUSS) aims at finding hot subdwarf stars with massive compact companions like massive white dwarfs ($M > 1.0$\,M$_\odot$), neutron stars, or stellar-mass black holes. The existence of such systems is predicted by binary evolution theory, and recent discoveries indicate that they exist in our Galaxy.\\
  We present orbital and atmospheric parameters and put constraints on the nature of the companions of 12 close hot subdwarf B star (sdB) binaries found in the course of the MUCHFUSS project. The systems show periods between $0.14$ and $7.4$\,days. In nine cases the nature of the companions cannot be constrained unambiguously whereas three systems most likely have white dwarf companions. We find that the companion to SDSS\,J083006.17$+$475150.3 is likely to be a rare example of a low-mass helium-core white dwarf. SDSS\,J095101.28$+$034757.0 shows an excess in the infrared that probably originates from a third companion in a wide orbit, which makes this system the second candidate hierarchical triple system containing an sdB star. SDSS\,J113241.58$-$063652.8 is the first helium deficient sdO star with a confirmed close companion. \\
  This study brings to 142 the number of sdB binaries with orbital periods of less than 30 days and with measured mass functions. We present an analysis of the minimum companion mass distribution and show that it is bimodal. One peak around $0.1$\,M$_\odot$ corresponds to the low-mass main sequence (dM) and substellar companions. The other peak around $0.4$\,M$_\odot$ corresponds to the white dwarf companions. The derived masses for the white dwarf companions are significantly lower than the average mass for single carbon-oxygen white dwarfs. In a $T_{\rm eff}$ -- $\log{g}$ diagram of sdB+dM companions, we find signs that the sdB components are more massive than the rest of the sample. 
  The full sample was compared to the known population of extremely low-mass white dwarf binaries as well as short-period white dwarfs with main sequence companions. Both samples show a significantly different companion mass distribution indicating either different selection effects or different evolutionary paths. We identified 16 systems where the dM companion will fill its Roche Lobe within a Hubble time and will evolve into a cataclysmic variable; two of them will have a brown dwarf as donor star. Twelve systems with confirmed white dwarf companions will merge within a Hubble time, two of them having a mass ratio to evolve into a stable AM\,CVn-type binary and another two which are potential supernova Ia progenitor systems. The remaining eight systems will most likely merge and form RCrB stars or massive C/O white dwarfs depending on the structure of the white dwarf companion.}
  \keywords{binaries: spectroscopic -- stars: subdwarf
               }
   \maketitle
%
\section{Introduction}
Hot subdwarf B stars (sdBs) are hot core helium-burning stars with masses around $0.5$\,M$_\odot$ and thin hydrogen layers (Heber 1986; for a recent review see Heber 2009\nocite{heb86,heb09}). It has been shown that a large fraction of sdBs are members of short-period binaries with periods below $\sim$10 days \citep{max01,nap04a}. For such short-period sdB binaries common envelope (CE) ejection is the only likely formation channel. One possible scenario is that two main sequence stars (MS) evolve in a binary system. The more massive one will evolve faster to become a red giant. Unstable mass transfer from the red giant to the companion will lead to a CE phase. Because of the friction in this phase, the two stellar cores lose orbital energy and angular momentum, which leads to a shrinkage of the orbit. This energy is deposited in the envelope which will finally be ejected. If the core reaches the mass required for the core-helium flash before the envelope is lost, a binary consisting of a core-helium burning sdB star and a main sequence companion is formed. In another possible scenario the more massive star evolves to become a white dwarf (WD) either through a CE phase or stable mass transfer onto the less massive companion. The less massive star evolves then to become a red giant. Unstable mass transfer will lead to a CE and once the envelope is ejected the red giant remnant starts burning helium, and a system consisting of an sdB with a WD companion is formed \citep{han02,han03}. 

If the red giant loses enough mass that the remnant is not massive enough to ignite helium the star will evolve directly to become a helium-core white dwarf. Helium-core white dwarfs with  masses below $0.3$\,M$_\odot$ are called extremely low-mass (ELM) white dwarfs \citep{bro10}. According to single star evolution, ELM-WDs should not exist as the evolutionary timescale to form them is much longer than the age of the universe. Therefore, significant mass transfer during the evolution is needed and most of the observed ELM-WDs indeed reside in close binary systems, usually with WD companions. Those systems are formed through the same CE-ejection process as the short-period sdB binaries, except that the envelope is ejected before the core reaches the mass required to ignite helium \citep{bro10}. Recent studies have increased the number of known ELM-WDs significantly (Brown et al. 2013 and references therein\nocite{bro13}). During their evolution they can spectroscopically appear as sdBs. However, they have lower masses compared to helium core-burning sdBs. Three low-mass sdBs, which evolve directly towards the ELM-WD phase are known so far. All have WD companions and lie below the Zero-Age Extreme Horizontal Branch \citep{heb03, oto06a,sil12}. Furthermore, hot He-WD progenitors in an earlier stage of evolution have been recently found orbiting intermediate-mass main sequence stars (EL\,CVn systems, Maxted et al. 2013).

\begin{table}[t!]
{\small \caption{Overview of the solved binary systems.} 
\label{tab:systems}
\begin{center}
\begin{tabular}{llll} 
\hline\hline
\noalign{\smallskip}
Short name & SDSS name & Other names \\
\hline
\noalign{\smallskip}
J01185$-$00254 & SDSS\,J011857.20$-$002546.5 & PB\,6373 \\
J03213$+$05384 & SDSS\,J032138.67$+$053840.0 & PG\,0319$+$055 \\
J08233$+$11364 & SDSS\,J082332.08$+$113641.8 & $-$ \\
J08300$+$47515 & SDSS\,J083006.17$+$475150.3 & PG\,0826$+$480 \\
J09523$+$62581 & SDSS\,J095238.93$+$625818.9 & PG\,0948$+$632 \\
J09510$+$03475 & SDSS\,J095101.28$+$034757.0 & PG\,0948$+$041 \\ 
J10215$+$30101 & SDSS\,J102151.64$+$301011.9 & $-$ \\
J11324$+$06365 & SDSS\,J113241.58$-$063652.8 & PG\,1130$-$063 \\
J13463$+$28172 & SDSS\,J134632.65$+$281722.7 & TON\,168 \\
J15082$+$49405 & SDSS\,J150829.02$+$494050.9 & SBSS\,1506$+$498 \\
J15222$-$01301 & SDSS\,J152222.14$-$013018.3 & $-$ \\
J18324$+$63091 & SDSS\,J183249.03$+$630910.5 & FBS\,1832$+$631 \\
\hline 
\end{tabular}
\end{center}}
\end{table}

Hot subdwarf binaries, as well as WDs with massive WD companions, turned out to be good candidates for SN\,Ia progenitors. Because of gravitational wave radiation, the orbit will shrink further and mass transfer from the sdB onto the WD will start once the sdB fills its Roche lobe. The Chandrasekhar limit might be reached either trough He accretion on the WD (e.g. Yoon \& Langer 2004 and references therein\nocite{yoo04}) or a subsequent merger of the sdB+WD system \citep{tut81,web84}. Two sdBs with massive WD companions have been identified to be good candidates for being SN\,Ia progenitors. KPD\,1930+2752 has a total mass of 1.47\,M$_\odot$ exceeding the Chandrasekhar limit and will merge within about 200 million years \citep{max00,gei07}. CD$-$30$^\circ$\,11223 has the shortest known orbital period of all sdB binaries ($P_{\rm orb}=0.048979$\,days) and is a good candidate to explode as an underluminous helium double-detonation SN\,Ia \citep{ven12,gei13}. The explosion of the massive WD companion in the eclipsing sdO+WD system HD\,49798 on the other hand may be triggered by stable mass transfer \citep{mer09}. 

Neutron star (NS) or even black hole (BH) companions are predicted by theory as well \citep{pod02,pfa03}. In this scenario two phases of unstable mass transfer are needed and the NS or the BH is formed in a supernova explosion. \citet{yun05} predicted that about 0.8\% of the short-period sdBs should have NS companions. In an independent study, \citet{nel10} showed that about 1\% of these systems should have NS companions whereas about 0.1\% should have BH companions.  

However, no NS/BH companion to an sdB has yet been detected unambiguously whereas a few candidates have been identified \citep{gei10}.
Follow-up observations have been conducted with radio and X-ray telescopes. \citet{coe11} did not detect any radio signals of a pulsar companion at the positions of four candidate systems from \citet{gei10}. \citet{mer11, mer14} searched for X-rays powered by wind accretion onto compact companions to sdBs using the \textsl{Swift} and \textsl{XMM} satellites, but did not detect any of those targets. 

NS+WD systems have been discovered amongst pulsars. \citet{fer10} showed that the peculiar system PSR\,J1802-2124 contains a millisecond pulsar and a low-mass C/O WD. This system may have evolved from an sdB+NS system. Most recently, \citet{kap13} discovered the close companion to the pulsar PSR\,J1816+4510 to be a He-WD progenitor with atmospheric parameters close to an sdB star ($T_{\rm eff}=16\,000\,{\rm K}$, $\log{g}=4.9$).
\begin{table*}[t!]
\caption{Summary of the follow-up observations in the course of the MUCHFUSS project.}
\label{tab:runs}
\begin{center}
\begin{tabular}{llcll} \hline\hline
\noalign{\smallskip}
Date & Telescope\,\&\,Instrument & Resolution [$\lambda$/$\Delta\lambda$] &  Coverage [\AA]  &  Observer\\ \hline
\noalign{\smallskip}
2009-May-27 -- 2009-May-31 & CAHA-3.5m+TWIN &  4000  &  3460 -- 5630  &  S. M.$^{\rm a}$ \\
2009-Nov-08 -- 2009-Nov-12 & ESO-NTT+EFOSC2 &  2200  &  4450 -- 5110  &  T. K. \\ 
2010-Feb-12 -- 2010-Feb-15 & SOAR+Goodman &    7000   & 3500 -- 6160  &  B. B. \\
2010-Aug-02 -- 2010-Aug-03 & SOAR+Goodman &    7000  & 3500 -- 6160  &  B. B. \\
February -- July 2011 & Gemini-North+GMOS-N &  1200   & 3610 -- 5000  &  Service \\
February -- July 2011 & Gemini-South+GMOS-S &  1200   & 3610 -- 5000  & Service \\
2011-Nov-15 -- 2011-Nov-19 & CAHA-3.5m+TWIN & 4000  &  3460 -- 5630  & S. G., P. B.$^{\rm a}$ \\
February -- July 2012 & Gemini-North+GMOS-N &  1200   & 3610 -- 5000  & Service \\
February -- July 2012 & Gemini-South+GMOS-S &  1200   & 3610 -- 5000  & Service \\
2012-May-25 -- 2012-May-27 & CAHA-3.5m+TWIN & 4000  &  3460 -- 5630  & C. H. \\
2012-Jul-09 -- 2012-Jul-12 & ING-WHT+ISIS &   4000  &  3440 -- 5270  & V. S. \\
2012-Oct-21 -- 2012-Oct-24 & SOAR+Goodman &  7000   & 3500 -- 6160  & B. B. \\
2012-Dec-14 -- 2012-Dec-18 & ING-WHT+ISIS &  4000  &  3440 -- 5270  & T. K., A. F.$^{\rm a}$ \\
2013-Jun-02 -- 2012-Jun-05 & ING-WHT+ISIS &  4000  &  3440 -- 5270  & C. H. \\
2013-Aug-11 -- 2013-Aug-13 & ING-WHT+ISIS & 4000  &  3440 -- 5270  &  T. K., M. S.$^{\rm a}$ \\ 
2014-Feb-01 -- 2014-Feb-05 & ESO-NTT+EFOSC2 & 2200  &  4450 -- 5110  & S. G., F. N.$^{\rm a}$ \\
\hline
\end{tabular}
\tablefoot{
$^{\rm a}$ see notes in the acknowledgements}
\end{center}
\end{table*}

\begin{table*}[t!]
{\small \caption{Derived orbital parameters.} 
\label{tab:orbits}
\begin{center}
\begin{tabular}{lllrrrrr} \hline\hline
\noalign{\smallskip}
Object & $T_{0}$ & $P$ &  \multicolumn{1}{c}{$\gamma$} & \multicolumn{1}{c}{$K$} & \multicolumn{1}{c}{$e_{\rm norm}$} & $\log{p_{\rm false}}[10\%]$ & $\log{p_{\rm false}}[1\%]$ \\
 & [$-$2\,450\,000] & [d] & \multicolumn{1}{c}{[${\rm km\,s^{-1}}$]} & \multicolumn{1}{c}{[${\rm km\,s^{-1}}$]} & \multicolumn{1}{c}{[${\rm km\,s^{-1}}$]} & &  \\ 
\hline
\noalign{\smallskip}
J08300$+$47515 & $6279.6067\pm0.0004$ & $0.14780\pm0.00007$ & $49.9\pm0.9$ & $77.0\pm1.7$ & $4.0$ & $<-4.0$ & $<-4.0$ \\
J08233$+$11364 & $6278.5729\pm0.0007$ & $0.20707\pm0.00002$ & $135.1\pm2.0$ & $169.4\pm2.5$ & $7.0$ & $<-4.0$ & $<-4.0$ \\
J10215$+$30101 & $6277.819\pm0.003$ & $0.2966\pm0.0001$ & $-28.4\pm4.8$ & $114.5\pm5.2$ & $6.4$ & $-1.9$ & $-1.9$ \\
J09510$+$03475 & $6693.666\pm0.003$ & $0.4159\pm0.0007$ & $111.1\pm2.5$ & $84.4\pm4.2$ & $2.8$ & $-2.0$ & $-0.8$ \\
J15222$-$01301 & $6516.632\pm0.005$ & $0.67162\pm0.00003$ & $-79.5\pm2.7$ & $80.1\pm3.5$ & $-$ & $-1.2$ & $-1.2$ \\
J15082$-$49405 & $6518.395\pm0.02$ & $0.967164\pm0.000009$ & $-60.0\pm10.7$ & $93.6\pm5.8$ & $6.0$ & $-0.9$ & $-0.9$ \\
J11324$-$06365 & $4583.06\pm0.01$ & $1.06\pm0.02$ & $8.3\pm2.2$ & $41.1\pm4.0$ & $8.6$ & $-1.1$ & $-0.2$ \\
J01185$-$00254 & $5882.000\pm0.008$ & $1.30\pm0.02$ & $37.7\pm1.8$ & $54.8\pm2.9$ & $6.9$ & $<-4.0$ & $-0.4$ \\
J13463$+$28172 & $6517.99\pm0.01$ & $1.96\pm0.03$ & $1.2\pm1.2$ & $85.6\pm3.4$ & $-$ & $-0.4$ & $-0.3$ \\
J18324$-$63091 & $6119.58\pm0.03$ & $5.4\pm0.2$ & $-32.5\pm2.1$ & $62.1\pm3.3$ & $2.7$ & $-0.9$ & $-0.1$ \\
J09523$+$62581 & $5210.23\pm0.08$ & $6.98\pm0.04$ & $-35.4\pm3.6$ & $62.5\pm3.4$ & $7.8$ & $-0.7$ & $-0.6$ \\
J03213$+$05384 & $6280.17\pm0.05$ & $7.4327\pm0.0004$ & $-16.7\pm2.1$ & $39.7\pm2.8$ & $4.7$ & $-1.8$ & $-1.8$ \\
\hline 
\end{tabular}
\end{center}}
\end{table*}

Many studies have been performed to identify short-period sdB binaries with periods from a few hours to more than ten days. Up to now, 142 short-period sdB binaries have measured radial velocity curves and mass functions with a peak in the period distribution between 0.5 to 1.0 days (e.g. Morales-Rueda et al. 2003; Geier et al. 2011; Copperwheat et al. 2011\nocite{mor03, gei11b, cop11}). In most cases it is hard to constrain the nature of the companion as most sdB binaries are single-lined systems. From the radial-velocity (RV) curve only lower mass limits can be derived. Most of the companions are either dMs or WDs. Only in a few examples could strong constraints be put on the nature of the companion. 

SdBs with main sequence companions are potential progenitors of detached WD+dM binaries, which possibly further evolve to become cataclysmic variables (CVs). Hence, there should be correlations between those two types of systems, even though a compact WD+dM system can be formed directly and does not necessarily evolve from an sdB+dM binary. Recent studies have increased the number of known WD+dMs significantly and population studies have been carried out (\citealt{neb11} and references therein).

In addition to the short-period sdB binaries, long period sdBs formed via stable Roche lobe overflow have been postulated \citep{han02,han03}. Between one third and half of sdB stars are found to show spectroscopic signature of a main sequence type F/G/K companion and associated infrared excess \citep{ree04}. Because of their long periods, these systems show only small radial velocity variations and radial velocity curves have not been measured for years. Just recently the first systems with orbits of several hundreds days were discovered \citep{dec12,ost12,vos12,vos13}. Eventually, these systems will evolve to become WD+main sequence binaries with periods of hundreds of days and might experience another phase of mass transfer, when the main sequence companion turns into a red giant. 

PG\,1253$+$284, the first triple system containing an sdB with a close companion and a wide main sequence companion was discovered by \citet{heb02}. This system shows an infrared excess caused by the wide component. However, the wide component is not involved in the formation of the sdB. The unseen close companion expelled the envelope of the sdB progenitor.

\begin{figure*}[t!]
\begin{center}
        \resizebox{6.0cm}{!}{\includegraphics{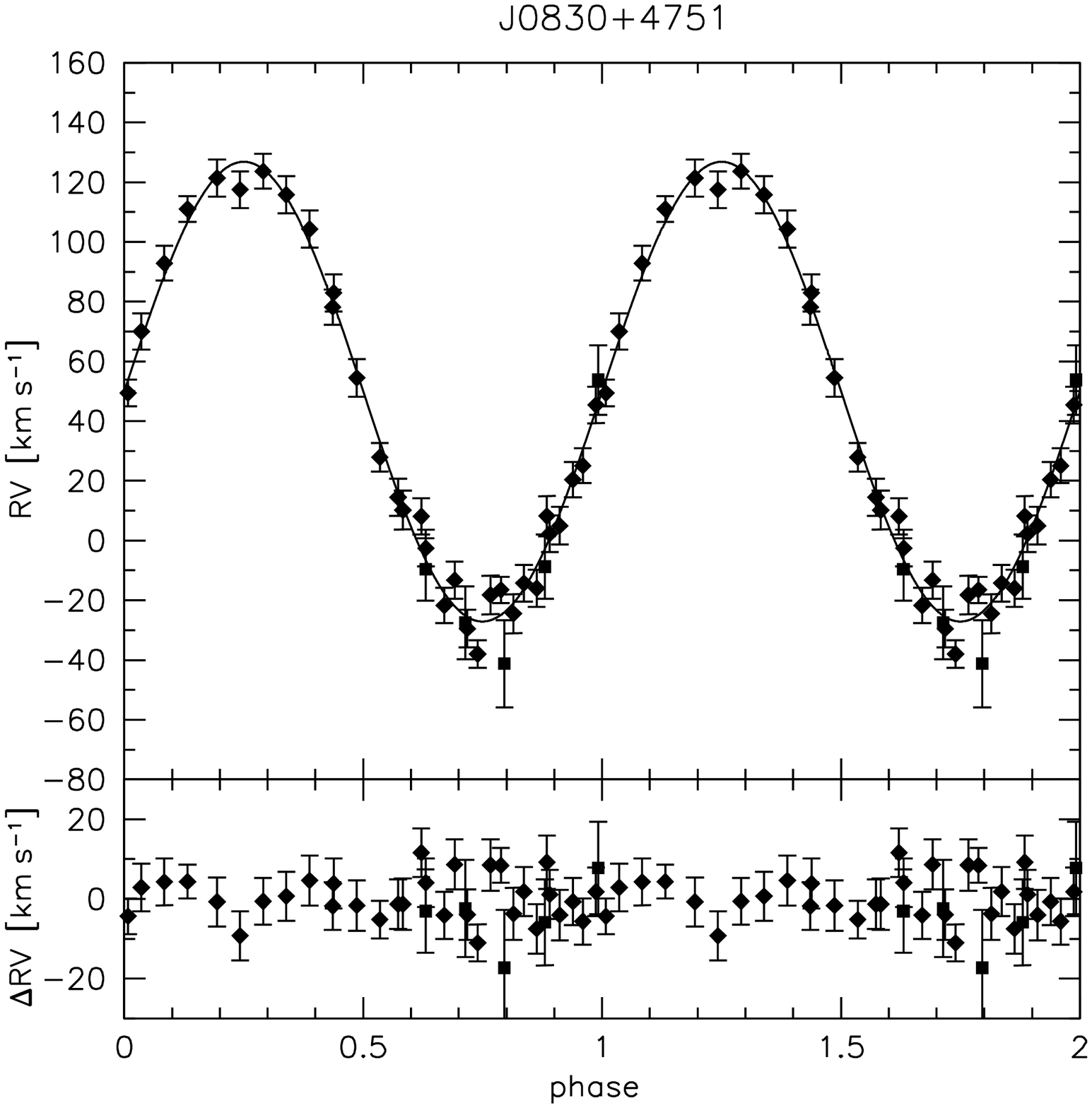}}
         \resizebox{6.0cm}{!}{\includegraphics{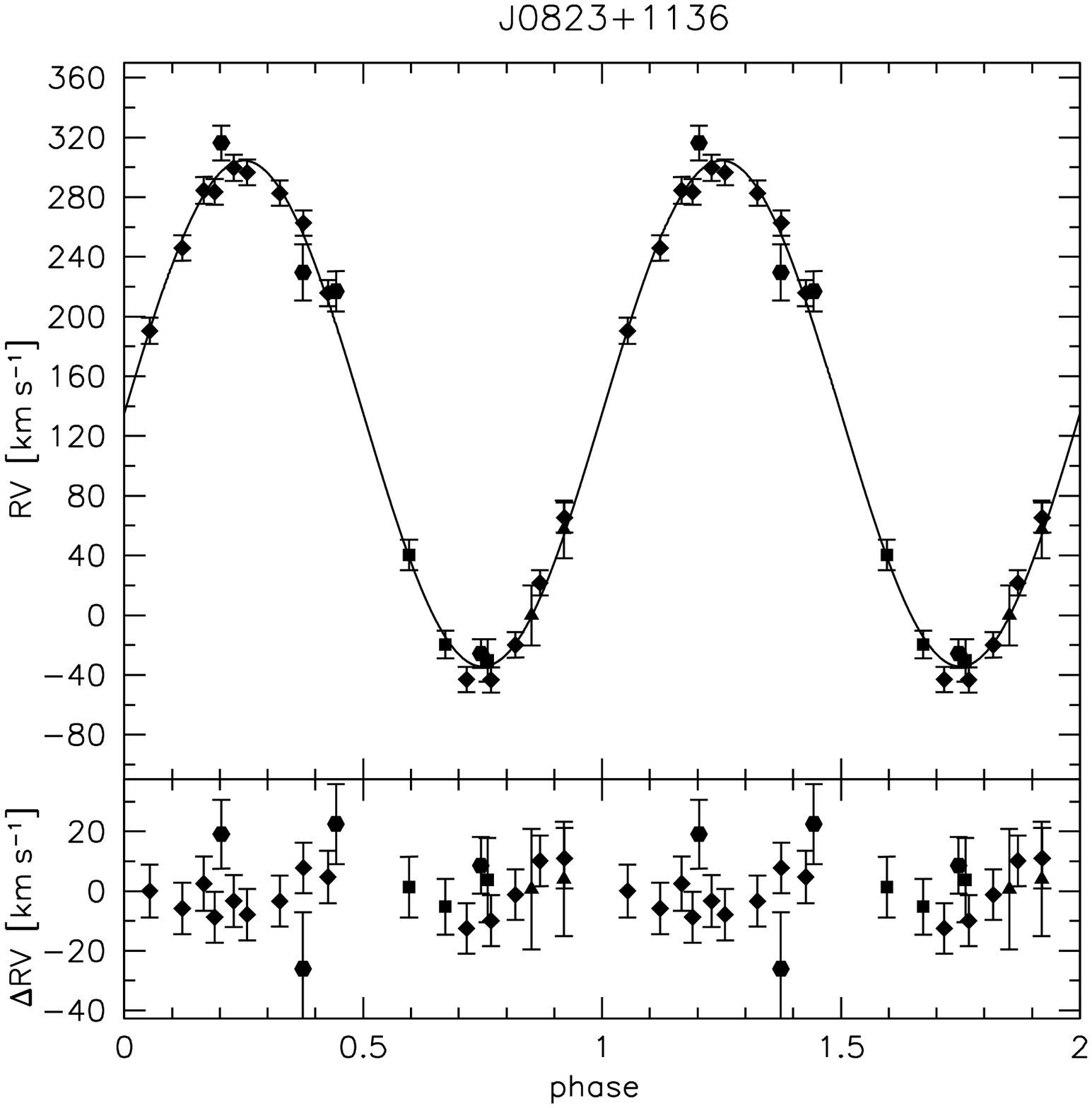}}
          \resizebox{6.0cm}{!}{\includegraphics{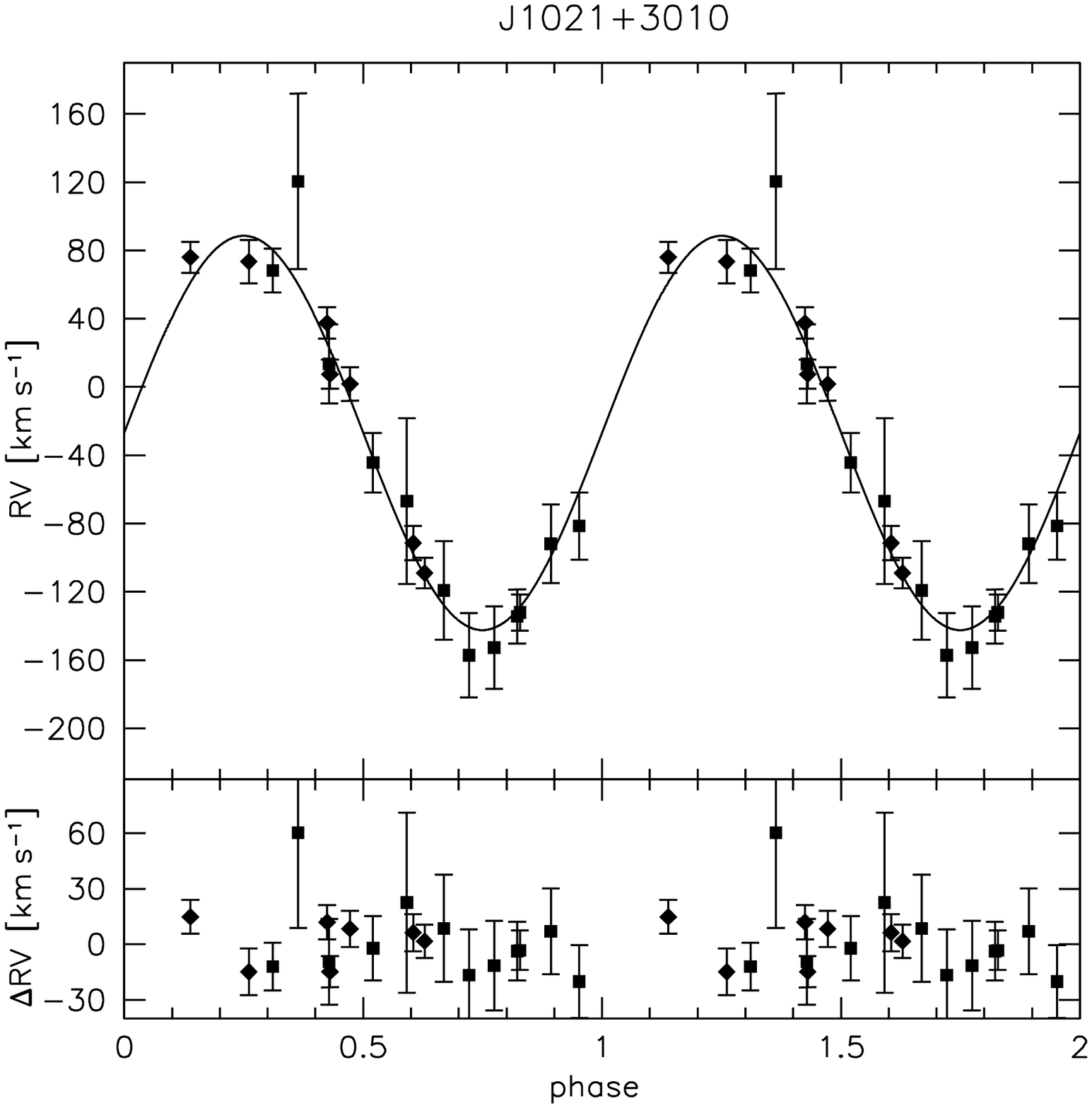}}
          \resizebox{6.0cm}{!}{\includegraphics{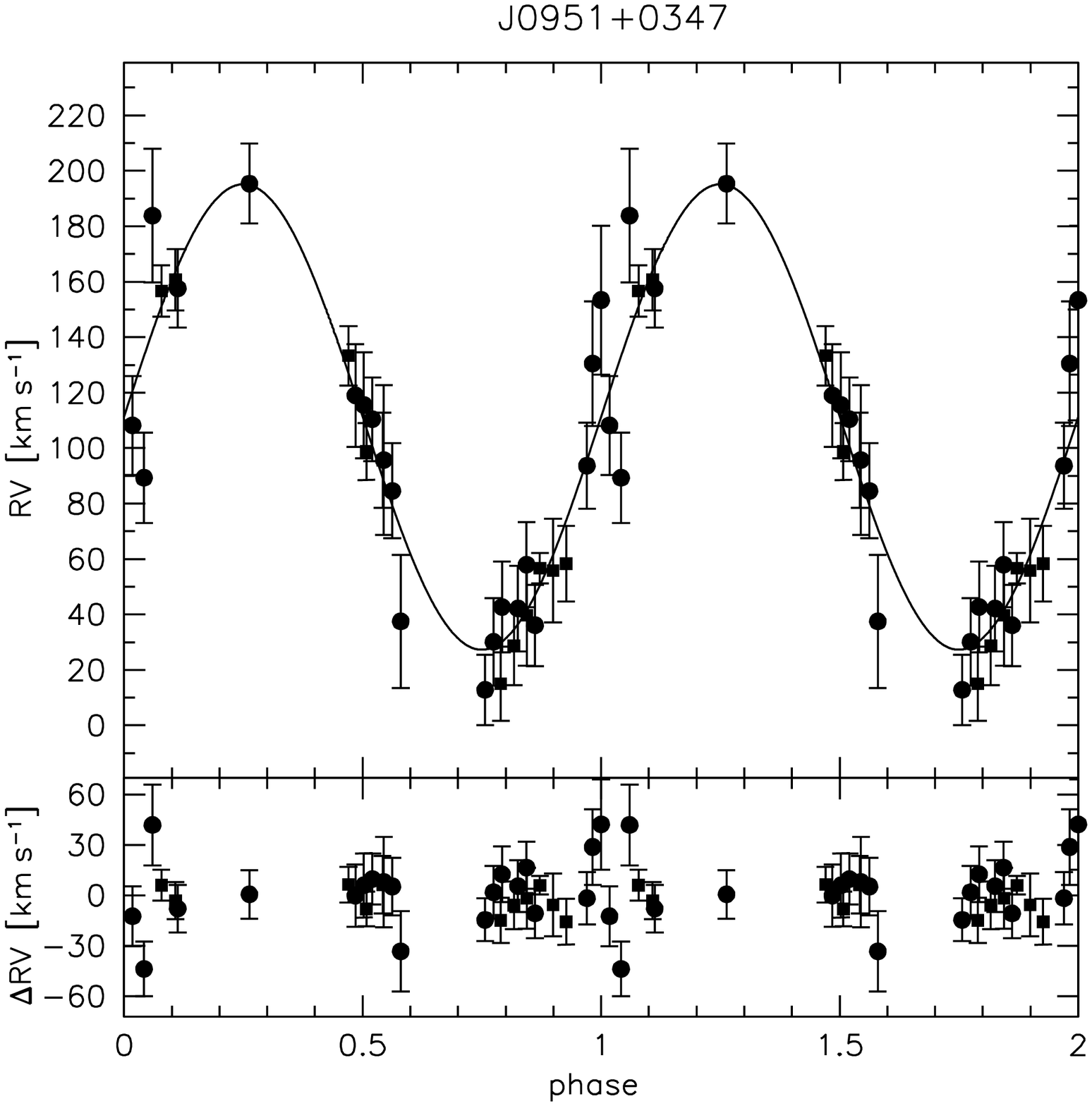}}
          \resizebox{6.0cm}{!}{\includegraphics{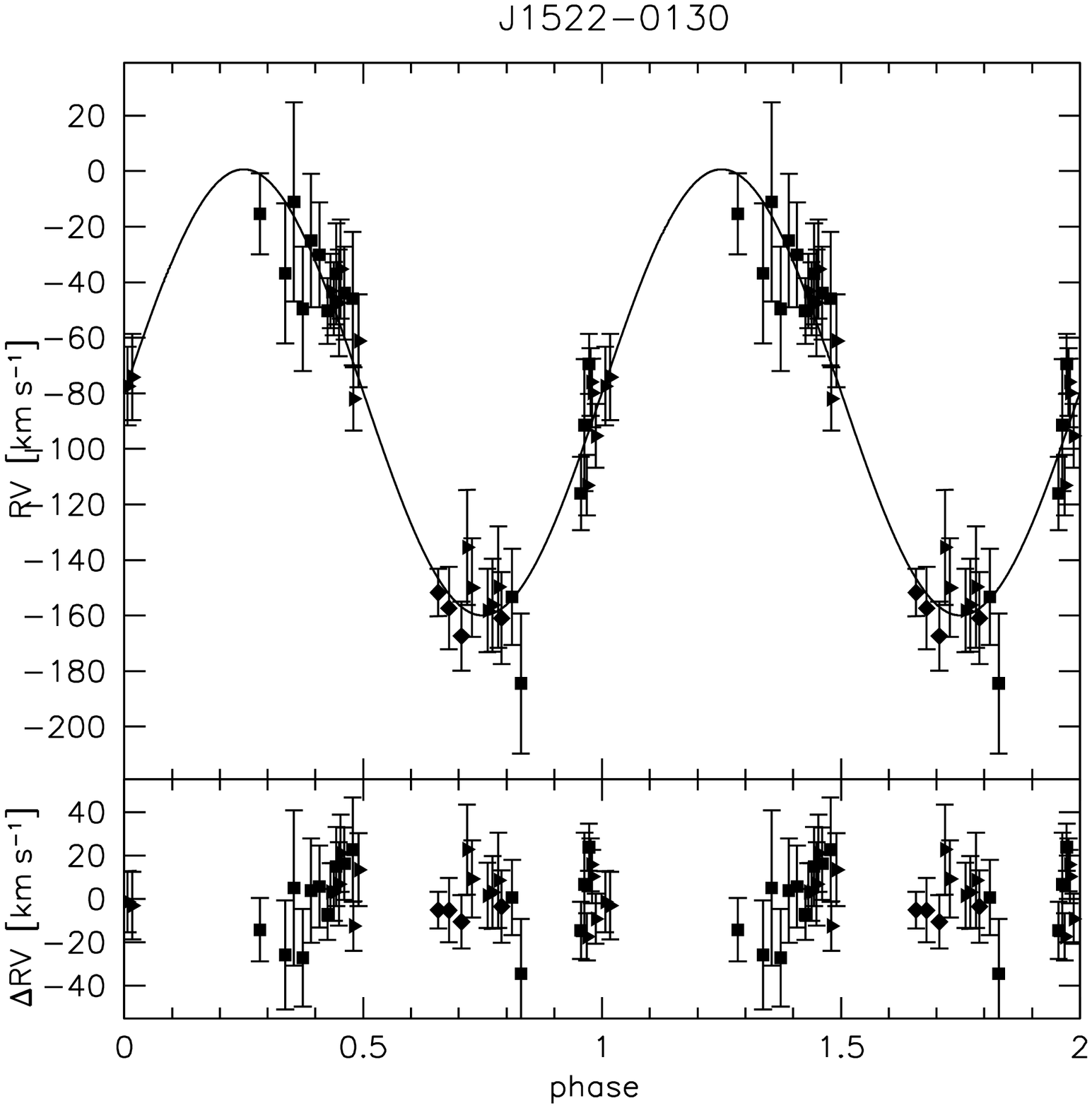}} 
            \resizebox{6.0cm}{!}{\includegraphics{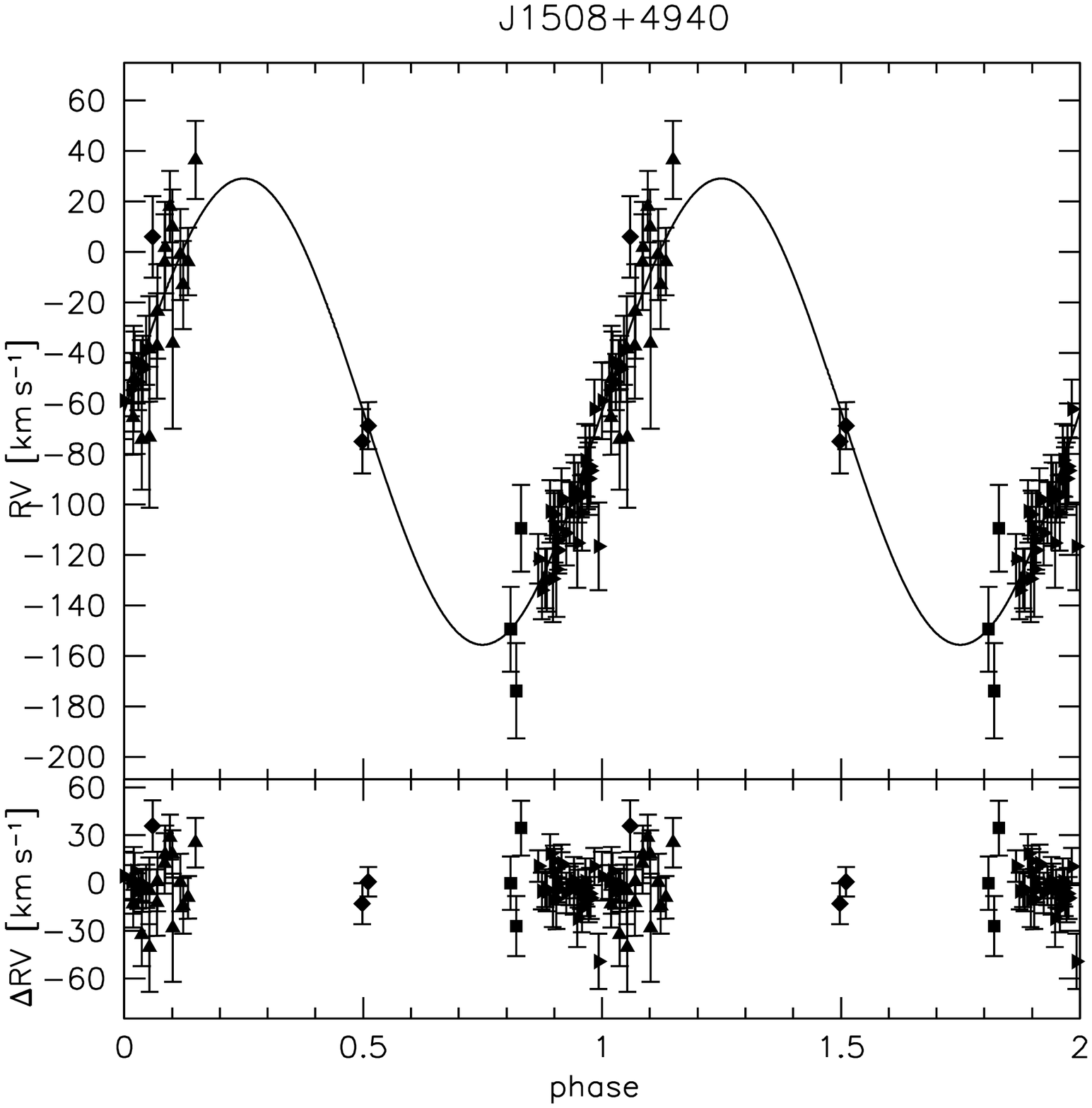}}
         \end{center}
\caption{Radial velocity plotted against orbital phase. The RV data were phase folded with the most likely orbital periods and are plotted twice for better visualisation. The residuals are plotted below. The RVs were measured from spectra obtained with SDSS (squares), CAHA3.5m/TWIN (upward triangles), WHT/ISIS (diamonds), Gemini/GMOS (triangles turned to the right), ESO-NTT/EFOSC2 (circles), and SOAR/Goodman (pentagons).}
\label{fig:rv1}
\end{figure*}       

Here we present orbital solutions for 12 new sdB binaries discovered in the course of the MUCHFUSS project. Sect.\,2 describes the status of the MUCHFUSS project. Sect.\,3 gives an overview of the observations and the data reduction. The derived orbital parameters, as well as the atmospheric parameters of the sdBs, are described in Sec.\,4 and 5. In Sec.\,6, we determine the minimum masses and put constraints on the nature of the unseen companions when no light variations were detected by searching for an infrared excess in a two-colour diagram. In addition, in Sec.\,7 we study the full population of sdBs in close binaries, discuss the period distributions, the companion mass distributions, as well as selection effects of the whole sample of sdB binaries with derived mass function. Sect.\,8 compares the full sample with the samples of known ELM-WD binaries and WD+dM binaries to gain insight in the formation history of hot subdwarfs. Summary and conclusions are given in Sec.\,9.

\section{The MUCHFUSS project} 
The project Massive Unseen Companions to Hot Faint Underluminous Stars from SDSS (MUCHFUSS) aims at finding hot subdwarf stars with massive compact companions like massive white dwarfs ($M > 1.0$\,M$_\odot$), neutron stars or stellar-mass black holes. Hot subdwarf stars were selected from the Sloan Digital Sky Survey by colour and visual inspection of the spectra. Objects with high radial velocity variations were selected as candidates for follow-up spectroscopy to derive the radial velocity curve and the mass function of the system. \footnote{Hot subdwarfs with a large but constant radial velocity were studied in the Hyper-MUCHFUSS project \citep{til11}.}

\citet{gei11a, gei12a} discuss the target selection and the follow-up strategy. A detailed analysis of seven sdB binaries discovered in the course of this project is presented in \citet{gei11b}. In addition, three eclipsing systems were detected, two of which host brown-dwarf companions. These are the first confirmed brown-dwarf companions to sdB stars \citep{gei12,sch14}. One sdB+dM system contains a hybrid pulsator and shows a strong reflection effect \citep{ost13}. Results from a photometric follow-up campaign of the MUCHFUSS targets will be described in detail in Schaffenroth et al. (in prep). During dedicated spectroscopic MUCHFUSS follow-up runs with unfavourable weather conditions, bright sdB binary candidates were observed \citep{gei13,gei14}. A full catalogue of all RV measurements is in preparation as well (Geier et al. in prep).

\section{Observations and data reduction}  
Follow-up medium resolution spectroscopy of 12 sdB binaries (see Table~\ref{tab:systems} for an overview) was obtained using different instruments including the CAHA-3.5m telescope with the TWIN spectrograph, the ESO-NTT telescope with the EFOSC2 spectrograph, the SOAR telescope with the Goodman spectrograph, the Gemini-North/South telescopes with the GMOS-N/S spectrographs and the William Herschel telescope (WHT) with the ISIS spectrograph. Table~\ref{tab:runs} gives an overview of all follow-up runs and the instrumental set-ups. 

All spectra were corrected with an average bias frame constructed from several individual bias frames as well as an average flat field constructed from several flat field lamps. Reduction was done either with the \texttt{MIDAS}\footnote{The ESO-MIDAS system provides general tools for data reduction with emphasis on astronomical applications including imaging and special reduction packages for ESO instrumentation at La Silla and the VLT at Paranal}, \texttt{IRAF}\footnote{IRAF is distributed by the National Optical Astronomy Observatories, which are operated by the Association of Universities for Research in Astronomy, Inc., under cooperative agreement with the National Science Foundation} or \texttt{PAMELA}\footnote{http://www2.warwick.ac.uk/fac/sci/physics/research/astro/people\\/marsh/software} and \texttt{MOLLY}$^{4}$ packages.

\section{Orbital parameters}
The radial velocities (RVs) were measured by fitting a set of mathematical functions matching the individual line shapes to the hydrogen Balmer lines as well as helium lines if present using the FITSB2 routine \citep{nap04}. Polynomials were used to match the continua and a combination of Lorentzian and Gaussian functions to match cores and wings of the lines.

\begin{figure*}[t!]
\begin{center}
     \resizebox{6.0cm}{!}{\includegraphics{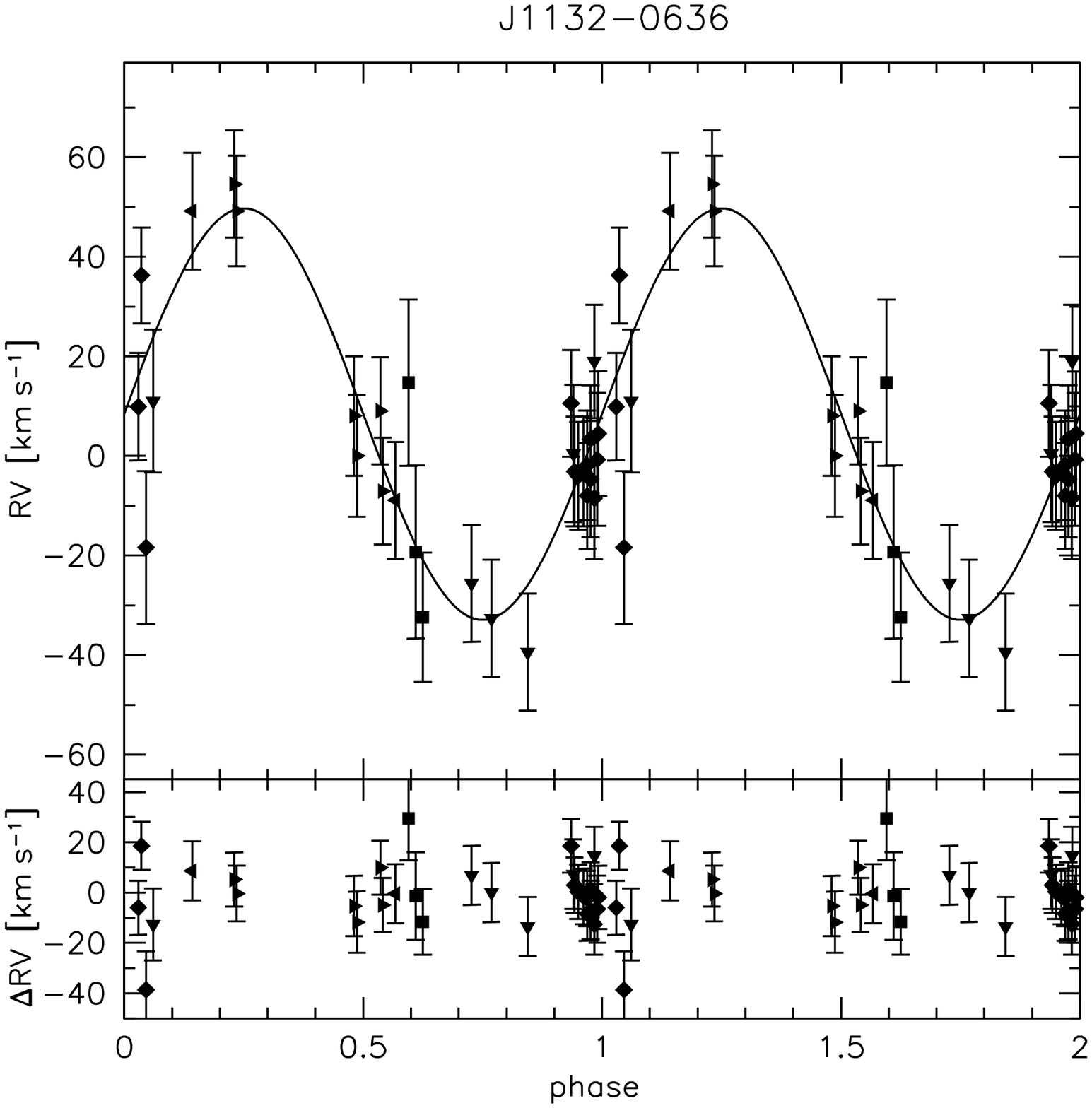}}
         \resizebox{6.0cm}{!}{\includegraphics{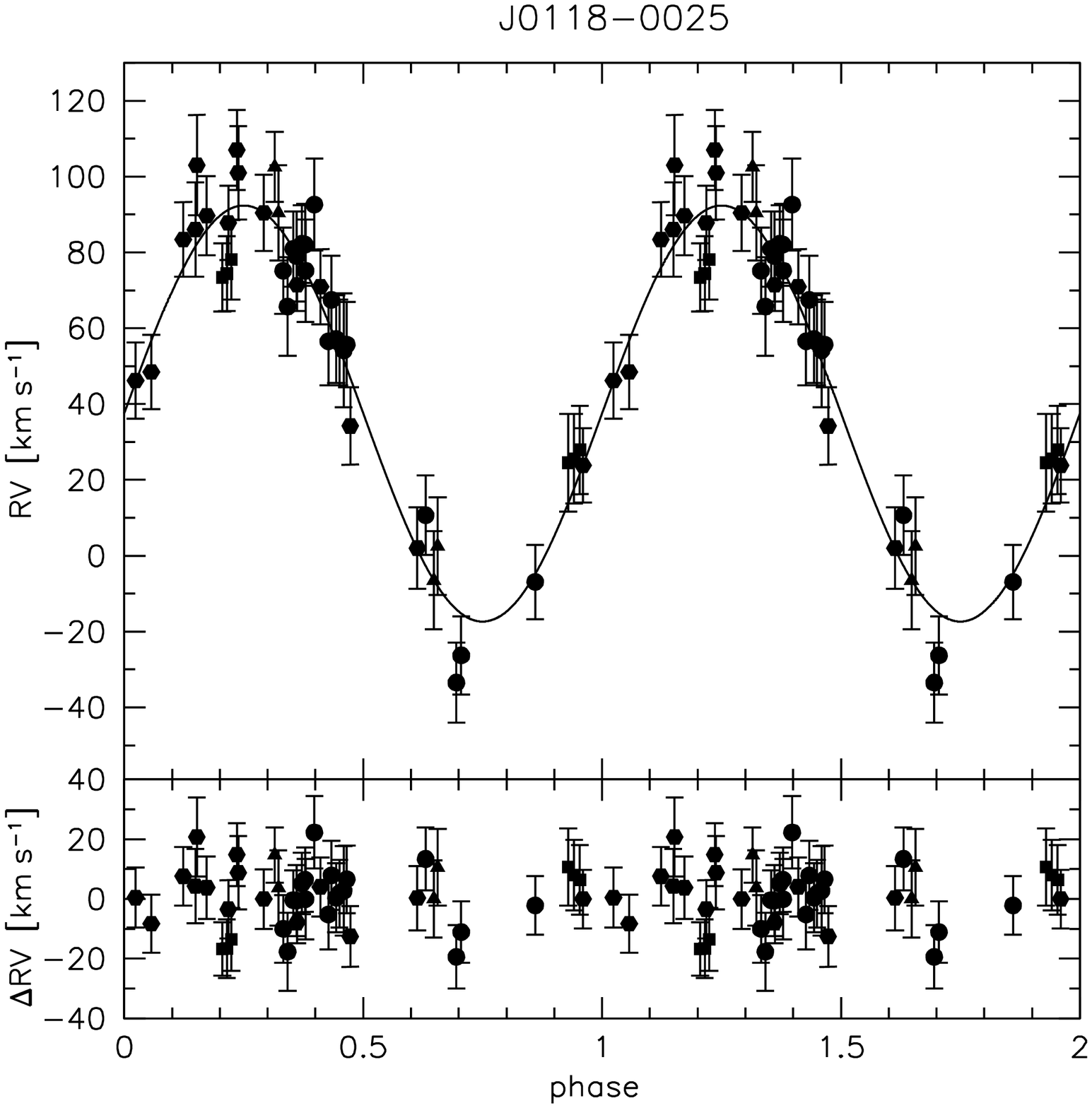}} 
          \resizebox{6.0cm}{!}{\includegraphics{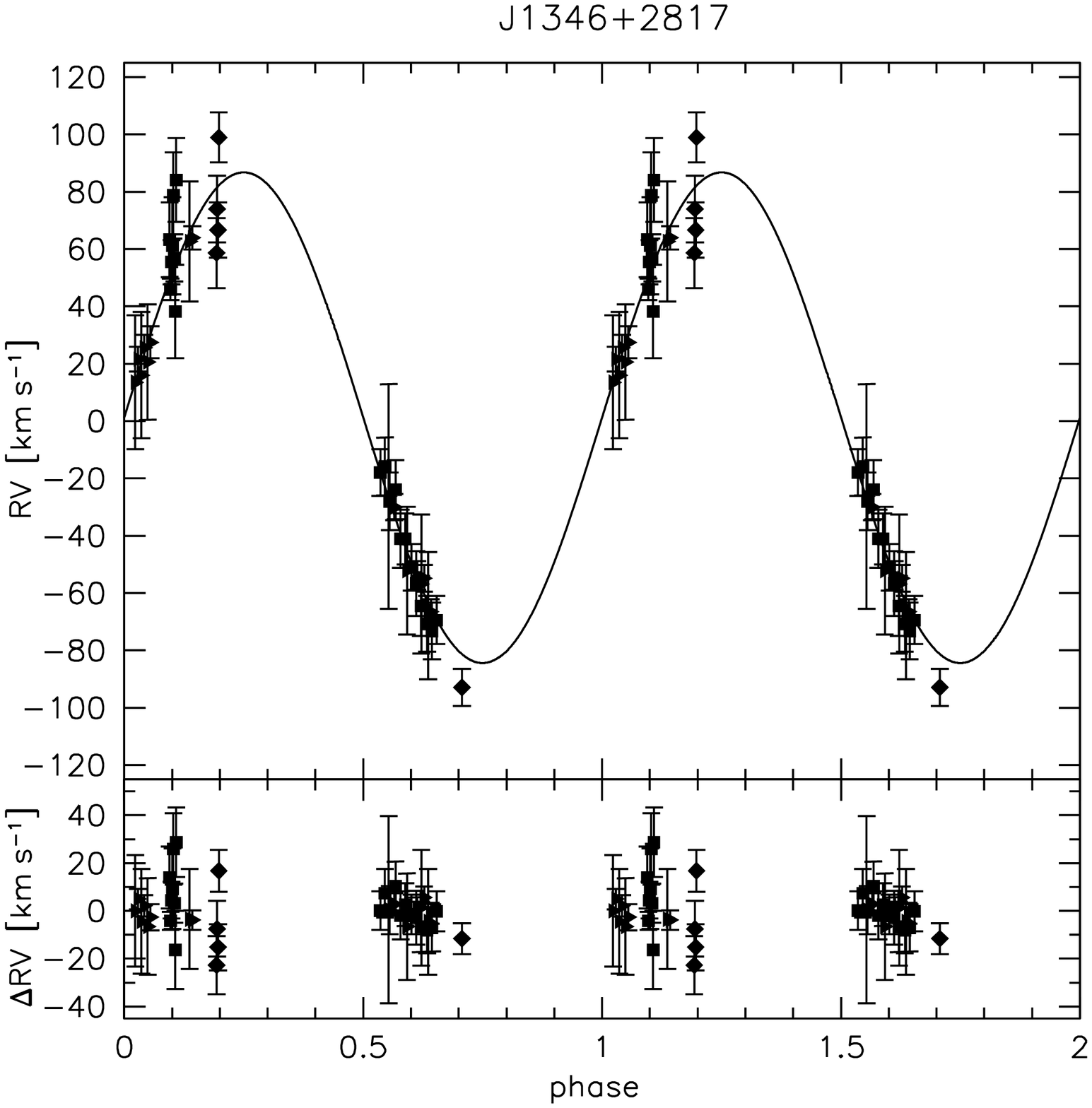}} 
          \resizebox{6.0cm}{!}{\includegraphics{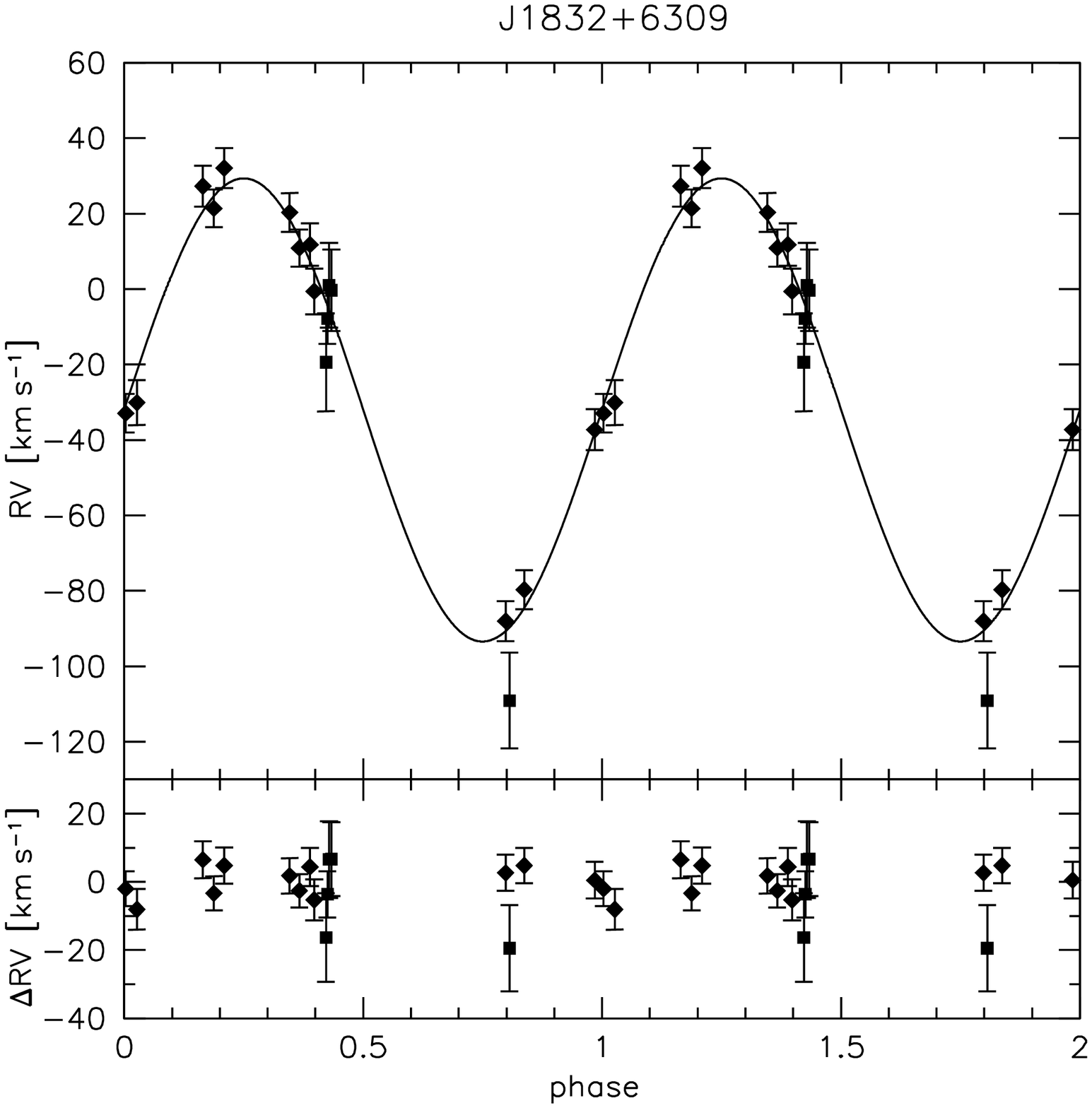}}
          \resizebox{6.0cm}{!}{\includegraphics{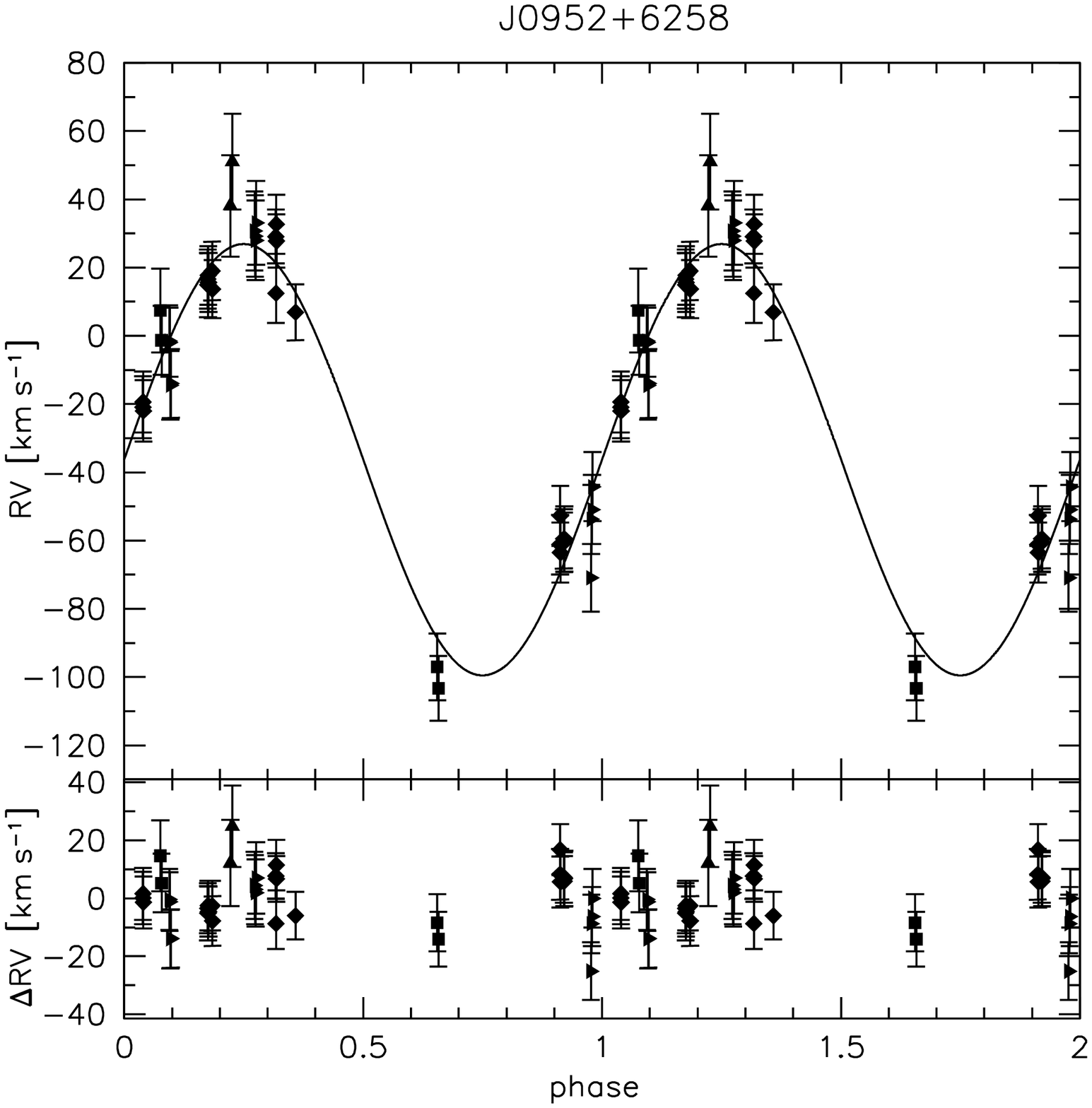}}
        \resizebox{6.0cm}{!}{\includegraphics{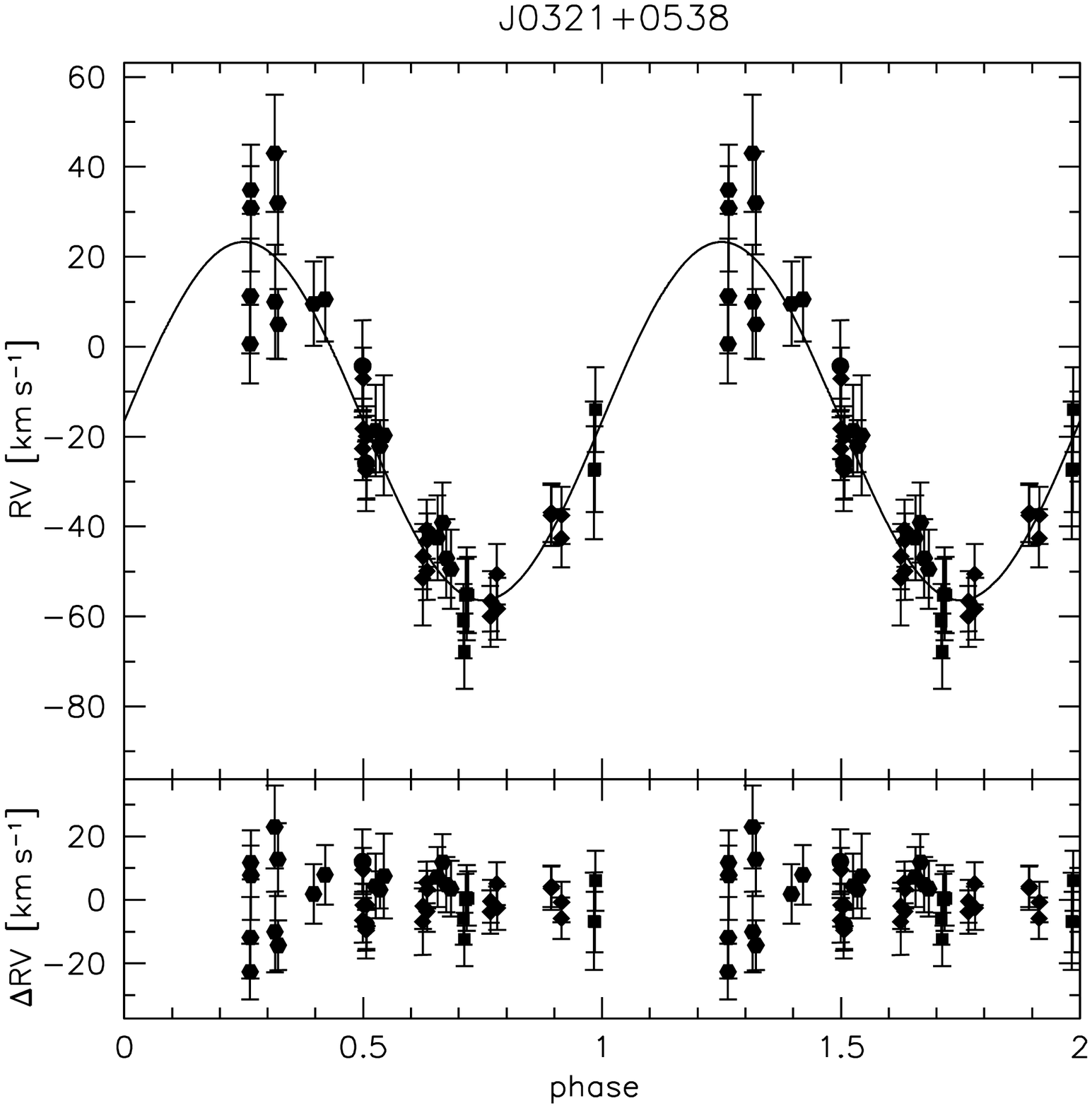}} 
                \end{center}
\caption{Radial velocity plotted against orbital phase (see Fig~\ref{fig:rv1}).}
\label{fig:rv2}
\end{figure*}     

\begin{figure*}[htp!]
\begin{center}
        \resizebox{6.0cm}{!}{\includegraphics{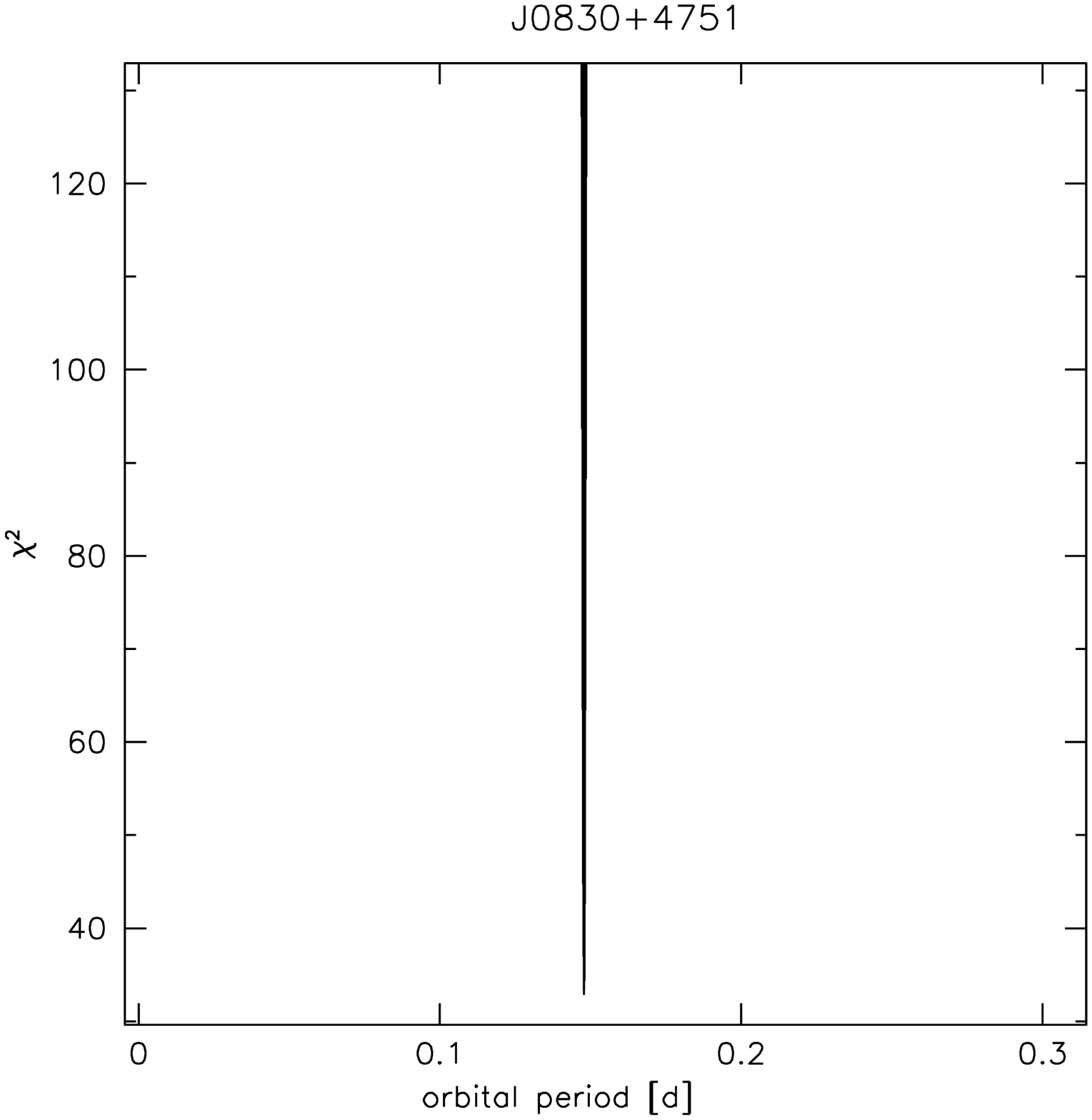}}
         \resizebox{6.0cm}{!}{\includegraphics{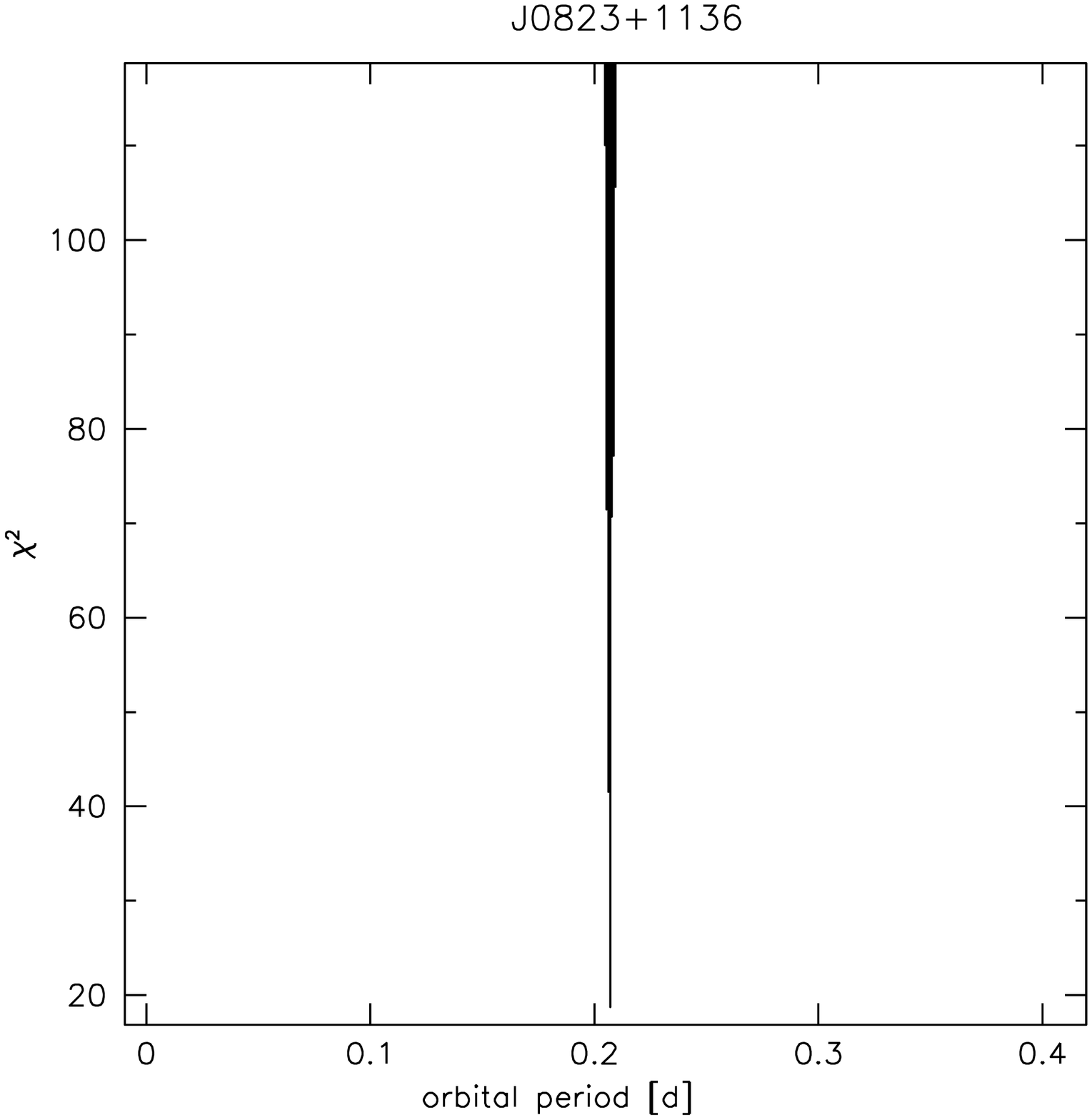}}
          \resizebox{6.0cm}{!}{\includegraphics{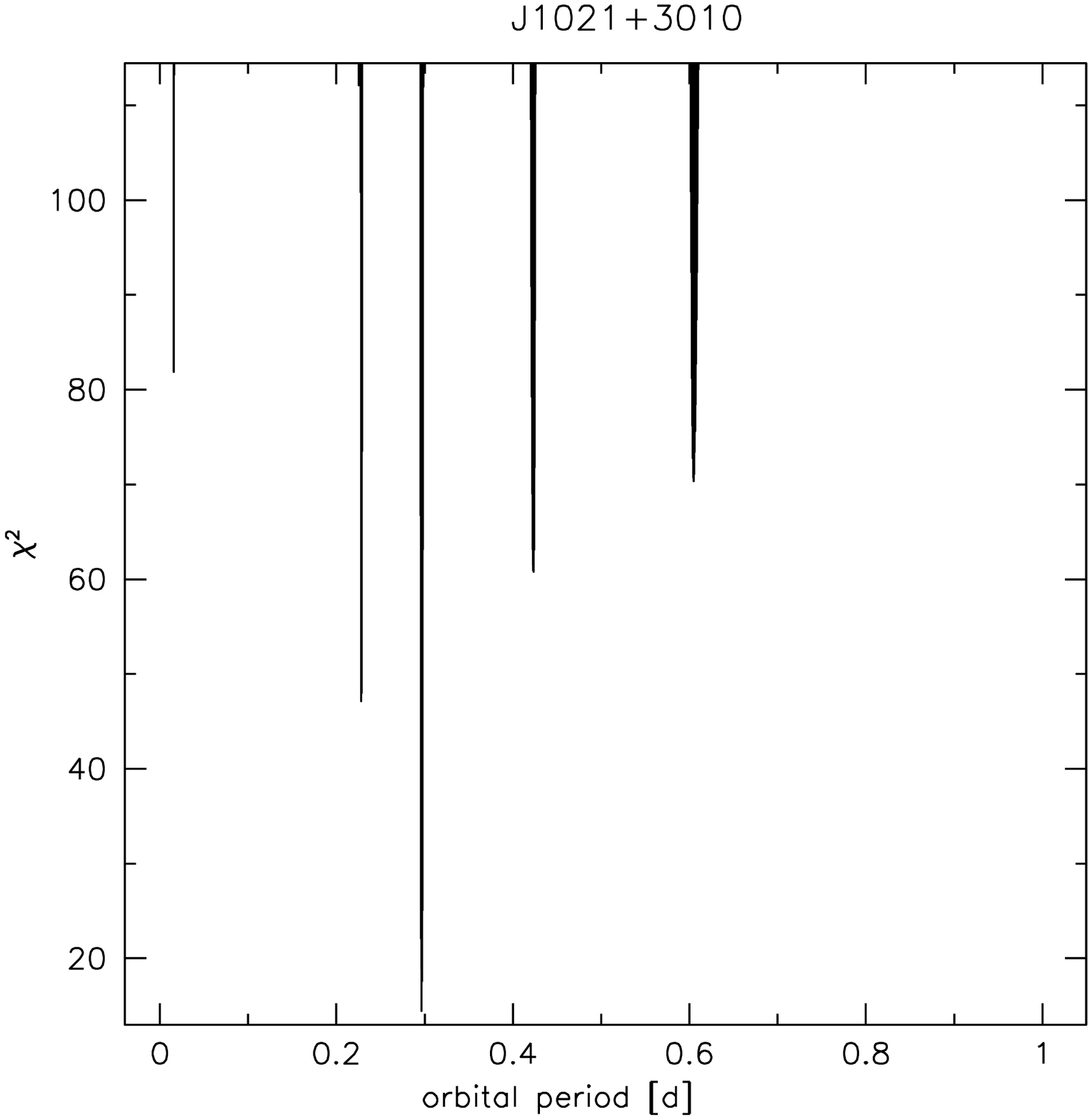}}
           \resizebox{6.0cm}{!}{\includegraphics{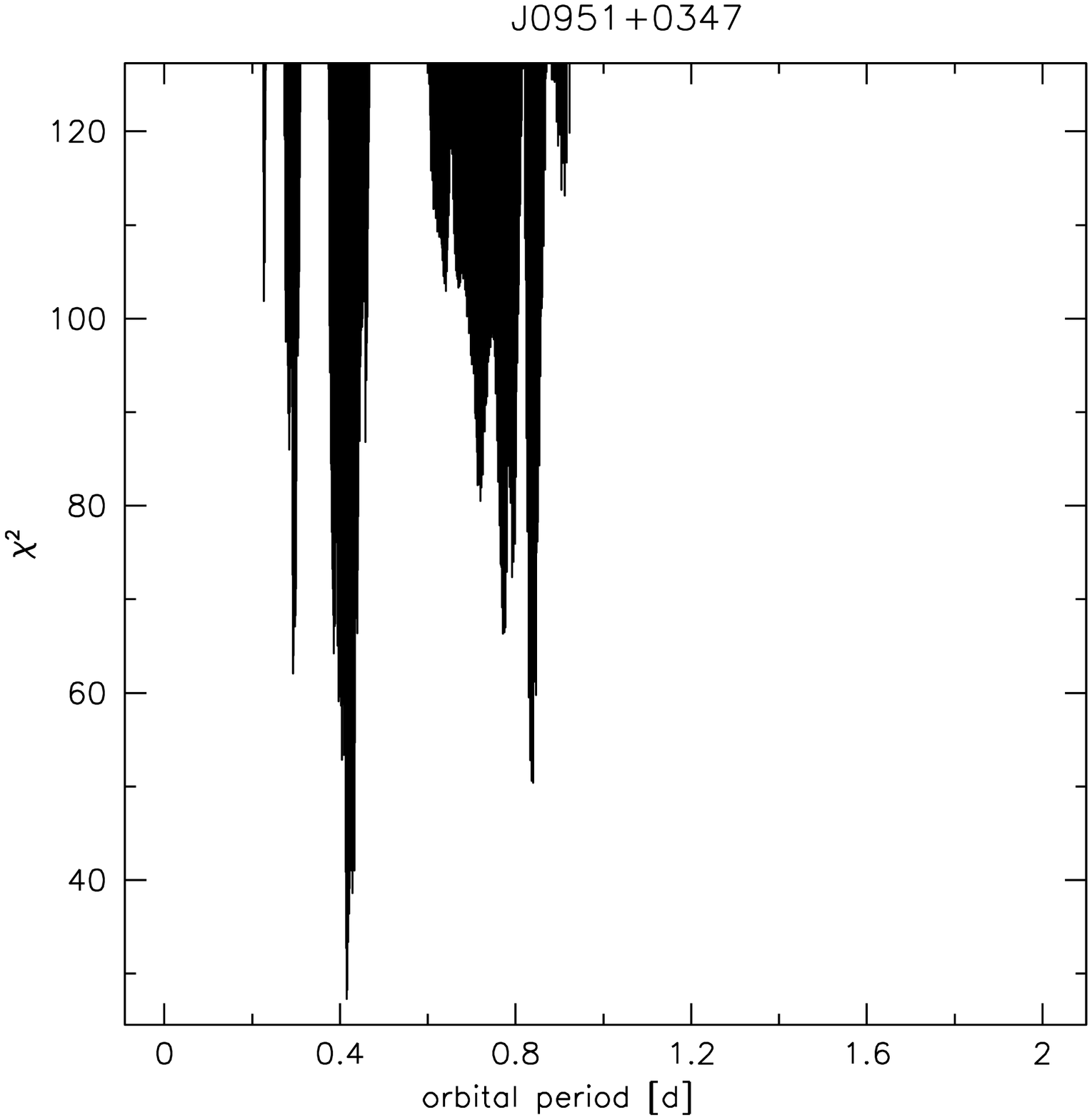}} 
             \resizebox{6.0cm}{!}{\includegraphics{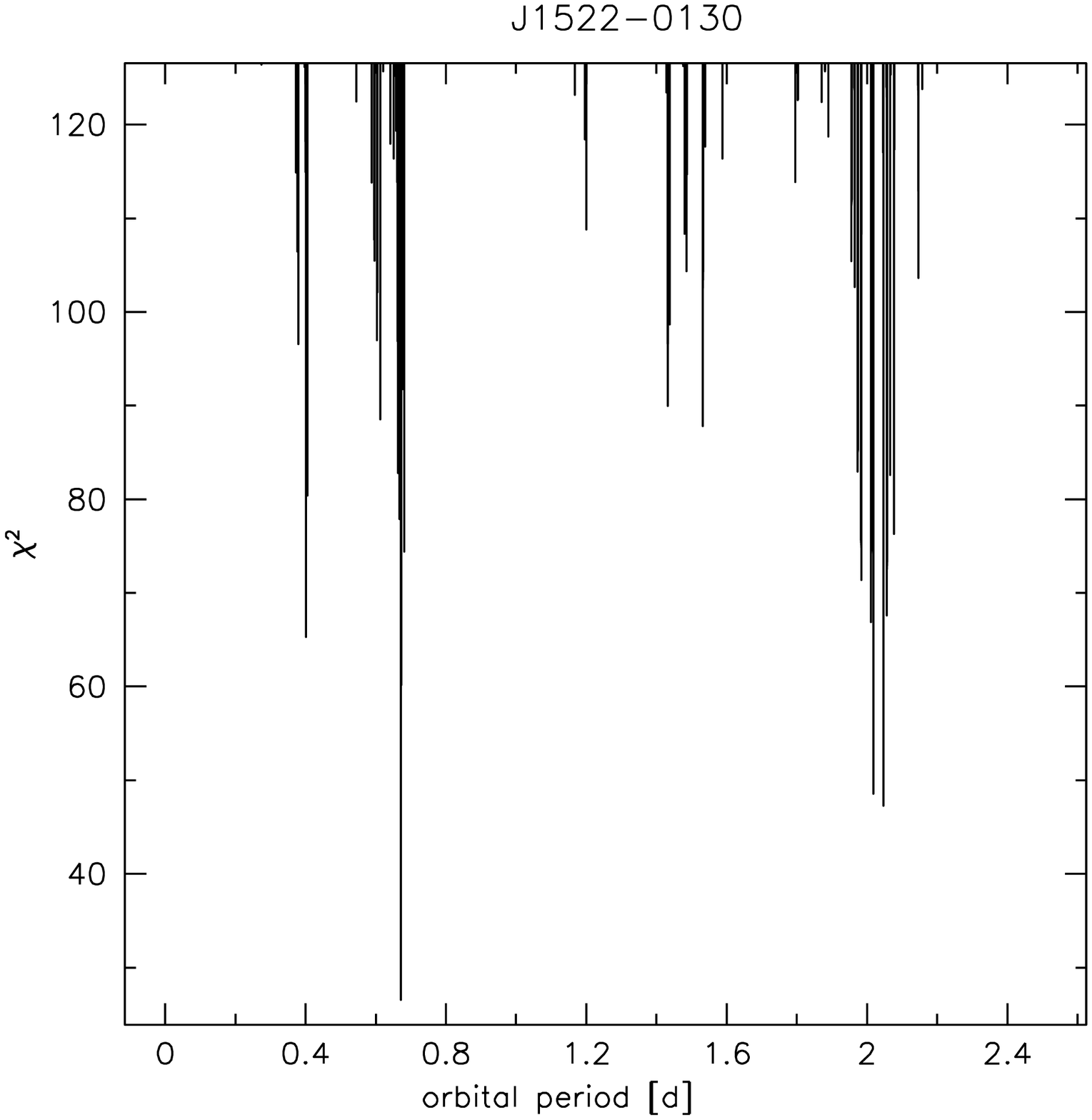}} 
           \resizebox{6.0cm}{!}{\includegraphics{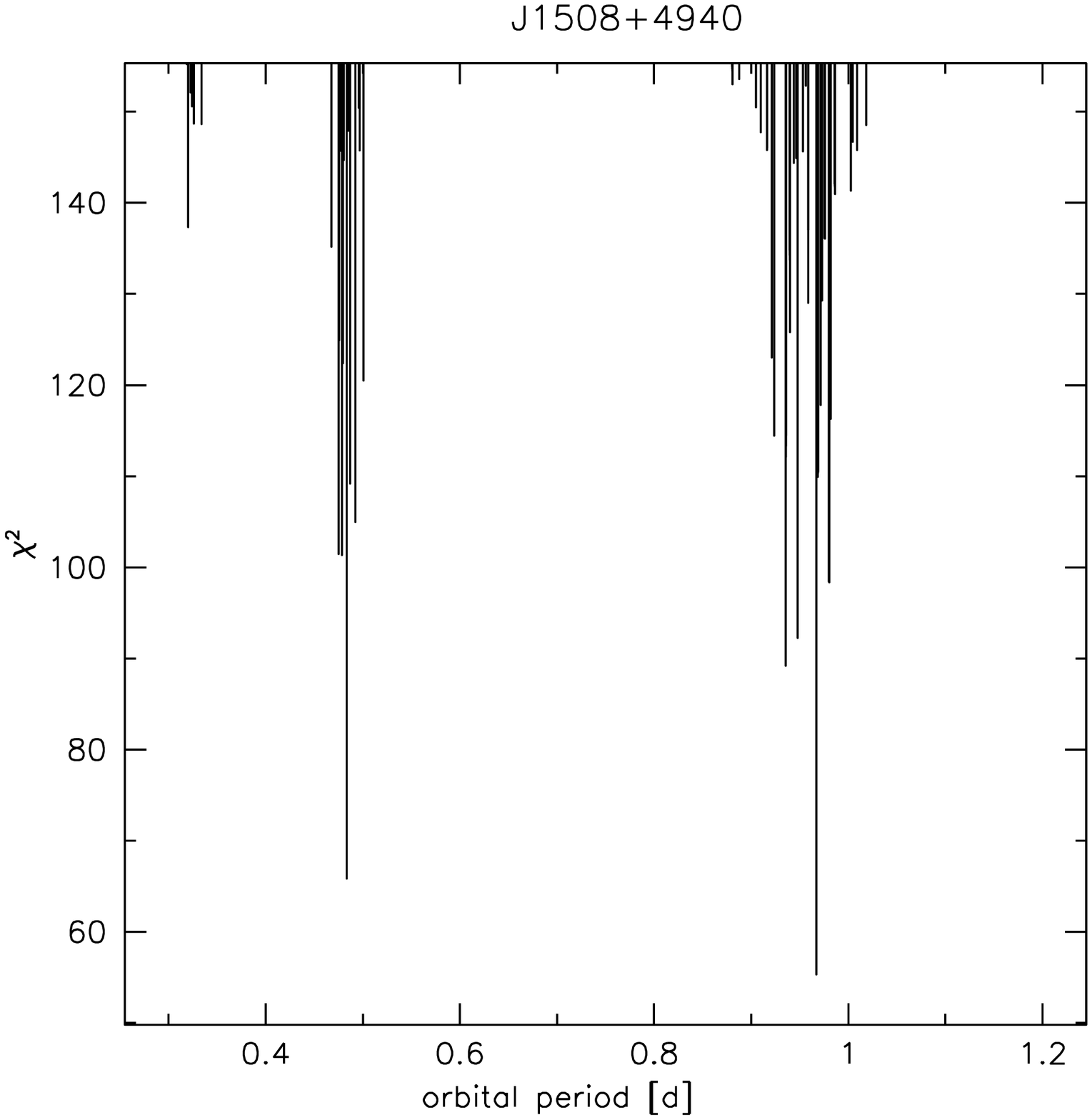}}
           \resizebox{6.0cm}{!}{\includegraphics{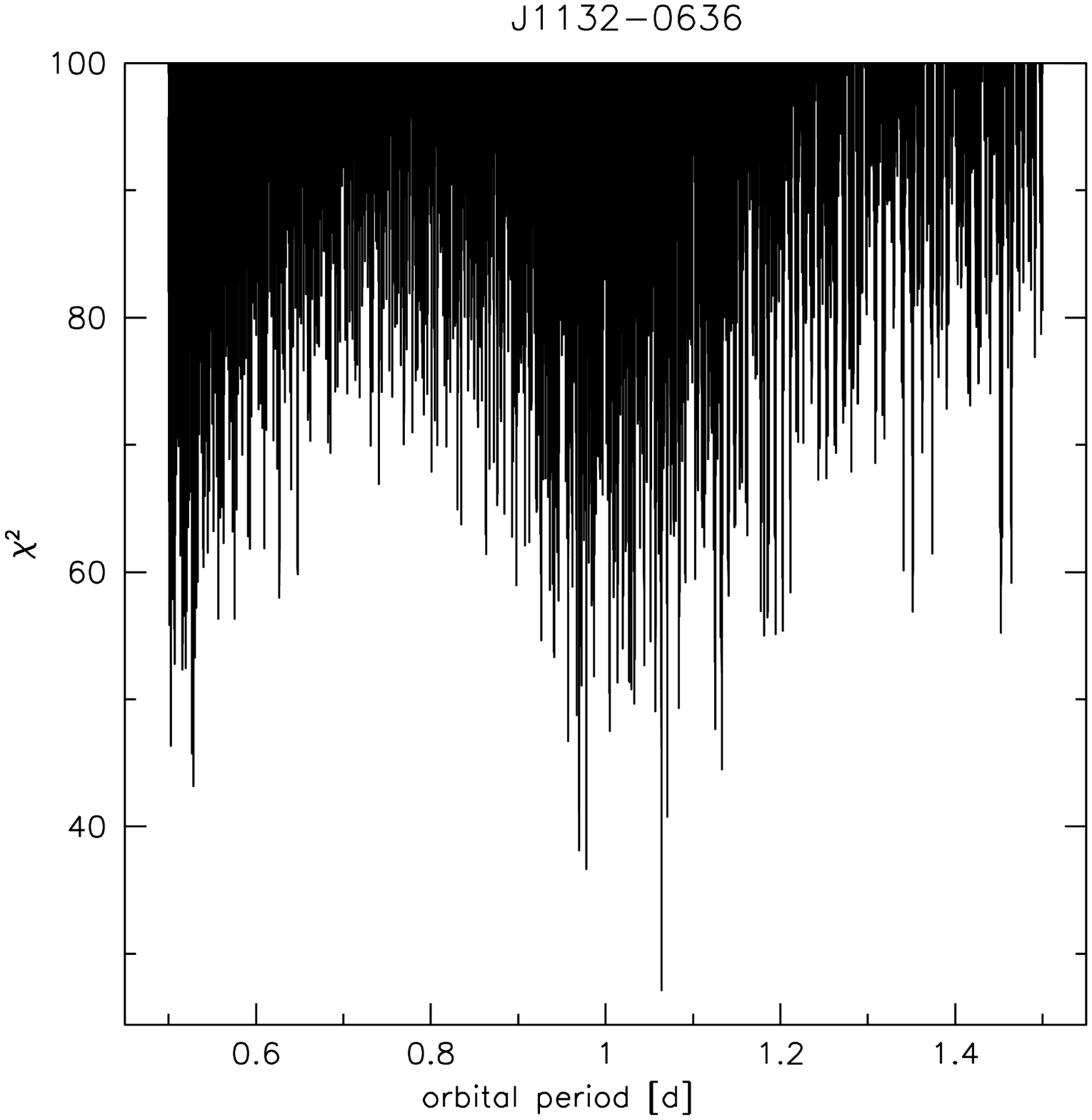}}
          \resizebox{6.0cm}{!}{\includegraphics{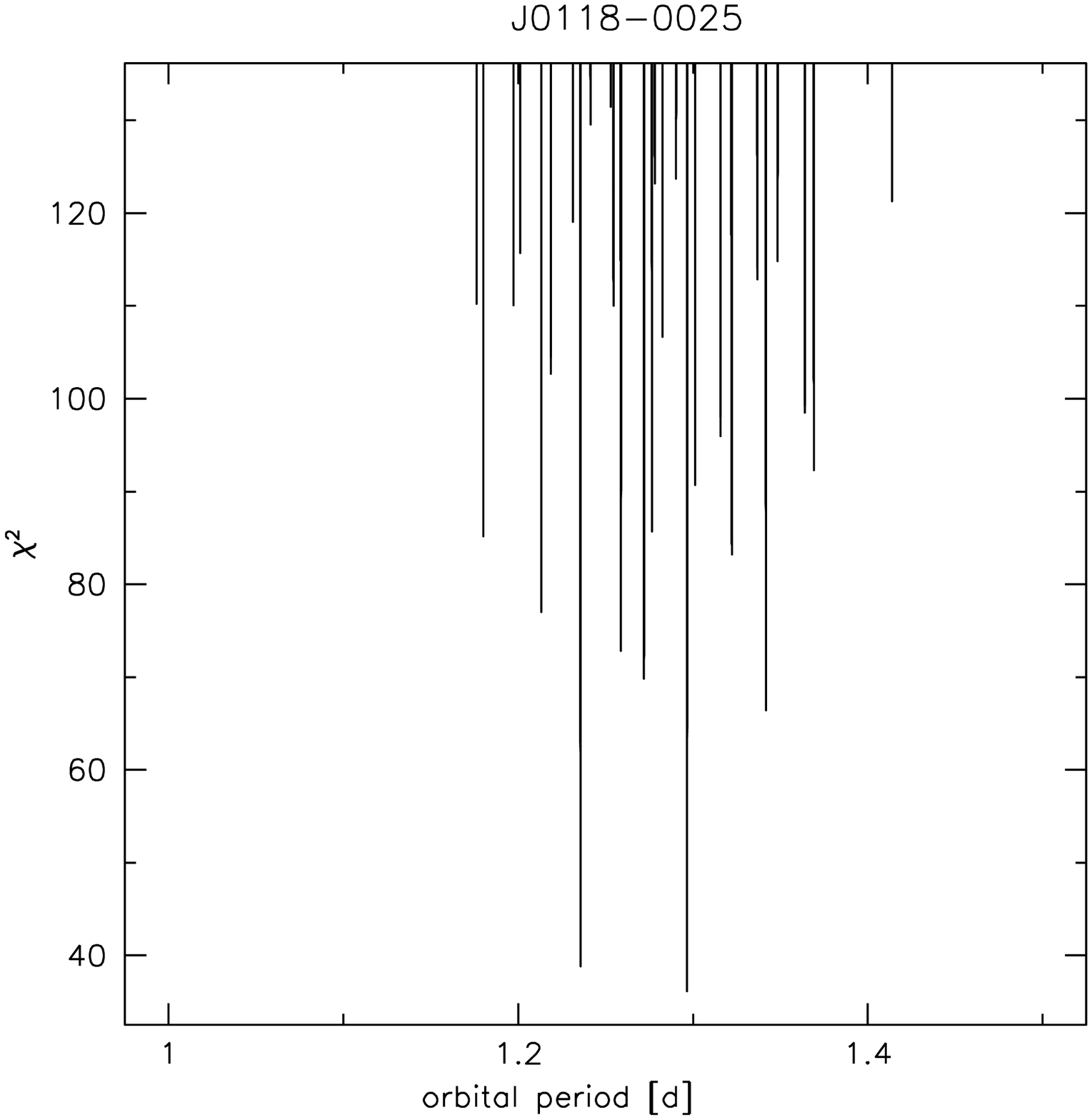}} 
           \resizebox{6.0cm}{!}{\includegraphics{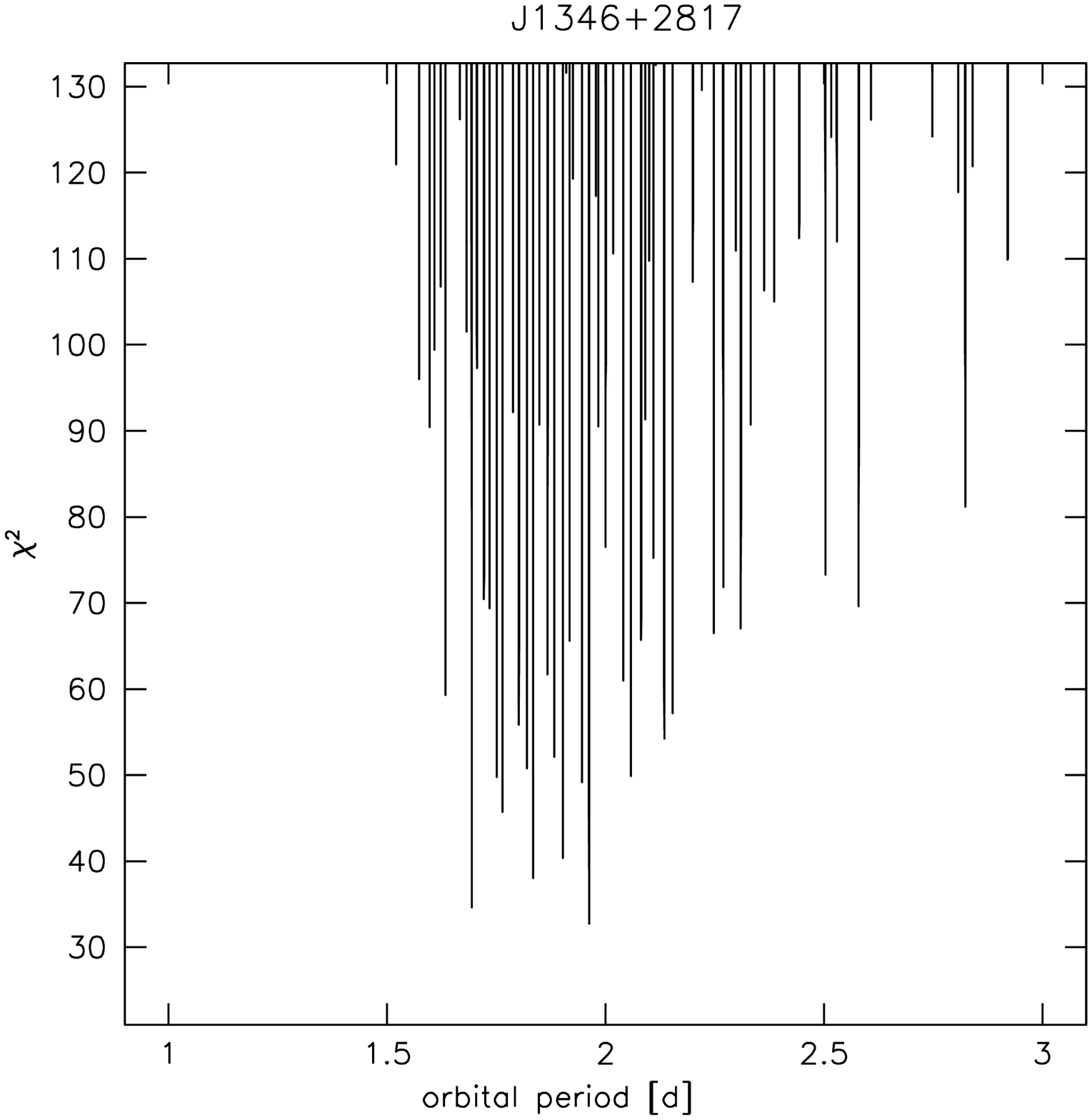}} 
           \resizebox{6.0cm}{!}{\includegraphics{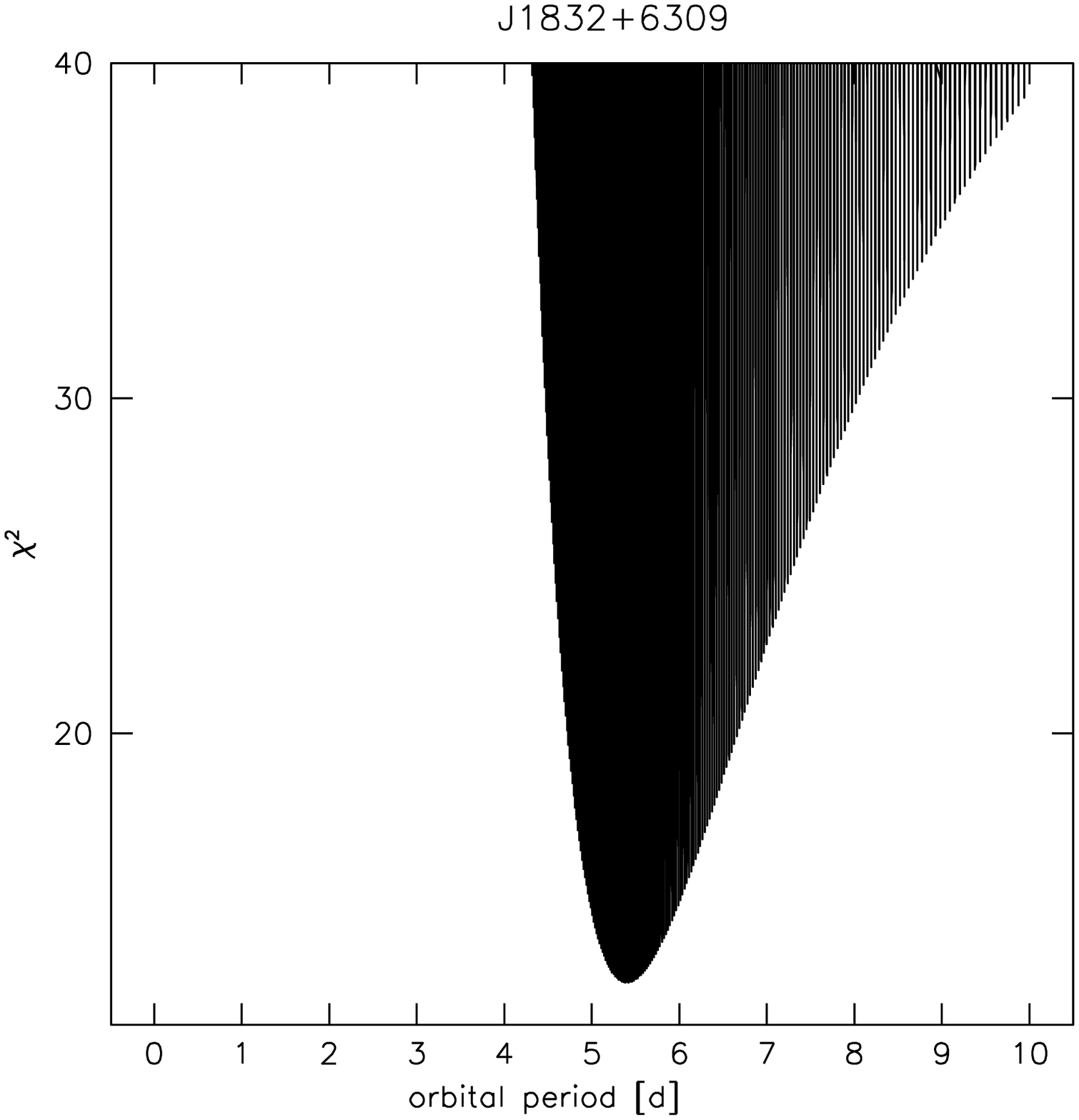}}
           \resizebox{6.0cm}{!}{\includegraphics{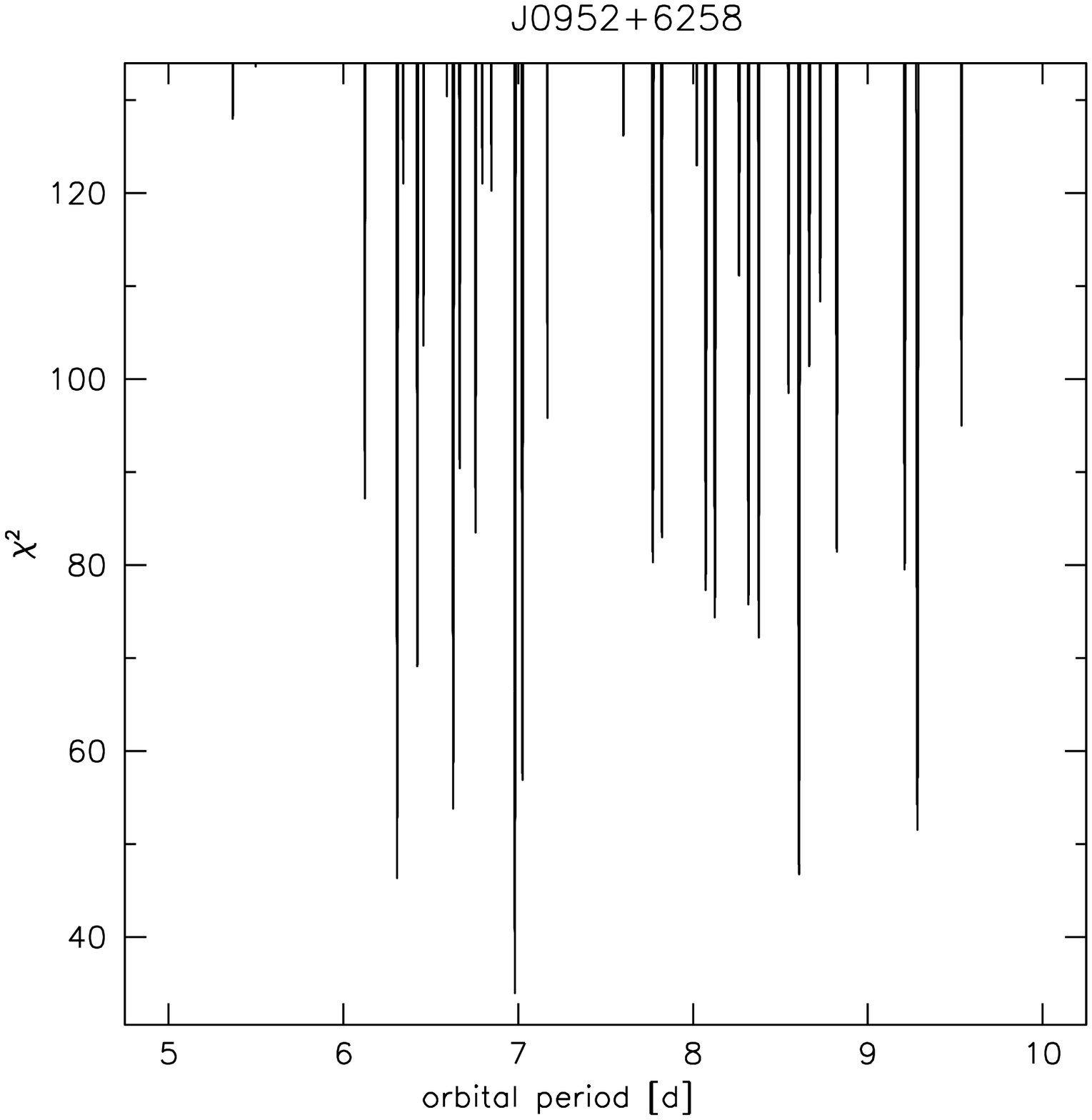}}
          \resizebox{6.0cm}{!}{\includegraphics{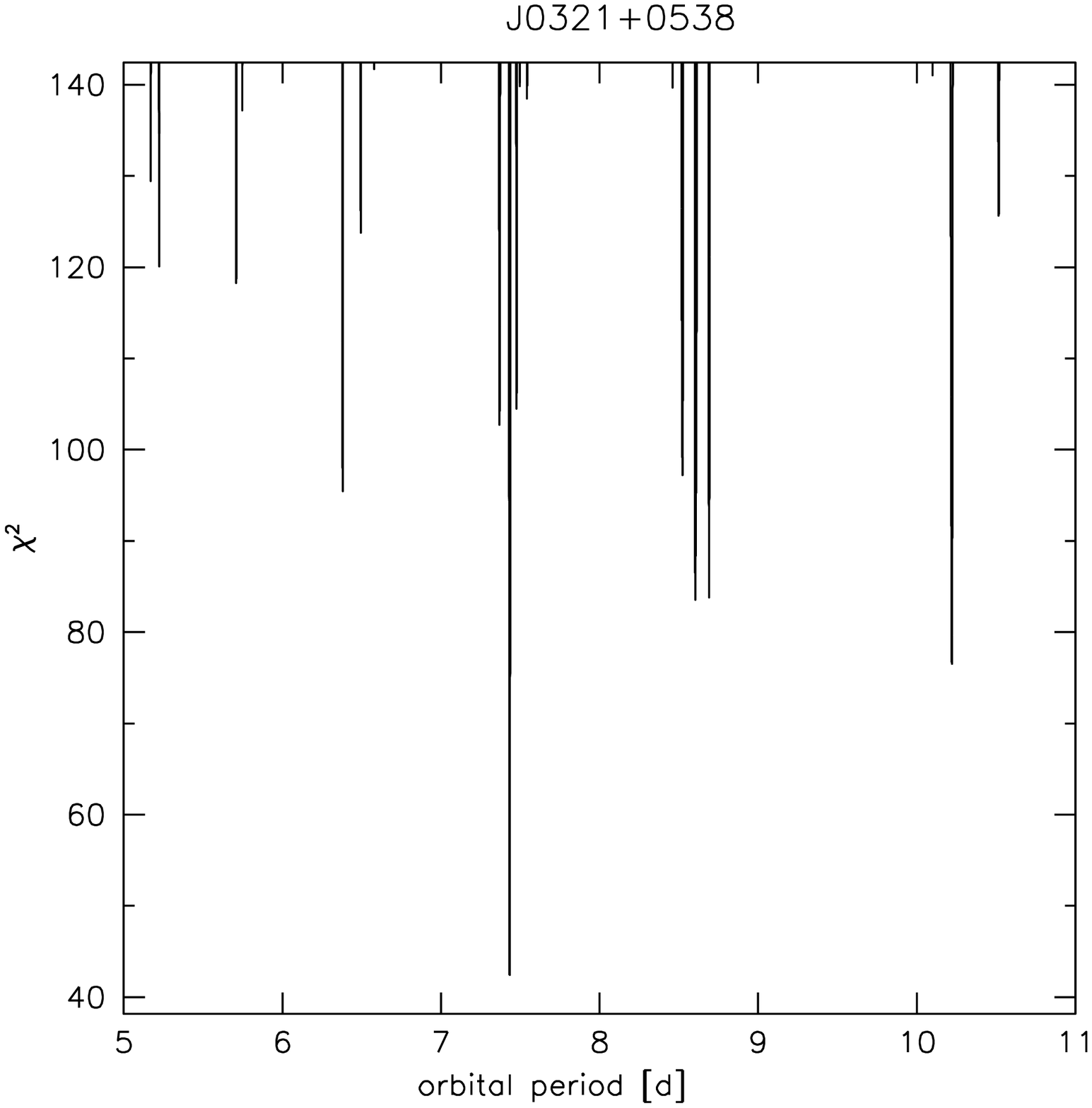}} 
                          \end{center}
\caption{$\chi^{2}$ plotted against orbital period. The lowest peak corresponds to the most likely solution.}
\label{fig:chi}
\end{figure*}  
The orbital parameter (orbital phase $T_{\rm 0}$\footnote{$T_{\rm 0}$ corresponds to the minimum distance of the sdB star from our Solar System}, period $P$, system velocity $\gamma$, and RV-semiamplitude $K$), as well as their uncertainties and associated false-alarm probabilities ($p_{\rm false}[1\%]$, $p_{\rm false}[10\%]$) are determined as described in \citet{gei11b,gei14}. To calculate the significance of the orbital solutions and to estimate contribution of systematic effects to the error budget, we modified the $\chi^{2}$ of the best solution by adding systematic errors $e_{\rm norm}$ in quadrature until the reduced $\chi^{2}$ reached $\sim\,1.0$. The phased RV curves for the best solutions are given in Fig.~\ref{fig:rv1} and \ref{fig:rv2}, the $\chi^{2}$-values plotted against orbital period in Fig.~\ref{fig:chi}. The minimum in $\chi^{2}$ corresponds to the most likely solution. The adopted systematic errors and false-alarm probabilities are given in Table~\ref{tab:orbits}. The probabilities that the adopted orbital periods are correct to within $10\%$ range from $80\%$ to $>99.99\%$. 

\section{Atmospheric parameters}
Atmospheric parameters have been determined by fitting appropriate model spectra to the hydrogen Balmer and helium lines in the way described in \citet{gei07}. For the hydrogen-rich and helium-poor ($\log{y}=\log(n($He$)/n($H$))<-1.0$) sdBs with effective temperatures below $30\,000\,{\rm K}$, a grid of metal line blanketed LTE atmospheres with solar metallicity was used \citep{heb00}. Helium-poor sdBs and sdOBs with temperatures ranging from $30\,000\,{\rm K}$ to $40\,000\,{\rm K}$ were analysed using LTE models with enhanced metal line blanketing \citep{oto06}. Metal-free NLTE models were used for the hydrogen-rich sdO J1132$-$0636 \citep{str07}.

Each spectrum was velocity corrected according to the orbital solution and co-added for the atmospheric fit. To account for systematic shifts introduced by the different instruments, atmospheric parameters for each star were derived separately from spectra taken with each instrument. Weighted means were calculated and adopted as the final solutions (see Table~\ref{tab:atmo}).

\section{The nature of the unseen companion}\label{sec:compnat}
All our objects appear to be single-lined. Therefore, only the binary mass function can be calculated: 
 \begin{equation}
 \label{equation-mass-function}
 f_{\rm m} = \frac{M_{\rm comp}^3 \sin^3i}{(M_{\rm comp} +  M_{\rm sdB})^2} = \frac{P K^3}{2 \pi G}.
 \end{equation}
From spectroscopy the orbital period $P$ and the RV semi-amplitude $K$ can be derived. Hence, the mass of the sdB ($M_{\rm sdB}$) and the companion ($M_{\rm comp}$) as well as the inclination angle $i$ remain free parameters. Assuming a canonical mass for the sdB $M_{\rm sdB}=0.47$\,M$_\odot$ (see Fontaine et al. 2012 \nocite{fon12} and references therein) and an inclination angle $i<90^\circ$, a minimum mass for the companion can be determined. If the derived minimum companion mass is higher than the Chandrasekhar limit, the NS/BH nature of the companion is proven without further constraints under the assumption that the sdB does not have a mass significantly lower than the canonical mass. 

All spectra were checked for contamination by a cool stellar companion. Typically, the Mg\,{\sc i} triplet around $5170\,\AA$ and the Ca\,{\sc ii} triplet around $8650\,\AA$ are the best indicators. None of our programme stars show obvious signs of a companion in the spectrum.

\begin{figure}[t]
\begin{center}
\includegraphics[width=0.49\textwidth]{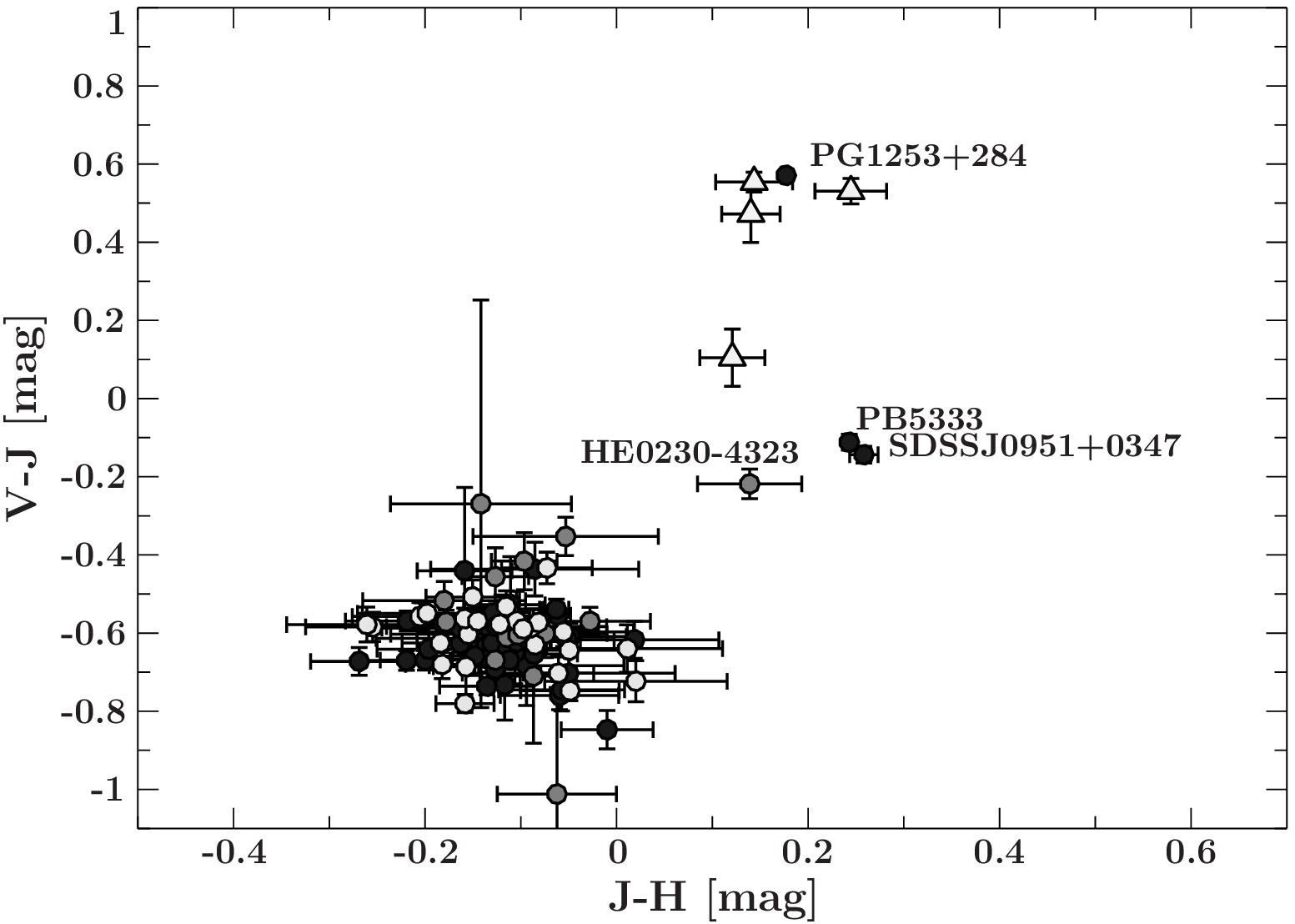}
\end{center}
\caption{Two-colour plots of $V$--$J$ vs. $J$--$H$ for all systems in our sample with 2MASS/UKIDSS colours, V magnitudes and low reddening \mbox{(E(B-V)<0.1}; light grey circles: WD companion, grey circles: dM companion, dark grey circles: unknown companion type). Most of the systems show no excess from a companion. The systems showing an infrared excess which indicates a cool companion are named. For comparison, 4 sdB binaries with confirmed G/K-type companions (light grey triangles) are shown \citep{vos12,vos13}. All colours were corrected for reddening.}
\label{fig:comp_color}
\end{figure}  

\begin{table}[t!]
\caption{Derived minimum masses and most probable nature of the companions.} 
\label{tab:rvmasses}
\begin{center}
\begin{tabular}{llll} \hline\hline
\noalign{\smallskip}
Object & $f(M)$ & $M_{\rm 2min}$ & Companion \\
 & [$M_{\rm \odot}$] & [$M_{\rm \odot}$] &  \\ 
\hline
\noalign{\smallskip}
J08300$+$47515 & $0.007$ & $0.14$ & WD$^{\rm lc}$ \\
J08233$+$11364 & $0.104$ & $0.44$ & MS/WD \\
J09510$+$03475 & $0.025$ & $0.23$ & MS/WD \\
J15222$-$01301 & $0.036$ & $0.27$ & MS/WD  \\
J10215$+$30101 & $0.046$ & $0.30$ & MS/WD \\
J15082$-$49405 & $0.08$ &  $0.39$ & MS/WD \\
J11324$-$06365 & $0.008$ & $0.14$ & MS/WD \\
J01185$-$00254 & $0.022$ & $0.22$ & MS/WD \\
J13463$+$28172 & $0.13$  & $0.49$ & WD \\
J18324$-$63091 & $0.13$ &  $0.50$ & MS/WD \\
J09523$+$62581 & $0.18$ &  $0.58$ & WD \\
J03213$+$05384 & $0.048$ & $0.31$ & MS/WD \\
\hline 
\end{tabular}
\tablefoot{
lc: companion type derived from the lightcurve}
\end{center}
\end{table}

A cool companion of spectral type $\sim\,M$1$-M$2 or earlier is detectable from an infrared excess even if the spectra in the optical range are not contaminated with spectral lines from the cool companion.
\citet{sta03} showed that two-colour diagrams can be used to detect unresolved late-type stellar companions using optical colours ($B$ and $V$) in combination with 2MASS colours ($J$ and $K_s$). \citet{ree04} convolved Kurucz models with appropriate 2MASS ($J$, $H$ and $K_s$) and $B$-filters and showed that companions of spectral type $M$2 and earlier would be separated from single sdBs in two-colour diagrams. \citet{gre06,gre08} created two-colour plots from $V$ band and 2MASS photometry ($J$ and $H$) of single-lined and composite sdB spectra that showed a clear separation between the composites and the single-lined spectra. Hence, the presence of a cool companion can be inferred by its infrared excess. 

We inspected each system with 2MASS/UKIDSS\footnote{for 2MASS; only colours flagged with quality A were used} ($J$ and $H$) and $V$ band colour information for an infrared excess to put tighter constraints on the nature of the companions. Figure~\ref{fig:comp_color} shows the two-colour diagram of the whole sample for systems with colour information and small reddening ($E(B-V)<0.1$). All colours were corrected for Galactic reddening using Table 6 with $R_v=3.1$ from \citet{sch11}. However, if a system does not show an excess in the infrared a cool companion can be excluded only when the minimum companion mass derived from the RV-curve is higher than the mass of a stellar companion which would cause an excess. To calculate the mass of the companion needed to cause an excess we used the following approach.

\begin{figure}[t!]
\begin{center}
\includegraphics[width=0.35\textwidth, angle=270]{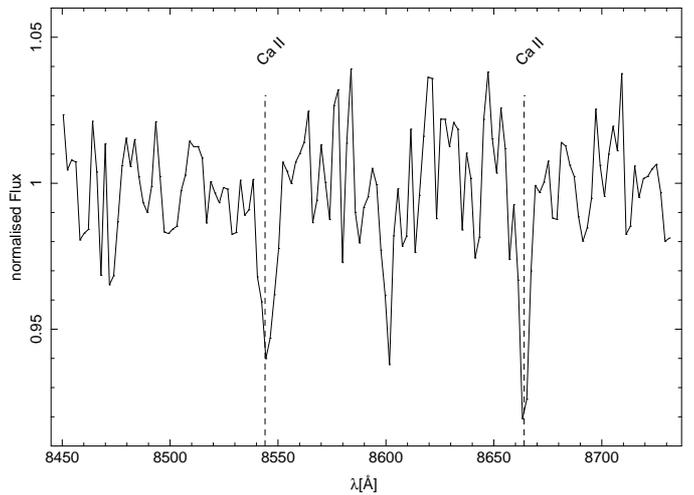}
\end{center}
\caption{Average SDSS spectrum of J09510$+$03475 showing evidence for weak absorption lines (two strongest components) of the infrared Ca\,{\sc ii} triplet originating very likely from a wide third component.}
\label{fig:0951}
\end{figure}  
First, we calculated the distance to each system using the reddening-corrected V magnitude, effective temperature and surface gravity as described in \citet{ram01}. The next step is to calculate the apparent magnitude in the $J$ band for different subclasses of dMs using the distance modulus
 \begin{equation}
 \label{equation-mass-function}
 m = 5\log_{10}(d)-5+M,
 \end{equation}
where $d$ is the distance in parsec and $M$ the absolute magnitude of the dM taken from Table 5 in \citet{kra07}. The calculated apparent magnitudes for each subclass in the $J$ band can be compared to the measured $J$ magnitudes of each individual system. Our assumption is that a cool companion would show up in the $J$ band if the calculated magnitude is 3 sigma above the measured $J$ magnitude. The calculated magnitude which would be visible in the $J$ band can be converted to the corresponding mass of the dM from Table 5 in \citet{kra07} using linear interpolation. If the derived mass is lower than the minimum companion mass derived from the RV-curve, a cool companion can be excluded because it would cause an excess in the infrared. In these systems a compact companion is most likely. If an excess is detected a cool companion is likely.

If time resolved photometry for the short-period sdB binaries is available, further constraints can be put even if the companion mass is inconclusive. The hemisphere of a cool low-mass main sequence companion facing the sdB is heated up by the significantly hotter sdB star. This causes a sinusoidal variation in the light curve. More(less) flux is emitted if the irradiated hemisphere of the cool companion is faced towards(away) from the observer. If this so-called reflection effect is detected, a compact companion can be excluded and a cool companion either a low-mass main sequence star of spectral type M or a brown dwarf is most likely. However, if the light curve of the short-period system shows no variation, a compact object like a WD is most likely to be the companion. Table~\ref{tab:rvmasses} gives an overview on the most likely companions of our sample.

\begin{figure}[t!]
\begin{center}
\includegraphics[width=0.49\textwidth]{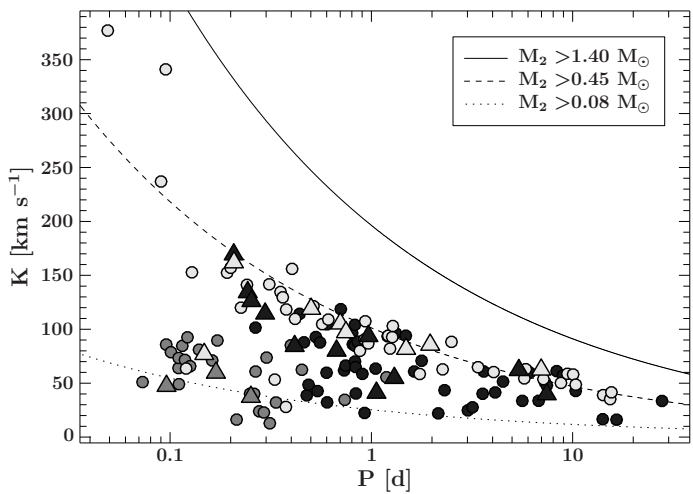}
                \end{center}
\caption{The RV semi-amplitudes of all known short-period sdB binaries with spectroscopic solutions plotted against their orbital periods (light grey: WD companions, grey: dM companion, dark grey: unknown companion type). The binaries from the MUCHFUSS programme are marked with triangles, binaries taken from the literature with circles. The lines mark the regions to the right where the minimum companion masses derived from the binary mass function (assuming $0.47$\,M$_\odot$ for the sdBs) exceed certain values.}
\label{fig:much_mass}
\end{figure}

\subsection{WD companions}
J09523$+$62581 and J13463$+$28172 have minimum companion masses obtained from the radial velocity curve higher than dM masses which would cause an infrared excess.  No sign of a companion is visible in the spectrum nor in the two-colour diagram. Therefore, the companion in both systems is most likely a WD. 

\subsection{J08300$+$47515 - a system with a possible ELM-WD companion}
J08300$+$47515 is a remarkable system. The minimum companion mass is only $0.14$\,M$_\odot$. Therefore, the nature of the companion cannot be constrained unambiguously from spectroscopy. The period of the system is $0.14780\pm0.00007$ days. This means that a cool main sequence companion would show a reflection effect in the light curve. However, a 2.14\,h light curve of J08300$+$47515 obtained with the CAHA-2.2m telescope using BUSCA shows no light variation with a standard deviation of $0.0063$ on the normalised lightcurve (see Schaffenroth et. al in prep). The companion might therefore be an ELM-WD. However, the inclination of the system cannot be constrained and it is still possible that the system is seen under low inclination. The maximum mass for an ELM-WD is $\sim\,0.3$\,M$_\odot$. If the sdB has the canonical mass of $\sim\,0.47$\,M$_\odot$ and J08300$+$47515 is seen under an inclination angle of $i<32.4^\circ$ the companion will be more massive than $\sim\,0.3$\,M$_\odot$ and not be a new ELM-WD companion. The probability of finding a system with $i<32.4^\circ$ is $\sim15$\,\%.



\subsection{J11324$-$06365 - the first helium deficient sdO with a close companion}
\citet{str07} studied the evolutionary status of 58 subdwarf O stars (sdOs) and concluded that the helium deficient sdOs are likely to be evolved sdBs. Indeed, evolution tracks by \citet{han02} and \citet{dor93} show that sdBs will become helium deficient sdOs as they evolve to higher temperatures. Since a significant fraction of short-period sdBs is found in compact binaries, helium deficient sdOs should have a similar binary fraction. Although close companions to some helium deficient, evolved sdOBs have been found \citep{alm12,kle11}, J1132$-$0636 is the first helium deficient highly evolved sdO with a confirmed close companion. The minimum companion mass derived from the RV-curve is well below the mass which would cause an infrared excess if the companion were a dM. Therefore, the nature of the unseen companion remains unclear.

\subsection{J09510$+$03475 - a hierarchical triple}\label{sec:0951}
J09510$+$03475 shows an excess in the J and H bands indicating a cool companion (see Fig.~\ref{fig:comp_color}). In addition, a combination of 7 SDSS spectra from DR9 shows the two strongest components of the Ca\,{\sc ii} triplet at around 8600\,\AA\, (Fig.~\ref{fig:0951}). Radial velocity measurements of the hydrogen lines confirm that 4 SDSS spectra were taken when the sdB moved through the minimum of the radial velocity curve. This means that a close companion should be observed in anti-phase around its maximum velocity which is expected to be $\sim250$\,km\,s$^{-1}$ depending on the mass ratio but certainly higher than the system velocity of  $\gamma=111.1\pm2.5$\,km\,s$^{-1}$. Using the same 4 SDSS spectra, an average velocity of $v_{\rm Ca}=86\pm16$\,km\,s$^{-1}$ for the calcium lines was measured which is just slightly below the system velocity ruling out a close companion. Therefore, the lines originate most likely from a third body in a wide orbit with a low RV amplitude. This makes this system the second candidate for a triple system after PG\,1253$+$284 \citep{heb02} which is an sdB star with one companion in a close orbit 
and another low-mass main sequence star in a wide orbit which causes the excess in the infrared and the Ca\,{\sc ii} lines. However, the nature of the close companion remains unclear.

\subsection{Unconstrained companions}
J18324$-$63091, J15222$-$01301, J01185$-$00254 and J03213$+$05384 have minimum companion masses derived from the RV-curve well below the mass which would cause an infrared excess. J10215$+$30101, J15082$+$49405 and J08233$+$11364 have no reliable infrared colours. Therefore, the nature of the unseen companion in those seven systems remains ambiguous. 

\subsection{The MUCHFUSS sample}
Fig.~\ref{fig:much_mass} shows the RV semi-amplitudes of all known short-period sdB binaries with orbital solutions plotted against their orbital periods. The dotted, dashed and solid lines mark the regions to the right where the minimum companion masses derived from the binary mass function (assuming $0.47$\,M$_\odot$ for the sdBs) exceed $0.08$\,M$_\odot$, $0.45$\,M$_\odot$ and $1.40$\,M$_\odot$.

The MUCHFUSS targets are marked with triangles. Most of the MUCHFUSS targets fall in the region with the highest minimum companion masses detected. However, the MUCHFUSS campaign discovered not only companions with the highest masses, but also detected the lowest-mass companions to sdBs \citep{gei12, sch14}. This shows that our target selection is efficient to find both the most massive companions known to sdB stars with periods of up to a few days and the least massive companions with short orbital periods of less then 3 hours. However, no NS or BH companion has been discovered yet.


\begin{figure}
 \centering
 \includegraphics[width=0.48\textwidth]{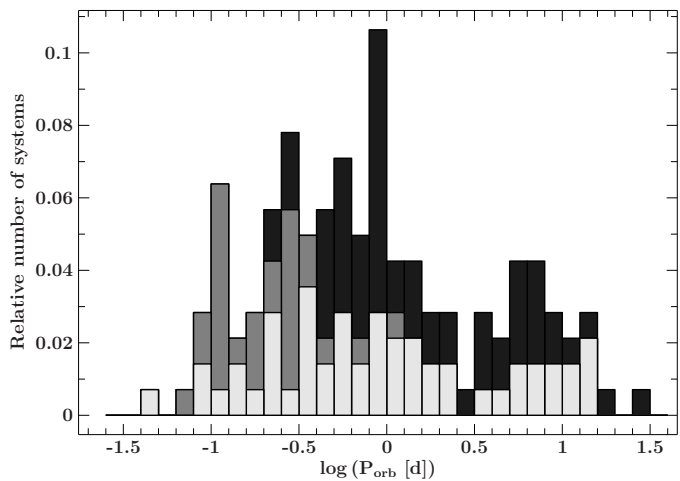}
         \caption{Period histogram of the full sample. Light grey: WD companions, grey: dM companion, dark grey: unknown companion type.}
\label{fig:period_histo}
\end{figure} 

\section{The population of close hot subdwarf binaries}
This study extends to 142, the sample of short-period sdB binaries that have measured mass functions. An overview is given in Tables\,\ref{tab:allbinaries} and \ref{tab:allbinaries1}.
 
In the following sections a canonical sdB mass of $0.47$\,M$_\odot$ will be adopted. All systems have unseen companions, but masses could only be determined for the eclipsing ones. Hence only a minimum companion mass could be derived for most of them. Many systems were pre-selected either from high RV variations between several single exposures or from light variations such as reflection effects, ellipsoidal variations and/or eclipses. Consequently, the distribution is by no means random but biased towards high inclination, both for RV variables (large amplitudes preferred) and light variables (reflection effect and/or eclipses detected). Therefore, statistically, the derived minimum companion masses are expected to be not far from the real companion mass.

\begin{figure}
\begin{center}
\includegraphics[width=0.48\textwidth]{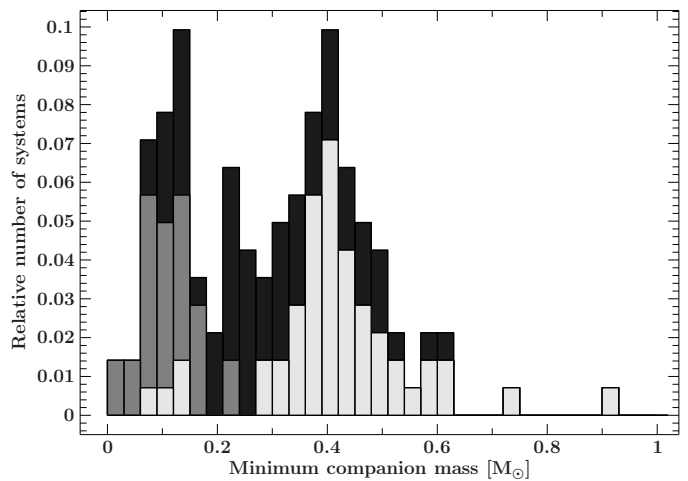}
\end{center}
\caption{Histogram of minimum companion masses (light grey: WD companions, grey: dM companion, dark grey: unknown companion type). Clearly visible are at least two separated populations. The first population peaks at around $0.1$\,M$_\odot$ and consists mainly of low mass main sequence companions. The second population peaks at around $0.4$\,M$_\odot$ and consists mainly of WD companions. The two high mass outliers belong to a population of massive WD companions.}
\label{fig:comp_mass}
\end{figure} 

We collected orbital and atmospheric parameters as well as $V$-band and infrared photometry for the full sample from the literature (see Table\,\ref{tab:allbinaries1}). Companion types for 82 systems were identified as described in Sec.\,\ref{sec:compnat}. Thirty low-mass stellar or substellar companions were identified from a reflection effect in the lightcurve. Twenty-three systems show either ellipsoidal variation, 
eclipses with no additional reflection effect or no light variations at all. For these systems a WD companion is most likely. Additionally, 29 systems could be confirmed as WD companions because the minimum companion mass is higher than the mass for a non-degenerate companion which would cause an infrared excess. None of those show an excess in the $J$ and $H$ band (Fig.\,\ref{fig:comp_color}). Therefore, the companions are most likely WDs as well.  

Four systems show an excess in the infrared and are good candidates for having a cool companion (see Fig.\,\ref{fig:comp_color}). Indeed, HE\,0230-4323 shows a reflection effect and a low-mass stellar companion is confirmed. 

Some systems may actually be triple as exemplified by PG1253$+$284, a radial velocity variable sdB with a dwarf companion which causes a strong infrared excess. The components were resolved by HST imaging  \citep{heb02}, which indicated that the dwarf companion is on a wide orbit. Nevertheless, RV variations of the sdB star were observed, which  must stem from another unresolved companion on a short-period orbit. Hence PG1253$+$284 is a triple system. Additional evidence of multiplicity (triples, quadruples) amongst sdB systems has recently been reported by \citet{bar14}. Hence it is worthwhile to search for triples amongst the four systems showing infrared excess.  In fact, there is one, J09510$+$03475, which shows signs that the system actually may be triple (see Sec.\,\ref{sec:0951} for a detailed discussion), while for HE\,0230-4323 and PB5333 there is no hint for a third companion.
 
In the following we shall discuss the distribution of periods and companion masses, compare the stars' positions in the $T_{\rm eff}$ -- $\log{g}$  plane to predictions from stellar evolution models, and discuss selection bias.
 
 \subsection{Distribution of orbital periods and minimum companion masses}
Fig.~\ref{fig:period_histo} shows the period distribution of the full sample. 

A wide peak near $P_{\rm orb}=0.3$\,days is found in the full sample. The majority of systems in this group are dM companions detected from reflection effects in the lightcurves. Beyond half a day the contribution from the confirmed dM companions decreases significantly, most likely because a reflection effect is much weaker and not easy to detect. Another peak can be found at around $0.8 - 0.9$\,days. Most of the systems here have unidentified companion type. At longer periods the number of systems goes down, but in this region the selection effects are stronger. 

Many of the WD companions were confirmed not only by the systems' lightcurves but also by the non-detection of an excess in the infrared (see Sec.~\ref{sec:compnat}) which is period independent. Therefore, in contrast to the dM companions, we find WD companions almost over the full period range. However, a gap near $3$\,days appears. We have no explanation for this but at the same time, with the present statistics, we cannot be sure that this gap is real.

 \begin{figure}[t!]
   \centering
    \includegraphics[width=0.48\textwidth]{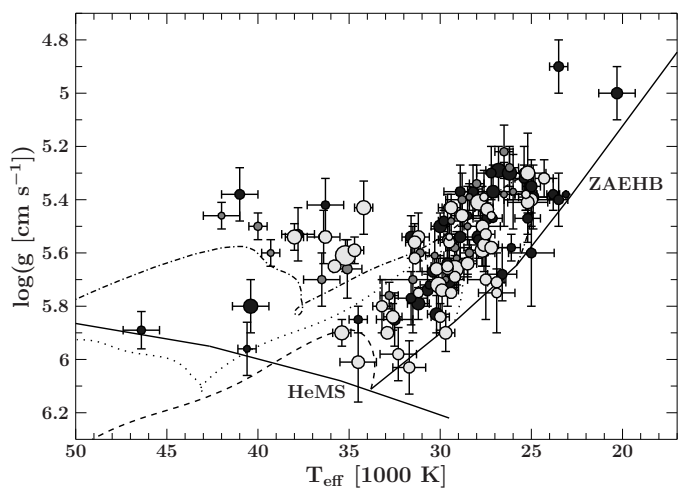}
       \caption{$T_{\rm eff}$ -- $\log{g}$ diagram of the full sample of binary sdB stars (light grey: WD companions, grey: M-dwarf companion, dark grey: unknown companion type). The size of the symbols corresponds to the minimum companion mass. The helium main sequence (HeMS) and the zero-age EHB (ZAEHB) are superimposed with EHB evolutionary tracks by \citet{han02} (dashed lines: $m_{\rm env}=0.000$\,M$_\odot$, dotted lines: $m_{\rm env}=0.001$\,M$_\odot$, dashed-dotted lines: $m_{\rm env}=0.005$\,M$_\odot$ using $0.45$\,M$_\odot$ models).}
       \label{fig:teff_logg_sep}
   \end{figure}

   \begin{figure}[t!]
   \centering
    \includegraphics[width=0.48\textwidth]{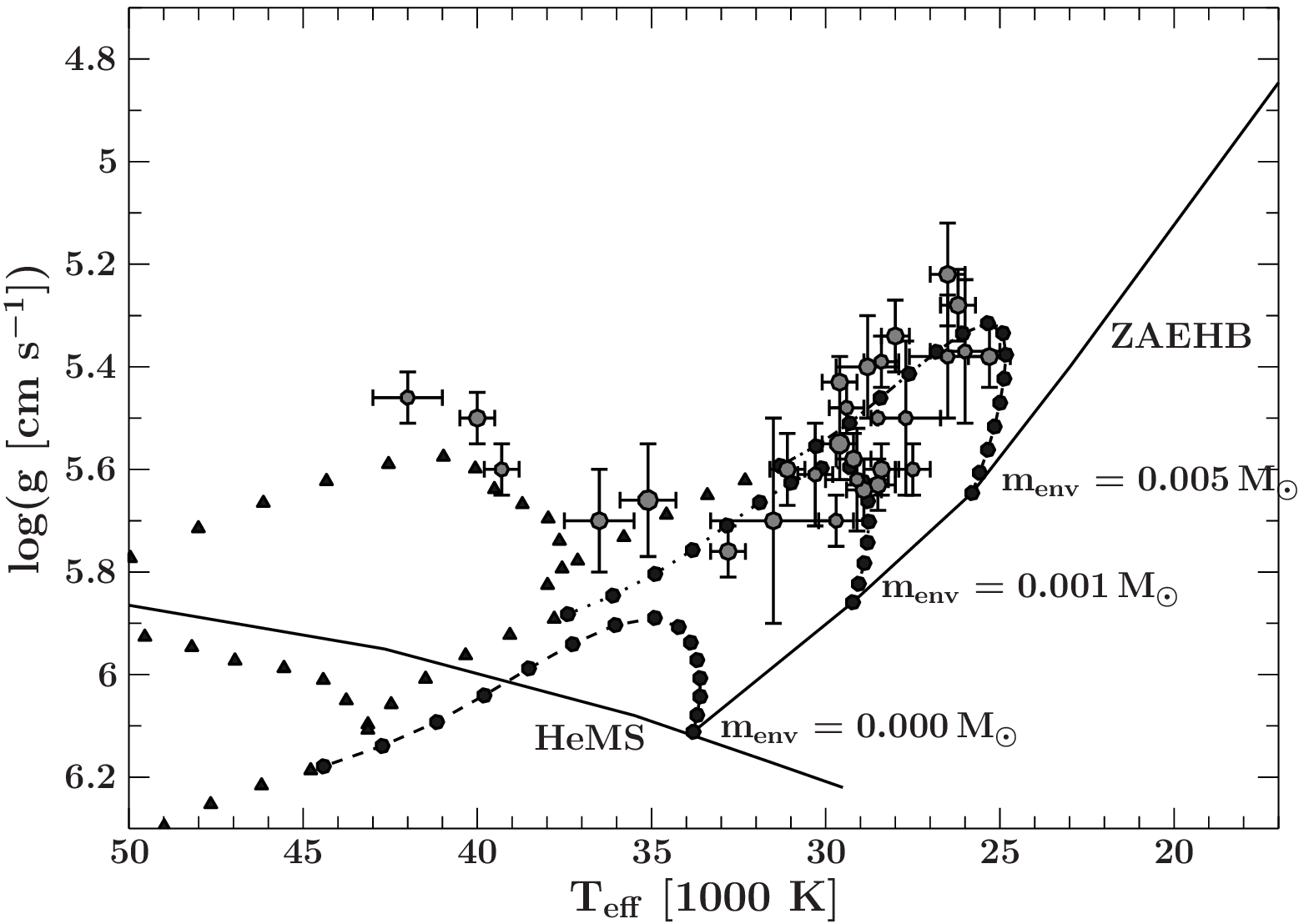}
    \includegraphics[width=0.48\textwidth]{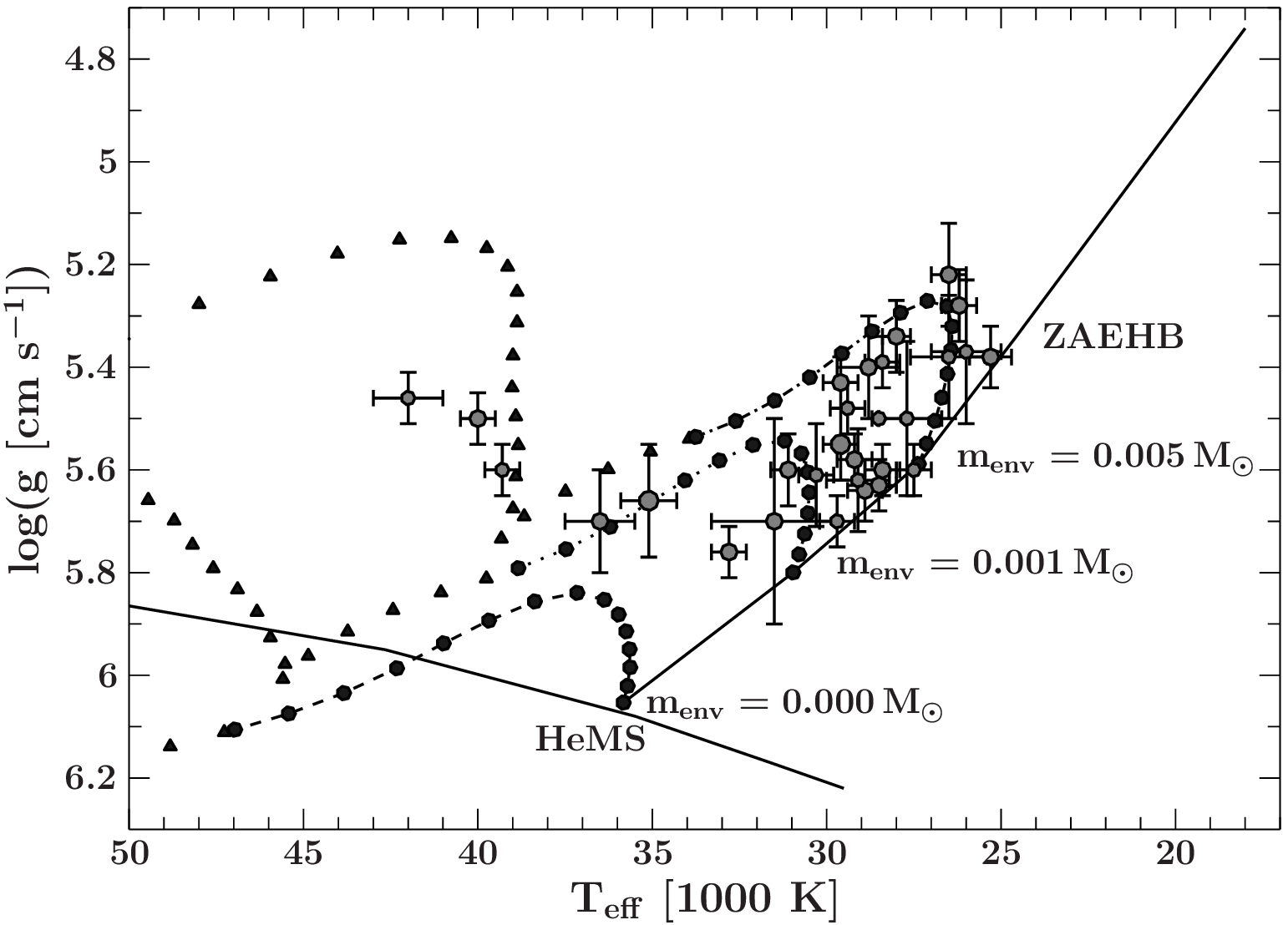}
       \caption{$T_{\rm eff}$ -- $\log{g}$ diagram of sdB stars with confirmed M-dwarf companions. The size of the symbols corresponds to the minimum companion mass. The helium main sequence (HeMS) and the zero-age EHB (ZAEHB) are superimposed with EHB evolutionary tracks by \citet{han02} using $0.45$\,M$_\odot$ (upper panel) and $0.50$\,M$_\odot$ (lower panel) models). The space between black symbols corresponds to equal times and shows that the sdB evolution speeds up once the sdB reaches its lowest gravity. Circles mark the region where the sdB star contains helium in the center of the core whereas triangles mark the region where the helium in the center of the core is completely burned.}
       \label{fig:teff_logg_sep1}
   \end{figure}
   

The distribution of the minimum masses of the companions is displayed in Fig.~\ref{fig:comp_mass}. 
We identify three separate populations.

1) The first population has an average minimum companion mass around $0.1$\,M$_\odot$, close to the hydrogen burning limit. Most of them were identified as either dMs or  brown dwarfs from the observation of their reflection effect. Only four WDs were found in this period regime, which could be ELM-WDs (M$<$0.3 M$_\odot$, see Sect. \ref{sec:elm}). 

2) The second population peaks around $0.4$\,M$_\odot$. Our analysis showed that the majority of this population are most likely WDs with an average minimum mass around $0.4$\,M$_\odot$, lower than the average mass of single WDs (see discussion in Sec.\,\ref{sec:WDcompmass}).

3) The third group are the high mass WD companions ($M_{\rm WD}>0.7$\,M$_\odot$). Systems with high companion masses stand out in radial velocity selected samples as they show higher RV variations compared to low-mass companions. However, only the eclipsing systems KPD1930+2752 and CPD-30$^\circ$11223 have confirmed companion masses above $0.7$\,M$_\odot$. This means that in our sample less than $2\%$ of the binaries with measured RV curves have such high mass WD companions.

\begin{figure}[t!]
   \centering
     \includegraphics[width=0.48\textwidth]{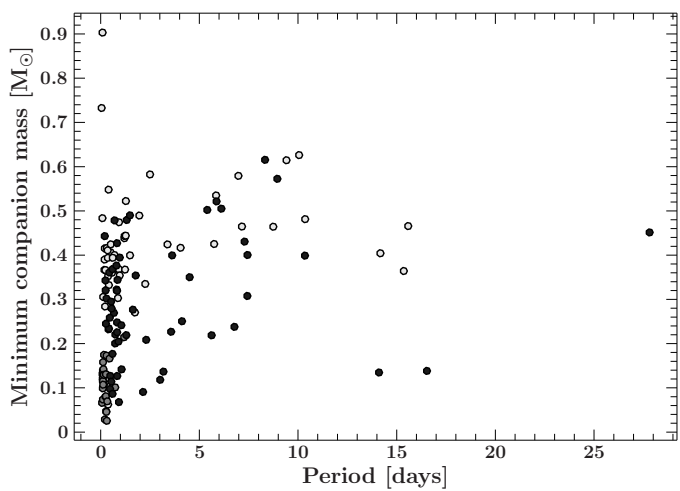}
      \caption{Minimum companion masses plotted against the period of the systems (light grey: WD companions, grey: M-dwarf companion, dark grey: unknown companion type). }
       \label{fig:separ_full}
   \end{figure}

\subsection{$T_{\rm eff}$ -- $\log{g}$ diagram}
Fig.~\ref{fig:teff_logg_sep} shows the $T_{\rm eff}$ -- $\log{g}$ diagram of the full sample with accurate atmospheric parameters. The size of the symbols represent the companion mass. Most of the stars populate the extreme horizontal branch (EHB) band all the way down to the helium main sequence while about 10\% of the sdB sample has already evolved off the EHB. The total evolution time on the EHB is $100$\,Myr, whereas post-EHB evolutionary timescales are lower by a factor of about 10. The theoretical tracks show a linear time-luminosity-relation while the star is in the EHB strip until it comes close to the terminal age EHB (TAEHB), where evolution starts to speed up. Hence, we would expect a homogeneous coverage of the EHB band as it is indeed observed. A more detailed comparison can be made using the cumulative luminosity function \citep[see][for details]{lis05}.  

In the next step we concentrate on the systems for which the companions have been classified and separate the distribution according to companion type, that is dMs and WDs, respectively. For WD companions, the sdBs populate the full EHB band homogeneously with a small fraction of sdBs having evolved off the EHB.  For sdB stars with dM companions the ratio of post-EHB to EHB stars is similar to that for sdB+WD systems. However, they appear not to cover the full EHB band. There is a lack of hot, high gravity sdBs close to the helium main sequence (Fig.~\ref{fig:teff_logg_sep1}). 
 
Most striking is that the width of the distribution of the sdB+dM systems is narrower than that of sdB+WD ones, in particular none of the sdB stars is found close to the zero age EHB (ZAEHB) if a sdB mass of 0.45\,M$_\odot$ is assumed. Because of the contamination of the sdB spectrum by light from the companion, their gravities could have been systematically overestimated and their effective temperatures could have been systematically underestimated, which would shift the systems away from the ZAEHB \citep{sch13}, which would shift them even further away from the ZAEHB. 
The location of the EHB band in the $T_{\rm eff}$ -- $\log{g}$ diagram depends on the adopted core mass. By increasing it, the EHB stars become more luminous and therefore the EHB band is shifted to lower gravities. This is demonstrated in Fig.~\ref{fig:teff_logg_sep1} by indicating the ZAEHB for a higher core mass of 0.5\,M$_\odot$ in addition to that of 0.45\,M$_\odot$ shown in all panels. 
The observed distribution of sdB stars is consistent with the 0.5\,M$_\odot$ ZAEHB, for which the EHB evolution timescales are shorter. Hence, adopting a higher core mass gives better agreement between observations and evolutionary tracks.

\subsection{Separation of the systems}
The details of the common envelope (CE) phase are still poorly understood \citep{iva13}. In a rather simplistic picture the orbital energy of the binary, which scales with the mass of the companion, is deposited in the envelope. If a more massive companion ejects the common envelope earlier, and therefore at a wider orbit than a less massive companion, a correlation between orbital period and minimum companion mass would the expected. Figure~\ref{fig:separ_full} shows the minimum companion masses plotted over the period of the systems: however, no obvious correlation can be seen in the sample with WD as well as dM companions (see Sec.\,\ref{sec:CE_evol} for further discussion). 

We also note that, even if the core masses of the sdB progenitors were very similar, their total masses (core + envelope) might have been quite different, implying different energies to expel the envelope and different final orbital separations. This may partially explain why we do not see any correlation between minimum companion mass and orbital period in Fig.~\ref{fig:separ_full}.

\subsection{Selection effects}
In order to compare the observed sample of close binary sdBs to population synthesis models (e.g. \citealt{han02,han03,cla12}) selection effects have to be taken into account. For the MUCHFUSS sample the target selection is well defined \citep{gei11a}. However, since the 142 solved binaries studied here are drawn from several different samples, it is impossible to come up with an unified description of selection bias.


All the sdBs studied here were initially discovered as faint blue stars from multi-band photometric survey data. However, spectral classification had to follow and the brighter limits of those spectroscopic observations have to be taken into account. 

\begin{figure}[t!]
 \centering
\includegraphics[width=0.48\textwidth]{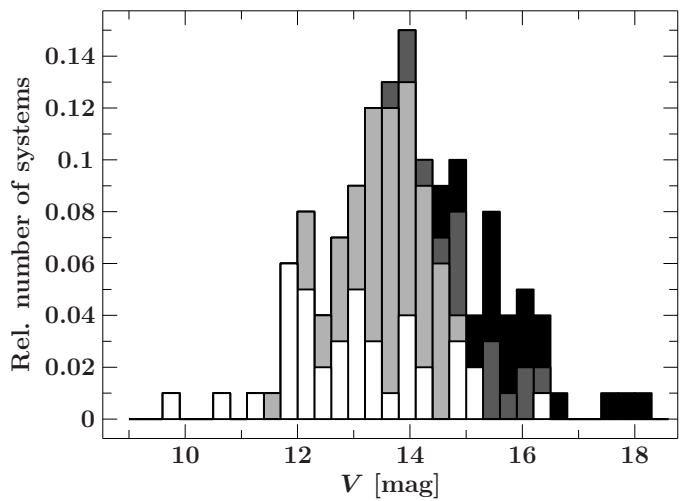}
\caption{Histogram of the magnitudes of the full sample. The systems selected from the SDSS surveys are marked in black and from the Hamburg/ESO (HE), Hamburg Schmidt (HS) are marked in dark grey. The systems selected from the Palomar Green (PG), Edinburgh Cape (EC) and Kitt Peak Downes (KPD) surveys are marked in light grey. The systems selected from smaller scale surveys (e.g. Feige...) are marked in white.}
\label{fig:histo_vmag}
\end{figure}

In the brightness distribution ($V$-band, Fig.~\ref{fig:histo_vmag}) of the whole sample, we can identify two sub-samples. One peaks around $14\,{\rm mag}$ and consists of binaries mostly discovered in the Palomar Green (PG, \citealt{gre86}), Edinburgh Cape (EC, \citealt{sto97}), Kitt Peak Downes (KPD, \citealt{dow86}) and some smaller scale surveys. The fainter subsample peaks around $15\,{\rm mag}$ and was selected mostly from the Hamburg/ESO (HE, \citealt{wis96}), the Hamburg Schmidt (HS, \citealt{hag95}) surveys and the SDSS \citep{yor00}.

\begin{figure}
 \centering
   \includegraphics[width=0.48\textwidth]{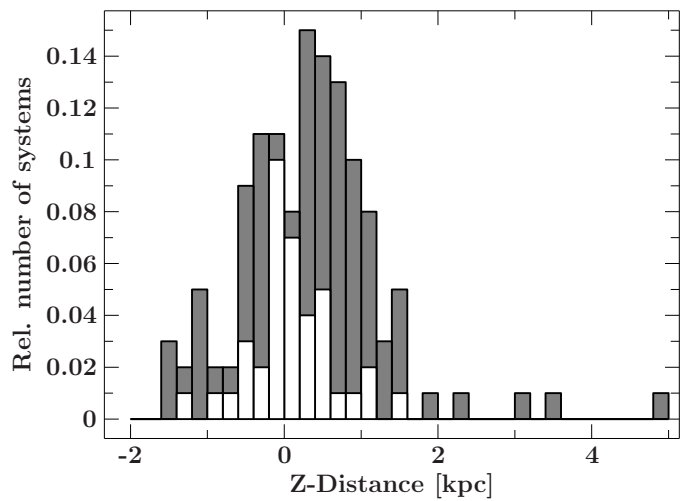} 
 \caption{Histogram of the spectroscopic z-distances above the Galactic plane. White are systems with a Galactic latitude $\vert b \vert<30^\circ$. Grey marked are the systems with  $\vert b \vert > 30^\circ$.}
      \label{fig:histo_zdist}
      \end{figure}

For 127 binaries we calculated the z-distances from the Galactic plane\footnote{We included near-Galactic plane objects neglecting reddening corrections as reddening is of little influence for calculating their z-distance.} assuming a canonical sdB mass of 0.47 M$_\odot$ (Fig.\,\ref{fig:histo_zdist}) and for 118 systems with sufficient data we calculated also the spectroscopic distances.



Except for four distant stars (at 2 to 5~kpc above the Galactic plane) all stars lie within 2~kpc below or above the Galactic plane. Their distribution is asymmetric, with an excess of objects in the northern Galactic hemisphere. The deficit of near-Galactic plane stars, as well as those below the Galactic plane, is due to the insufficient depth of near-plane and southern surveys. The most distant systems are likely to be halo stars and may be considerably older than the bulk. A significant fraction belongs to the thick disc, which on average is older than the thin disc. Because the age of the progenitor population is an important ingredient for binary population synthesis, it is crucial to assign each system to one of the stellar populations via an investigation of its kinematic and thus derive age estimates. 
 


The search for binarity of the targets has either been done by photometric or spectroscopic follow-up observations.
Either the star shows light variations indicative of a close companion or the star RV shifts become apparent. Both discovery methods introduce different selection effects. 

Only 20 of the binaries in the sample have been discovered photometrically (see Table\,\ref{tab:allbinaries1}). Short-period sdB stars with cool dM or BD companions show a reflection effect and often also eclipses. 14 binaries with periods of less than $0.3-0.4\,{\rm d}$ and a peak period of $0.1\,{\rm d}$ have been discovered in this way. Close sdB+WD binaries can show ellipsoidal variations and sometimes very shallow eclipses. Of the six binaries discovered in this way, two have periods of less than $0.09\,{\rm d}$ and one has a period of $0.3\,{\rm d}$. The remaining three long-period systems ($>3\,{\rm d}$) have been discovered by the Kepler mission, which has a much higher sensitivity than ground-based telescopes. In general, photometric selection is clearly biased towards the shortest-period systems at high inclinations.

The remaining 122 systems have been discovered from RV shifts, most of them from medium-resolution spectra with an RV accuracy of $\sim10-20\,{\rm km\,s^{-1}}$ (e.g. \citealt{max01}). The binaries studied by \citet{ede05} and in the course of the ESO Supernova Ia Progenitor Survey (SPY, \citealt{nap01, kar06, gei10a, gei11c}) have been discovered using high-resolution spectra with an RV accuracy better than $5\,{\rm km\,s^{-1}}$.

To our knowledge, well-defined cuts of RV-shifts were only used in the MUCHFUSS target selection (see \citealt{gei11a}). In general, short-period systems with high RV shifts and therefore high inclinations are the easiest ones to solve within a few nights of observations. This introduces a selection in favour of such systems. It is unlikely that a significant population of binaries with periods longer than one day and RV semi-amplitudes higher than $100\,{\rm km\,s^{-1}}$ has been missed in the high-galactic latitude population of hot subdwarf stars covered by the SDSS. The missing population of close sdB binaries with periods from a few days to a few tens of days most likely consists of systems with small RV semiamplitudes and rather low-mass companions ($<0.5\,M_{\rm \odot}$). 


\section{Comparison with related binary populations}
\subsection{The population of helium-core WD binaries}\label{sec:elm}
The formation of helium-core WD binaries with masses $<0.45$\,M$_\odot$ is expected to occur in a similar way as the formation of sdB+WD binaries discussed in this paper. Both systems survive two phases of mass transfer with the second phase where the helium-core WD/sdB is formed being a common envelope phase. The helium-core WDs start transferring mass already when the progenitor evolves on the red giant branch (RGB) and lose so much mass that they are not able to ignite helium in the core. The sdBs start mass transfer on or near the tip of the RGB and are massive enough to ignite helium. 
 
\begin{figure}
\begin{center}
\includegraphics[width=0.49\textwidth]{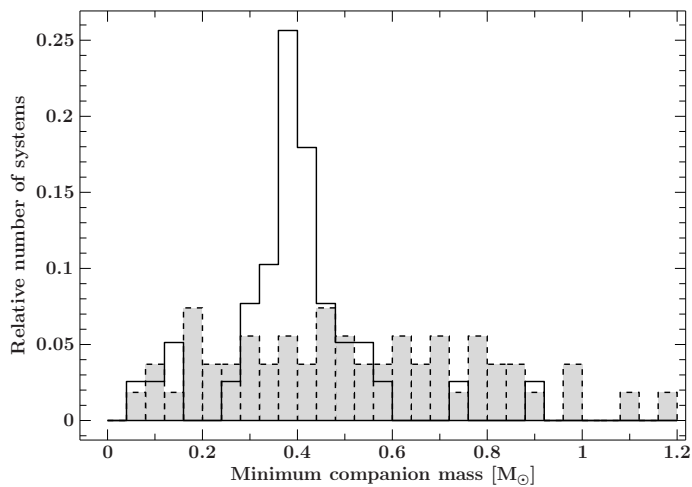}
                \end{center}
\caption{Comparison of minimum companion masses of sdB binaries with confirmed WD companions to the ELM-WD binaries (grey shaded area) taken from \citet{gia14}. }
\label{fig:elm_mass}
\end{figure}

Orbital parameters of 55 helium-core WD binaries were selected from \citet{gia14}. All the companions are WDs. Figure~\ref{fig:elm_mass} shows the minimum companion mass histogram of the sample compared to the sdB sample. ELM companions cover a wider range of masses, extending to low as well as high masses and indicating a different evolutionary path. The distribution does not show a clear separate population with a peak at $0.4$\,M$_\odot$ like the confirmed WD companions to sdB stars. 

\begin{figure}
\begin{center}
\includegraphics[width=0.49\textwidth]{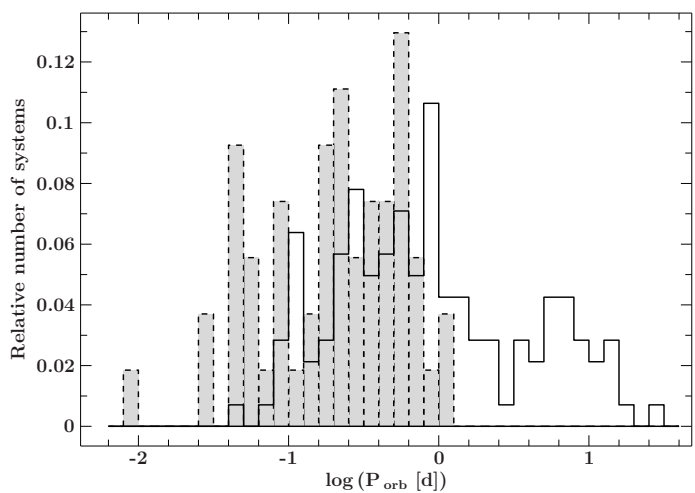}
                \end{center}
\caption{Comparison of orbital periods of sdB binaries with confirmed WD companions to the known ELM-WD binaries with orbital solutions (grey shaded area) taken from \citet{gia14}. }
\label{fig:elm_period}
\end{figure}

Figure~\ref{fig:elm_period} shows the orbital period distribution of the helium-core WDs compared to our sample. Between an orbital period of $0.1$ and $1.0$\,days both distributions look very similar. However, below $0.25$\,days helium-core WDs are more numerous compared to the sdB+WD systems. On the other hand at longer periods sdB+WD systems are more numerous which indicates that helium-core WD binaries are formed preferentially with shorter periods.

\subsection{The population of compact WD+dM systems}
Compact WD+dM binaries are also the product of CE evolution and so we might expect that the properties of these binaries are similar to the sdB+dM binaries. Indeed, some WD+dM systems may have been created as sdB stars with dM companions that have since evolved to become white dwarfs with masses close to $0.47$\,M$_\odot$. 

\begin{figure}[t!]
\begin{center}
\includegraphics[width=0.49\textwidth]{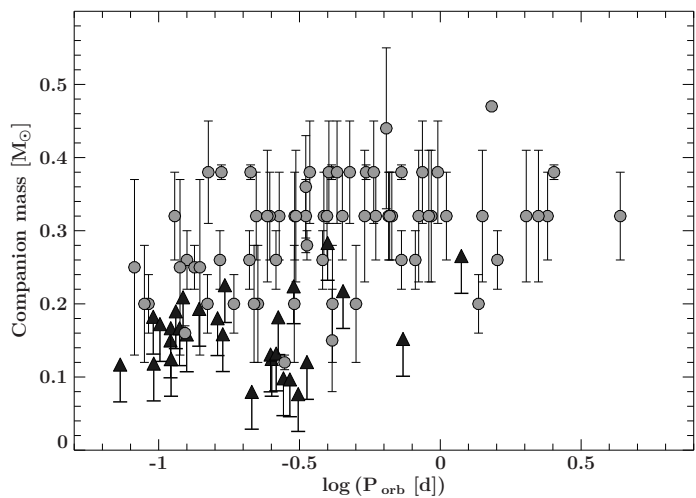}
                \end{center}
\caption{Companion mass plotted against the orbital period. Grey circles mark derived companion masses of known WD+dM binaries with orbital solutions taken from \citet{zor11}, \citet{neb11}, \citet{pyr12} and \citet{par12}. Black arrows mark minimum companion masses of the known sdB+dM binaries.}
\label{fig:wdms_period}
\end{figure}

Orbital parameters of 68 post-common envelope binaries consisting of a WD+dM were selected from the literature \citep{zor11, neb11, pyr12, par12}. The systems cannot be compared in the same way as the ELM-WD binaries because all WD+dM systems have confirmed companion masses whereas, for sdB+dMs, only minimum companion masses are known. Figure~\ref{fig:wdms_period} shows the orbital period vs. the companion masses of the dM companions to WDs compared to the confirmed dM companions to sdB stars. The plot shows dM companion masses larger than $\sim$0.2\,M$_\odot$ for the WD+dM systems, while the minimum masses of dM companions to sdBs peak well below, near $\sim$\,0.1\,M$_\odot$. To increase the companion masses of the sdB+dM systems to an sdB mass larger than 0.2\,M$_\odot$ the systems have to be observed with inclination angles below $30^\circ$. This is not very likely as the majority of the systems were selected from photometry by eclipses and/or reflection effects which are hardly detectable in systems with small inclinations. 

The WD+dM systems show a wide spread in orbital period whereas the majority of the sdB+dM systems were found with periods below $0.3$\,days. A possible reason might be that WD+dM systems are usually identified spectroscopically because features of the dM dominate the red part of the composite spectra. In contrast to that, almost all sdB+dM systems were identified from the reflection effect in the lightcurves. Longer period systems show much weaker reflection effects and therefore are much harder to detect. 
 
\section{Discussion}

\subsection{Distribution of sdB masses}
Several previous studies discussed the sdB mass distribution. \citet{fon12} collected sdB masses of a sample of 22 sdBs (15 derived from asteroseismology and 7 from resolved binaries), and found a sharp peak at \mbox{$M_{\rm sdB}=0.47$\,M$_\odot$}. \citet{han03} discussed the sdB mass distribution formed via different phases of mass transfer. Figure~12 in \citet{han03} showed a sharp peak at $M_{\rm sdB}=0.46$\,M$_\odot$ for sdBs formed after a common envelope phase.

In our analysis of the companion mass distribution shown in Fig.\,\ref{fig:comp_mass} we apply the assumption that the sdBs have all canonical masses of \mbox{$M_{\rm sdB}=0.47$\,M$_\odot$} because of the results of previous studies (see Fontaine et al. 2012 \nocite{fon12} and references therein). The distribution of the minimum companion masses (Fig.\,\ref{fig:comp_mass}) shows two quite narrow peaks. If the distribution of the sdB masses would be much more smeared out than predicted, those two peaks would have to be smeared out as well. We therefore conclude that the width of the sdB mass distribution is of the order of $0.2$\,M$_\odot$ at most, which is consistent with the prediction from theory. We note that from our analysis we cannot claim that \mbox{$M_{\rm sdB}=0.47$\,M$_\odot$} is the canonical mass for sdBs because adopting a higher (lower) average sdB mass would also increase (decrease) the companion masses but the distinct peaks in the companion mass distribution would persist.

\subsection{WD companion masses}\label{sec:WDcompmass}
The majority of minimum companion masses of confirmed WD companions are located around $0.4$\,M$_\odot$, which is significantly below the average mass for single (DA) WDs of $\sim0.59$\,M$_\odot$ (e.g. \citealt{kle13}).  Because of projection effects and selection biases the detection of high inclination systems should be favoured, which means that the derived limits should be on average close to the companion masses. Since the minimum masses of the WD companions are significantly smaller than the average mass of single C/O-WDs, we test this hypothesis by computing the inclination angles for all sdB+WD binaries assuming that all companion WDs have an average mass of $0.6$\,M$_\odot$. 

\begin{figure}
\begin{center}
\includegraphics[width=0.49\textwidth]{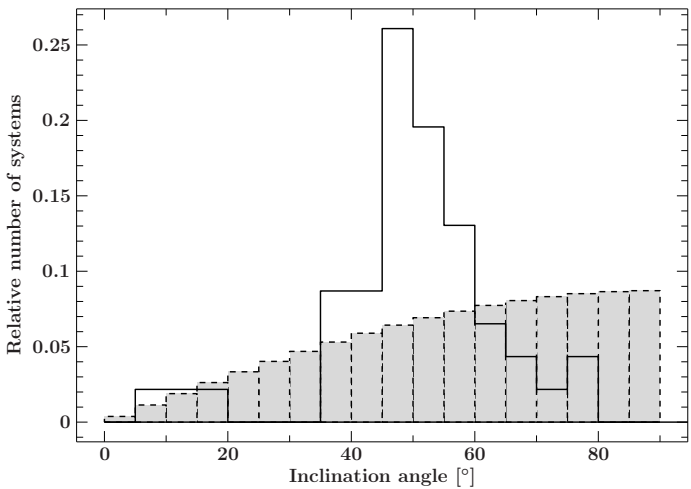}
\end{center}
\caption{Comparison of computed inclination angles of the confirmed sdB+WD systems to match companion masses of $0.6$\,M$_\odot$ to a theoretical inclination angle distribution assuming randomly distributed inclination angles.}
\label{fig:incl60deg}
\end{figure}

Figure\,\ref{fig:incl60deg} shows a comparison between the computed distribution of inclination angles and the one expected for randomly distributed inclinations taking into account projection effects. We do not include any selection biases, but want to point out that they would in any case lead to even higher probabilities of seeing the systems at high inclinations. One can clearly see that the inclination distribution is not consistent with the one expected for a population of $0.6$\,M$_\odot$ C/O-WD companions. Hence, it is likely that a significant fraction of the sdB binaries host WDs of masses below $0.6$\,M$_\odot$.
 
\begin{table}[t!]
\caption{Derived times when the dM will fill its Roche Lobe and start accreting onto the primary to form a cataclysmic variable. The derived $M_{\rm Comp}$ are minimum companion masses which means that the time when the dM fills its Roche Lobe are upper limits}. 
\label{tab:rochefill}
\begin{center}
\begin{tabular}{llll} \hline\hline
\noalign{\smallskip}
Object & Period & $M_{\rm Comp}$  &  Time  \\
       & [days] & [M$_\odot$] &  [Gyr]   \\ 
\hline
\noalign{\smallskip}
HS\,2333$+$3927      & 0.172 & 0.174  &  3.44 \\
J192059$+$372220   & 0.169 & 0.107  &  6.09 \\
2M1533$+$3759      & 0.162 & 0.129  &  4.21 \\
ASAS102322$-$3737  & 0.139 & 0.142  &  2.15  \\
2M1938$+$4603      & 0.126 & 0.107  &  2.38  \\
BULSC16335       & 0.122 & 0.158  &  0.95  \\
EC10246$-$2707     & 0.119 & 0.115  &  1.70  \\
HW\,Vir            & 0.115 & 0.138  &  0.97  \\
HS\,2231$+$2441      & 0.111 & 0.073  &  3.05  \\
NSVS14256825     & 0.110 & 0.115  &  1.27  \\
UVEX\,0328$+$5035    & 0.110 & 0.098  &  1.77  \\
PG\,1336$-$018       & 0.101 & 0.121  &  0.72  \\
J082053$+$000843   & 0.096 & 0.067  &  2.22  \\ 
HS\,0705$+$6700      & 0.096 & 0.131  &  0.34  \\ 
PG\,1017$-$086       & 0.073 & 0.066  &  1.01  \\
J162256$+$473051   & 0.069 & 0.060  &  0.95 \\
\hline
\end{tabular}
\end{center}
\end{table}

\subsection{Triple systems}
Only one sdB system was known to be triple up to recently. We found another sdB binary, J09510$+$03475, with a third component in a wide orbit. 
\citet{bar14} studied 15 sdB binaries to detect long period companions. 
At least one has a visual companion well separated from the sdB. However, RV measurement show that the orbital period of the system is below $10$\,days, indicating a close companion in addition to the wide companion.  


The properties of J09510$+$03475 also imply that a fraction of sdB binaries showing an excess in the infrared might be triples systems. If we assume all 3 systems to be triples we find a fraction of $2.1\,\%$ in our sample. However, if the wide companion is too faint to show an excess in the infrared then it would be hidden in our sample and the fraction of triple systems might be significantly higher. Some of the sdB triples might have formed from solar type triples. \citet{rag10} found a fraction of $9\pm2\%$ for solar type triples which is not in disagreement with our findings. 

\subsection{Massive companions}
Hot subdwarf + WD binaries are potential supernova Ia progenitors if their masses are sufficiently large. However, only a very small fraction of massive WD companions ($<2\,\%$) were detected. KPD1930+2752 and CD-30$^\circ$11223 are the only systems with companion masses $>\,0.7$\,M$_\odot$. This implies that only a specific evolutionary path can form such systems. \citet{wan13} showed that a system like CD-30$^\circ$11223 is formed from a young stellar population which can only be found in the Galactic disc. Indeed, CD-30$^\circ$11223 is a confirmed member of the Galactic disc population \citep{gei13}. No NS or BH was detected in our sample but previous studies concentrated on high Galactic latitudes. Based on the non-detection of a NS or BH in our sample with 142 systems, we find that $<\,0.7\,\%$ of the close sdB binaries contain a NS or BH companion which is a small fraction but still consistent with the predictions from binary evolution calculations \citep{yun05, nel10}. We encourage a systematic search for compact sdB binaries at low Galactic latitudes.

\begin{table}[t!]
\caption{Derived merger timescales of the confirmed sdB+WD systems. The derived $M_{\rm Comp}$ are minimum companion masses which means that the merger timescales are upper limits to merger times.}
\label{tab:merge}
\begin{center}
\begin{tabular}{lllll}\hline\hline
 \noalign{\smallskip}
Object & Period & $M_{\rm Comp}$  &  Time & Merger \\
       & [days] & [M$_\odot$] &  [Gyr]  & result  \\ 
\hline
\noalign{\smallskip}
PG\,0941$+$280         & 0.311 & 0.415  &  10.27 & WD/RCrB \\
PG\,2345$+$318         & 0.241 & 0.366  &  5.78  & WD/RCrB \\
PG\,1432$+$159         & 0.225 & 0.283  &  6.01  & WD/RCrB \\
J113840$-$003531     & 0.208 & 0.415  &  3.49  & WD/RCrB \\
HS\,1741$+$2133        & 0.20  & 0.389  &  3.34  & WD/RCrB \\
HE\,1414$-$0309        & 0.192 & 0.366  &  3.16  & WD/RCrB \\
J083006$+$475150     & 0.148 & 0.137  &  3.77  & AM\,CVn \\ 
EC00404$-$4429       & 0.128 & 0.305  &  1.26  & WD/RCrB \\
PG\,1043$+$760         & 0.120 & 0.101  &  2.88  & AM\,CVn  \\
KPD\,1930$+$2752       & 0.095 & 0.903  &  0.23  & SN\,Ia \\
KPD\,0422$+$5421       & 0.090 & 0.483  &  0.33  & WD/RCrB \\
CD$-$30$^\circ$11223 & 0.049 & 0.732  &  0.05  & SN\,Ia \\
\hline 
\end{tabular}
\end{center}
\end{table}
 
\subsection{Implications for the common envelope phase}\label{sec:CE_evol}
A remarkable result of our analysis is that we find clearly distinct populations. The majority of confirmed WD companions have minimum companion masses strongly peaked at $\sim\,0.4$\,M$_\odot$. This is much lower than the average mass of single WDs and leads to the conclusion that the WDs need to lose a significant amount of mass during the evolution either during the first phase of mass transfer when the WD is formed or during the common envelope phase when the sdB is formed. The first phase can either be stable Roche lobe overflow or also a common envelope phase depending on the initial separation and the mass ratio of the system. White dwarf masses of $\sim\,0.4$\,M$_\odot$ are on the border between a WD with a helium core and a C/O core and a significant fraction of white dwarf companions might be helium-core WDs. 

In comparison, ELM-WD binaries show a much wider companion mass distribution starting at very low masses up to high masses close to the Chandrashekar limit. Either these systems form in a different way or sdB binaries need a special WD companion mass to lose the right amount of mass and form an sdB.

%



The dM companions were found to have minimum companion masses of $\sim\,0.1$\,M$_\odot$ close to the hydrogen burning limit. These systems have experienced one phase of mass transfer, namely the CE phase when the sdB was formed, and are direct progenitors of WD+dM systems. However, in comparison with the known population of WD+dM systems we find that the main sequence companions in WD systems are significantly more massive than the main sequence companions in sdB systems. This shows on the one hand that only a small number of WD+dM systems evolved from sdB+dM systems and on the other hand that sdBs might be formed preferentially by low mass main sequence companions whereas WD are preferably formed with higher mass main sequence companions. The other possible way to form a compact WD+dM system without forming an sdB first is the formation of the WD directly during a CE phase when the WD progenitor evolves on the asymptotic giant branch.  


In addition we found no correlation of the orbital separation of the sdB binaries with companion mass (see Fig.~\ref{fig:separ_full}) which means that the red giant progenitors of the sdB must have had different envelope masses. This could be tested, if we were able to identify systems that had similar envelope masses prior to envelope ejection. In this respect the halo population of sdB binaries would be of great interest because they are expected to form from systems where the sdB progenitor has a mass of $\sim\,0.8$\,M$_\odot$. A detailed kinematic analysis to identify the halo population of compact sdB binaries is crucial. The majority of the MUCHFUSS sample is faint and therefore they might be the best candidates to be member of the halo population and a good starting point for an extended kinematic analysis of the complete sample of compact sdB binaries.

\subsection{Future evolution: Pre-CV vs. Merger}
For systems with main sequence companions we calculate the time when the dM will fill its Roche Lobe and starts accretion. As approximation for the Roche radius the Eggleton equation was used \citep{egg83}, with $q$ being the mass ratio $q=M_{\rm comp}/M_{\rm sdB}$:
 \begin{equation}
 \label{equation egg}
 r_L = \frac{0.49q^{2/3}}{0.6q^{2/3}+\ln(1+q^{1/3})}.
 \end{equation}
We used the minimum companion mass for the dM and calculated the corresponding radius using Table 1 in \citet{kal09} by linear interpolation. Once the companion fills its Roche Lobe mass accretion starts and the systems becomes a cataclysmic variable (CV). We assumed that only gravitational wave radiation brings the to components closer. The time until the system starts accretions was calculated from the gravitational wave timescale, equation 7 from \citet{pir11}:
    \begin{equation}
 \label{equationpirlo}
 \tau_{\rm GW}=P\left\vert\frac{dP}{dt}\right\vert^{-1}=\frac{5}{96}\frac{c^5}{G^{5/3}}\frac{M_{\rm total}^{1/3}}{M_{\rm sdb}M_{\rm dM}}\left(\frac{P}{2\pi}\right)^{8/3}.
 \end{equation}
Table\,\ref{tab:rochefill} shows the 16 systems which will become a CV and start accreting within a Hubble time. J0820+0008 and J1622+4730 have confirmed brown dwarf companions \citep{gei12,sch14}. Therefore at least two systems (J0820+0008 and J1622+4730) will have brown dwarf (BD) donor stars. The dM companion in HS0705+6700 will fill its Roche Lobe in about $340$\,Myr, beeing the first system of our sample. At this stage the sdB is already evolved and turned into a C/O-WD. Therefore, all 16 systems of our sample will appear as WD+dM/BD with a low-mass companion ($M_{\rm Comp}<\sim0.17$\,M$_\odot$) before they become a CV. The currently known population of WD+dMs lacks such low-mass main sequence companions (see Fig.\,\ref{fig:wdms_period}). However, our findings show that low mass dM companions to WDs should exist as well. 
 
Merger timescales were calculated for systems of the full sample which have a confirmed WD companion and will merge within a Hubble time using equation 9 in \citet{pac67}:
    \begin{equation}
 \label{equation pac}
 T_0 ({\rm years}) = 3.22\cdot 10^{-3}\frac{(M_{\rm sdB}+M_{\rm Comp})^{1/3}}{M_{\rm sdB}M_{\rm Comp}}P_{\rm orb}^{8/3}.
 \end{equation}
We identified 12 systems of the full sample which will merge within a Hubble time. Only CD-30$^\circ$11223 will merge before the sdB turns into a WD. \citet{gei13} showed that this system will most likely explode as a subluminous SN\,Ia. All other systems will evolve and turn into a C/O WD before they merge. Depending on the mass ratio the systems either merge ($q>$2/3) or form an AM\,CVn type binary ($q<$2/3). For a helium-core white dwarf companion the merger might form an RCrB star, whereas a C/O-WD companion forms a massive single C/O-WD. If the system reaches the Chandrashekar mass it might explode as a SN\,Ia (e.g. \citealt{web84}).

PG1043+760 and J0830+4751 have low minimum companion masses and a mass ratio $q<$2/3. The companions in those systems are most likely helium-core white dwarfs. Both systems are therefore good candidates to have stable mass transfer and form an AM\,CVn type binary. KPD1930+2752 has a massive WD companion. The combined mass is close to the Chandrashekar limit. Thus, this is a good system to explode as a SN\,Ia. The other 8 systems have mass ratios $q>$2/3 and therefore are potential progenitors for mergers. Depending on the structure of the companion, the merger with a helium-core white dwarf might form an RCrB star, whereas a C/O-WD companion might form a massive single C/O-WD.

This analysis shows that the majority of sdB binaries with white dwarf companions will not merge within a Hubble time and only a small number of systems have periods and companion masses to either merge, form an AM\,CVn type binary or explode as a supernova Ia.

\section{Summary}
In this paper we have presented atmospheric and orbital parameters of 12 new close sdB binaries discovered by the MUCHFUSS project. Three of them have most likely WD companions. We found the first helium deficient sdO with a compact companion, a good candidate for an ELM-WD companion  and confirmed the second known hierarchical triple amongst the known sdBs.

This study increases the number of hot subdwarf binaries with orbital periods less than 30 days and measured mass functions to 142 systems. The companion mass distribution of the full sample shows two separate peaks. The confirmed dM/BD companions are concentrated around $0.1\,$M$_\odot$ whereas the majority of the WD companions peak at around $0.4$\,M$_\odot$ showing that WDs in compact hot subdwarf binaries have significantly lower masses than single WDs. The $T_{\rm eff}$ -- $\log{g}$ diagram of the sdB+dM systems indicates that in these systems the sdBs might have higher masses compared to the rest of the sample.

Close hot subdwarf binaries are expected to be formed in a similar way as the compact ELM-WD binaries or the WD+dM pre-CV systems. However, both samples show significantly different companion mass distributions indicating either selection biases or differences in their evolutionary paths.

We discussed possible implications for the common envelope phase, but also found that the progenitor stars of the sdB in our sample might have had a rather broad mass distribution. More insights in the formation process of field sdB stars can be gained, if they can be clearly assigned to their parent populations, either the thin disc, the thick disc or the Galactic halo. Accurate distances and kinematics are crucial for such an analysis. The {\it GAIA} space mission will provide accurate distances, luminosities and kinematics for most of the known sdB stars and will also cover the Galactic disc region, which has been avoided by previous surveys because of reddening.

This data will make it possible to derive sdB masses, identify different sdB populations and allow us to put constraints on the evolution history and the common envelope phase which forms the sdBs in close binaries.


 \begin{acknowledgements}
TK acknowledges support by the Netherlands Research School
 of Astronomy (NOVA). The research leading to these results has received funding from the European Research Council under the European Union’s Seventh Framework Programme (FP/2007-2013) / ERC Grant Agreement n. 320964 (WDTracer). BTG was supported in part by the UK’s Science and Technology Facilities Council (ST/I001719/1). CH is supported by the Deutsche Forschungsgemeinschaft (DFG) through grant HE1356/62-1. VS is supported by Deutsches Zentrum f\"ur Luft- und Raumfahrt (DLR) under grant 50OR1110.  We acknowledge that some observations used in this paper were carried out by Sebastian M\"uller, Patrick Br\"unner, Anna Faye McLeod, Markus Schindewolf and Florian Niederhofer.
 \\
Based on observations at the Paranal Observatory of the European Southern Observatory for programme number 165.H-0588(A). Based on observations at the La Silla Observatory of the European Southern Observatory for programmes number 079.D-0288(A), 080.D-0685(A), 084.D-0348(A) and 092.D-0040(A). Based on observations collected at the Centro Astron\'omico Hispano Alem\'an (CAHA) at Calar Alto, operated jointly by the Max-Planck Institut f\"ur Astronomie and the Instituto de Astrof\'isica de Andaluc\'ia (CSIC). Based on observations with the William Herschel Telescope operated by the Isaac Newton Group at the Observatorio del Roque de los Muchachos of the Instituto de Astrofisica de Canarias on the island of La Palma, Spain. Based on observations with the Southern Astrophysical Research (SOAR) telescope operated by the U.S. National Optical Astronomy Observatory (NOAO), the Ministério da Ciencia e Tecnologia of the Federal Republic of Brazil (MCT), the University of North Carolina at Chapel Hill (UNC), and Michigan State University (MSU). Based on observations obtained at the Gemini Observatory, which is operated by the Association of Universities for Research in Astronomy, Inc., under a cooperative agreement with the NSF on behalf of the Gemini partnership: the National Science Foundation (United States), the Science and Technology Facilities Council (United Kingdom), the National Research Council (Canada), CONICYT (Chile), the Australian Research Council (Australia), Ministério da Ciência e Tecnologia (Brazil)  and Ministerio de Ciencia, Tecnología e Innovación Productiva (Argentina).
 \end{acknowledgements}
 
\bibliographystyle{aa} 
\bibliography{refs}

\newpage
\begin{appendix}

\section{}

\begin{table}[h!] 
{\small \caption{Atmospheric parameters} 
\label{tab:atm}
\begin{center}
\begin{tabular}{lllll}
\hline\hline
\noalign{\smallskip}
Object & $T_{\rm eff}$ & $\log{g}$ & $\log{y}$ & Instrument \\
       & [K] &  &  &  \\ 
\noalign{\smallskip}
\hline
\noalign{\smallskip}
J01185$-$00254 & $26700\pm1000$ & $5.36\pm0.15$ & $-3.0$ & SDSS \\
             & $27700\pm600$  & $5.55\pm0.09$ & $<-3.0$ & TWIN \\
             & $28000\pm350$  & $5.55\pm0.05$ & $<-3.0$ & Goodman \\
             & $27900\pm600$  & $5.55\pm0.07$ & $<-3.0$ & adopted \\
\noalign{\smallskip}
\hline
\noalign{\smallskip}
J03213$+$05384 & $30200\pm500$ & $5.74\pm0.11$ & $-2.4$ & SDSS \\
             & $30700\pm100$ & $5.73\pm0.02$ & $-2.3$ & ISIS \\
             & $31200\pm300$ & $5.74\pm0.05$ & $-2.5$ & Goodman \\
             & $30700\pm500$  & $5.74\pm0.06$ & $-2.4\pm0.1$ & adopted \\
\noalign{\smallskip}
\hline
\noalign{\smallskip}
J08233$+$11364 & $31300\pm600$ & $5.78\pm0.12$ & $-1.9$ & SDSS \\
             & $31100\pm200$ & $5.78\pm0.03$ & $-2.0$ & ISIS \\
             & $31200\pm400$ & $5.80\pm0.06$ & $-2.0$ & Goodman \\
             & $31200\pm600$ & $5.79\pm0.06$ & $-2.0\pm0.1$ & adopted \\
\noalign{\smallskip}
\hline
\noalign{\smallskip}
J08300$+$47515 & $25200\pm500$ & $5.30\pm0.05$ & $-3.3\pm0.7$ & SDSS$\dag$ \\
             & $25400\pm200$ & $5.45\pm0.02$ & $<-3.0$ & ISIS \\
             & $25300\pm600$ & $5.38\pm0.06$ & $<-3.0$ & adopted \\
\noalign{\smallskip}
\hline
\noalign{\smallskip}
J09510$+$03475& $29800\pm300$ & $5.48\pm0.04$ & $-2.8\pm0.3$ & SDSS$\dag$ \\
             & $29800\pm300$ & $5.48\pm0.04$ & $-2.8\pm0.3$ & adopted \\             
\noalign{\smallskip}
\hline
\noalign{\smallskip}
J09523$+$62581 & $27800\pm500$ & $5.61\pm0.08$ & $-2.6$       & SDSS$\dag$ \\
             & $27600\pm200$ & $5.56\pm0.02$ & $-2.5$       & ISIS \\
             & $27700\pm600$ & $5.59\pm0.06$ & $-2.6\pm0.1$ & adopted \\
\noalign{\smallskip}
\hline
\noalign{\smallskip}
J10215$+$30101 & $30700\pm500$ & $5.71\pm0.06$ & $<-3.0$ & SDSS$\dag$ \\
             & $30000\pm200$ & $5.63\pm0.02$ & $-2.5$ & ISIS \\
             & $30400\pm600$ & $5.67\pm0.06$ & $-2.6\pm0.1$ & adopted \\
\noalign{\smallskip}
\hline
\noalign{\smallskip}
J1132$-$0636 & $46400\pm1900$ & $5.83\pm0.11$ & $-2.7$ & SDSS \\
             & $46400\pm500$  & $5.94\pm0.03$ & $-3.0$ & ISIS \\
             & $46400\pm1000$ & $5.89\pm0.07$ & $-2.9\pm0.2$ & adopted \\
\noalign{\smallskip}
\hline
\noalign{\smallskip}
J13463$+$28172 & $28000\pm800$  & $5.38\pm0.12$ & $-2.7$ & SDSS \\
             & $29500\pm200$  & $5.54\pm0.02$ & $-2.4$ & GMOS \\
             & $28800\pm600$  & $5.46\pm0.07$ & $-2.6\pm0.2$ & adopted \\
\noalign{\smallskip}
\hline
\noalign{\smallskip}
J15082$+$49405 & $28200\pm600$ & $5.34\pm0.09$ & $-2.0\pm0.2$ & SDSS$\dag$ \\
             & $27000\pm1100$ & $5.28\pm0.19$ & $-2.2$ & GMOS \\             
             & $29500\pm600$ & $5.76\pm0.10$ & $-2.3$ & TWIN \\
             & $29600\pm300$ & $5.70\pm0.05$ & $-2.3$ & ISIS \\         
             & $29600\pm600$ & $5.73\pm0.07$ & $-2.3\pm0.1$ & adopted \\         
\noalign{\smallskip}
\hline
\noalign{\smallskip}
J15222$-$01301 & $24800\pm1000$ & $5.52\pm0.15$ & $-2.6\pm0.5$ & SDSS$\dag$ \\
             & $25600\pm500$  & $5.41\pm0.07$ & $<-3.0$ & ISIS \\
             & $25200\pm700$  & $5.47\pm0.09$ & $<-3.0$ & adopted \\
\noalign{\smallskip}
\hline
\noalign{\smallskip}
J18324$+$63091 & $26700\pm1100$ & $5.26\pm0.17$ & $-2.5$ & SDSS \\
             & $26900\pm200$  & $5.32\pm0.03$ & $-2.7$ & ISIS \\
             & $26800\pm700$  & $5.29\pm0.09$ & $-2.6\pm0.1$ & adopted \\
\noalign{\smallskip}
\hline
\end{tabular}
\tablefoot{
$\dag$ Parameters taken from \citet{gei11a}
}
\label{tab:atmo}
\end{center}}
\end{table}

\newpage
\onecolumn
{\small 
\begin{longtab}
\begin{longtable}{llllllc}
\caption{Orbital parameters of all published helium burning hot subdwarf binaries}\tabularnewline
\hline\hline
\noalign{\smallskip}
Object  &  T$_{\rm eff}$ [K] &  log\,$g$ &  Period [days]   &    $\gamma$ [${\rm km\,s^{-1}}$] &   K [${\rm km\,s^{-1}}$]   &  References \\
\hline
\noalign{\smallskip}
\endfirsthead
\caption{continued.}\tabularnewline
\hline\hline
\noalign{\smallskip}
Object  & T$_{\rm eff}$    &  log\,$g$  &  Period [days]   &    $\gamma$ [${\rm km\,s^{-1}}$] &   K [${\rm km\,s^{-1}}$]   &   References \\
\hline
\noalign{\smallskip}
\endhead
\hline
\endfoot
PG0850+170      & 27100$\pm$1000 &5.37$\pm$0.10 & 27.815$\pm$0.005          &   32.2$\pm$2.8 & 33.5$\pm$3.3 & [1,47]  \\
EGB5            & 34500$\pm$500  &5.85$\pm$0.05 & 16.532$\pm$0.003          &   68.5$\pm$0.7 & 16.1$\pm$0.8 & [2]  \\
PG0919+273      & 32900   &5.90   & 15.5830$\pm$0.00005       &  -68.6$\pm$0.6 & 41.5$\pm$0.8 & [3]  \\
PG1619+522      & 32300$\pm$1000 &5.98$\pm$0.10 & 15.3578$\pm$0.0008        &  -52.5$\pm$1.1 & 35.2$\pm$1.1 & [1,47]  \\
KIC7668647      & 27680$\pm$310  &5.50$\pm$0.03 & 14.1742$\pm$0.0042        &  -27.4$\pm$1.3 & 38.9$\pm$1.9 & [4]  \\
CS1246          & 28500$\pm$700  &5.46$\pm$0.11 & 14.105$\pm$0.011          &   67.2$\pm$1.7 & 16.6$\pm$0.6 & [5]  \\
LB1516          & 25200$\pm$1100 &5.41$\pm$0.12 & 10.3598$\pm$0.00005       &   14.3$\pm$1.1 & 48.6$\pm$1.4 & [6]  \\
PG1558-007      & 20300$\pm$1000 &5.00$\pm$0.10 & 10.3495$\pm$0.00006       &  -71.9$\pm$0.7 & 42.8$\pm$0.8 & [3,48]  \\
KIC11558725     & 27900$\pm$500  &5.41$\pm$0.05 & 10.0545$\pm$0.0048        &  -66.1$\pm$1.4 & 58.1$\pm$1.7 & [7]  \\
PG1110+294      & 30100$\pm$1000 &5.72$\pm$0.10 & 9.4152$\pm$0.0002         &  -15.2$\pm$0.9 & 58.7$\pm$1.2 & [1,47]  \\
EC20260-4757    & -              & -            & 8.952$\pm$0.0002          &   56.6$\pm$1.6 & 57.1$\pm$1.9 & [3]  \\
Feige108        & 34500$\pm$1000 &6.01$\pm$0.15 & 8.74651$\pm$0.00001       &   45.8$\pm$0.6 & 50.2$\pm$1.0 & [8,49]  \\
PG0940+068      &   -            & -            & 8.330$\pm$0.003           &  -16.7$\pm$1.4 & 61.2$\pm$1.4 & [9]  \\
PHL861          & 30000$\pm$500  &5.50$\pm$0.05 & 7.44$\pm$0.015            &  -26.5$\pm$0.4 & 47.9$\pm$0.4 & [10]  \\
{\bf J032138+053840}  & 30700$\pm$500  &5.74$\pm$0.06 & 7.4327$\pm$0.0004         &  -16.7$\pm$2.1 & 39.7$\pm$2.8 &  This work \\
HE1448-0510     & 34700$\pm$500  &5.59$\pm$0.05 & 7.159$\pm$0.005           &  -45.5$\pm$0.8 & 53.7$\pm$1.1 & [10   \\
PG1439-013      &   -            & -            & 7.2914$\pm$0.00005        &  -53.7$\pm$1.6 & 50.7$\pm$1.5 & [3]  \\
{\bf J095238+625818} & 27700$\pm$600  &5.59$\pm$0.06 & 6.98$\pm$0.04             &  -35.4$\pm$3.6 & 62.5$\pm$3.4 &  This work  \\
PG1032+406      & 31600$\pm$900  &5.77$\pm$0.10 & 6.7791$\pm$0.0001         &   24.5$\pm$0.5 & 33.7$\pm$0.5 & [1,47]  \\
PG0907+123      & 26200$\pm$900  &5.30$\pm$0.10 & 6.11636$\pm$0.00006       &   56.3$\pm$1.1 & 59.8$\pm$0.9 & [1,47]  \\
HE1115-0631     & 40400$\pm$1000 &5.80$\pm$0.10 & 5.87$\pm$0.001            &   87.1$\pm$1.3 & 61.9$\pm$1.1 & [11,50]  \\
CD-24731        & 35400$\pm$500  &5.90$\pm$0.05 & 5.85$\pm$0.003            &   20.0$\pm$5.0 & 63.0$\pm$3.0 & [12,51]  \\
PG1244+113      & 36300   &5.54   & 5.75211$\pm$0.00009       &    7.4$\pm$0.8 & 54.4$\pm$1.4 &  [3] \\
PG0839+399      & 37800$\pm$900  &5.53$\pm$0.10 & 5.6222$\pm$0.0002         &   23.2$\pm$1.1 & 33.6$\pm$1.5 &  [1] \\
{\bf J183249+630910}  & 26800$\pm$700  &5.29$\pm$0.09 & 5.4$\pm$0.2               &  -32.5$\pm$2.1 & 62.1$\pm$3.3 & This work \\
EC20369-1804    &   -            &  -           & 4.5095$\pm$0.00004        &    7.2$\pm$1.6 & 51.5$\pm$2.3 & [3]  \\
TONS135         & 25000$\pm$1250 &5.60$\pm$0.20 & 4.1228$\pm$0.0008         &   -3.7$\pm$1.1 & 41.4$\pm$1.5 & [12,52]  \\
PG0934+186      & 35800   &5.65   & 4.051$\pm$0.001           &    7.7$\pm$3.2 & 60.3$\pm$2.4 & [3]  \\
PB7352          & 25000$\pm$500  &5.35$\pm$0.10 & 3.62166$\pm$0.000005      &   -2.1$\pm$0.3 & 60.8$\pm$0.3 & [12,53]  \\
KPD0025+5402    & 28200$\pm$900  &5.37$\pm$0.10 & 3.5711$\pm$0.0001         &   -7.8$\pm$0.7 & 40.2$\pm$1.1 & [1]  \\
KIC10553698     & 27423$\pm$293  &5.436$\pm$0.024 & 3.387$\pm$0.014         &   52.1$\pm$1.5 & 64.8$\pm$2.2 & [65] \\
PG0958-073      & 26100$\pm$500  &5.58$\pm$0.05 & 3.18095$\pm$0.000007      &   90.5$\pm$0.8 & 27.6$\pm$1.4 & [3,54]  \\
PG1253+284      &   -            & -            & 3.01634$\pm$0.000005      &   17.8$\pm$0.6 & 24.8$\pm$0.9 & [3]  \\
TON245          & 25200$\pm$1000 &5.30$\pm$0.15 & 2.501$\pm$0.000           &   - & 88.3& [1,49]  \\
PG1300+279      & 29600$\pm$900  &5.65$\pm$0.10 & 2.25931$\pm$0.0001        &   -3.1$\pm$0.9 & 62.8$\pm$1.6 & [1,47]  \\
CPD-201123      & 23500$\pm$500  &4.90$\pm$0.10 & 2.3098$\pm$0.0003         &   -6.3$\pm$1.2 & 43.5$\pm$0.9 & [13]  \\
NGC188/II-91    &  -             & -            & 2.15             &    - & 22.0& [14] \\
{\bf J134632+281722}  & 28800$\pm$600  &5.46$\pm$0.07 & 1.96$\pm$0.03             &    1.2$\pm$1.2 & 85.6$\pm$3.4 &  This work \\
V1093Her        & 27400$\pm$800  &5.47$\pm$0.10 & 1.77732$\pm$0.000005      &   -3.9$\pm$0.8 & 70.8$\pm$1.0 &  [1,47] \\
PG1403+316      & 31200   &5.75   & 1.73846$\pm$0.000001      &   -2.1$\pm$0.9 & 58.5$\pm$1.8 &  [3] \\
HD171858        & 27200$\pm$800  &5.30$\pm$0.10 & 1.63280$\pm$0.000005      &   62.5$\pm$0.1 & 60.8$\pm$0.3 &  [12,53]  \\
{\bf J002323-002953}  & 29200$\pm$500  &5.69$\pm$0.05 & 1.4876$\pm$0.0001         &   16.4$\pm$2.1 & 81.8$\pm$2.9 &  [15] \\
KPD2040+3955    & 27900   &5.54   & 1.482860$\pm$0.0000004    &  -16.4$\pm$1.0 & 94.0$\pm$1.5 &  [3] \\
HE2150-0238     & 30200$\pm$500  &5.83$\pm$0.07 & 1.321$\pm$0.005           &  -32.5$\pm$0.9 & 96.3$\pm$1.4 &  [10,55] \\
{\bf J011857-002546}  & 27900$\pm$600  &5.55$\pm$0.07 & 1.30$\pm$0.02             &   37.7$\pm$1.8 & 54.8$\pm$2.9 &  This work \\
UVO1735+22      & 38000$\pm$500  &5.54$\pm$0.05 & 1.278$\pm$0.001           &   20.6$\pm$0.4 &103.0$\pm$1.5 &  [12,53] \\
PG1512+244      & 29900$\pm$900  &5.74$\pm$0.10 & 1.26978$\pm$0.000002      &   -2.9$\pm$1.0 & 92.7$\pm$1.5 &  [1,47]  \\
PG0133+114      & 29600$\pm$900  &5.66$\pm$0.10 & 1.23787$\pm$0.000003      &   -0.3$\pm$0.2 & 82.0$\pm$0.3 &  [12,1] \\
HE1047-0436     & 30200$\pm$500  &5.66$\pm$0.05 & 1.21325$\pm$0.00001       &   25.0$\pm$3.0 & 94.0$\pm$3.0 &  [16] \\
PG2331+038      & 27200   &5.58   & 1.204964$\pm$0.0000003    &   -9.5$\pm$1.1 & 93.5$\pm$1.9 & [3]  \\
HE1421-1206     & 29600$\pm$500  &5.55$\pm$0.07 & 1.188$\pm$0.001           &  -86.2$\pm$1.1 & 55.5$\pm$2.0 &  [17,55] \\
{\bf J113241-063652}  & 46400$\pm$000  &5.89$\pm$0.07 & 1.06$\pm$0.02             &    8.3$\pm$2.2 & 41.1$\pm$4.0 &  This work \\
PG1000+408      & 36400$\pm$900  &5.54$\pm$0.10 & 1.049343$\pm$0.0000005    &   56.6$\pm$3.4 & 63.5$\pm$3.0 &  [3,47] \\
{\bf J150829+494050}  & 29600$\pm$600  &5.73$\pm$0.07 & 0.967164$\pm$0.000009     &  -60.0$\pm$10.7& 93.6$\pm$5.8 &  This work \\
PG1452+198      & 29400   &5.75   & 0.96498$\pm$0.000004      &   -9.1$\pm$2.1 & 86.8$\pm$1.9 & [3]  \\
HS2359+1942     & 31400$\pm$500  &5.56$\pm$0.07 & 0.93261$\pm$0.00005       &  -96.1$\pm$6.0 &107.4$\pm$6.8 & [6,55]  \\
PB5333          & 40600$\pm$500  &5.96$\pm$0.10 & 0.92560$\pm$0.0000012     &  -95.3$\pm$1.3 & 22.4$\pm$0.8 & [8,54]  \\
HE2135-3749     & 30000$\pm$500  &5.84$\pm$0.05 & 0.9240$\pm$0.0003         &   45.0$\pm$0.5 & 90.5$\pm$0.6 &  [10] \\
EC12408-1427    &  -             & -            & 0.90243$\pm$0.000001      &  -52.2$\pm$1.2 & 58.6$\pm$1.5 & [3]  \\
PG0918+029      & 31700$\pm$900  &6.03$\pm$0.10 & 0.87679$\pm$0.000002      &  104.4$\pm$1.7 & 80.0$\pm$2.6 &  [1,47] \\
PG1116+301      & 32500$\pm$1000 &5.85$\pm$0.10 & 0.85621$\pm$0.000003      &   -0.2$\pm$1.1 & 88.5$\pm$2.1 & [1,47]  \\
PG1230+052      & 27100   &5.47   & 0.837177$\pm$0.0000003    &  -43.1$\pm$0.7 & 40.4$\pm$1.2 &  [3] \\
EC21556-5552    &   -            &   -          & 0.8340$\pm$0.00007        &   31.4$\pm$2.0 & 65.0$\pm$3.4 &  [3] \\
V2579Oph        & 23500$\pm$500  &5.40$\pm$0.10 & 0.8292056$\pm$0.0000014   & -54.16$\pm$0.27& 70.1$\pm$0.13&  [18,53] \\
EC13332-1424    &   -            &  -           & 0.82794$\pm$0.000001      &  -53.2$\pm$1.8 &104.1$\pm$3.0 &  [3] \\
TONS183         & 27600$\pm$500  &5.43$\pm$0.05 & 0.8277$\pm$0.00002        &   50.5$\pm$0.8 & 84.8$\pm$1.0 &  [12,53] \\
KPD2215+5037    & 29600   &5.64   & 0.809146$\pm$0.0000002    &   -7.2$\pm$1.0 & 86.0$\pm$1.5 &  [3] \\
EC02200-2338    &  -             &  -           & 0.8022$\pm$0.00007        &   20.7$\pm$2.3 & 96.4$\pm$1.4 &  [3]  \\
{\bf J150513+110836}  & 33200$\pm$500  &5.80$\pm$0.10 & 0.747773$\pm$0.00005      &  -77.1$\pm$1.2 & 97.2$\pm$1.8 &  [15] \\
PG0849+319      & 28900$\pm$900  &5.37$\pm$0.10 & 0.74507$\pm$0.000001      &   64.0$\pm$1.5 & 66.3$\pm$2.1 &  [1,47] \\
JL82            & 26500$\pm$500  &5.22$\pm$0.10 & 0.73710$\pm$0.00005       &   -1.6$\pm$0.8 & 34.6$\pm$1.0 &  [12,53] \\
PG1248+164      & 26600$\pm$800  &5.68$\pm$0.10 & 0.73232$\pm$0.000002      &  -16.2$\pm$1.3 & 61.8$\pm$1.1 &  [1,47] \\
EC22202-1834    &   -            &  -           & 0.70471$\pm$0.000005      &   -5.5$\pm$3.9 &118.6$\pm$5.8 &  [3] \\
{\bf J225638+065651}  & 28500$\pm$500  &5.64$\pm$0.05 & 0.7004$\pm$0.0001         &   -7.3$\pm$2.1 &105.3$\pm$3.4 & [15]  \\
{\bf J152222-013018}  & 25200$\pm$700  &5.47$\pm$0.09 & 0.67162$\pm$0.00003       &  -79.5$\pm$2.7 & 80.1$\pm$3.5 &  This work \\
PG1648+536      & 31400   &5.62   & 0.6109107$\pm$0.00000004  &  -69.9$\pm$0.9 &109.0$\pm$1.3 &  [3] \\
PG1247+554      &  -             &  -           & 0.602740$\pm$0.000006     &   13.8$\pm$0.6 & 32.2$\pm$1.0 &  [9] \\
PG1725+252      & 28900$\pm$900  &5.54$\pm$0.10 & 0.601507$\pm$0.0000003    &  -60.0$\pm$0.6 &104.5$\pm$0.7 &  [1,47] \\
EC20182-6534    &  -             &  -           & 0.598819$\pm$0.0000006    &   13.5$\pm$1.9 & 59.7$\pm$3.2 &  [3] \\
PG0101+039      & 27500$\pm$500  &5.53$\pm$0.07 & 0.569899$\pm$0.000001     &    7.3$\pm$0.2 &104.7$\pm$0.4 &  [19] \\
HE1059-2735     & 41000$\pm$1000 &5.38$\pm$0.10 & 0.555624         &  -44.7$\pm$0.6 & 87.7$\pm$0.8 &  [11,50] \\
PG1519+640      & 30600   &5.72   & 0.54029143$\pm$0.0000000025 &  0.1$\pm$0.4 & 42.7$\pm$0.6 &  [8,3] \\
PG0001+275      & 25400$\pm$500  &5.30$\pm$0.10 & 0.529842$\pm$0.0000005    &  -44.7$\pm$0.5 & 92.8$\pm$0.7 &  [12,53] \\
PG1743+477      & 27600$\pm$800  &5.57$\pm$0.10 & 0.515561$\pm$0.0000001    &  -65.8$\pm$0.8 &121.4$\pm$1.0 &  [1] \\
{\bf J172624+274419}  & 32600$\pm$500  &5.84$\pm$0.05 & 0.50198$\pm$0.00005       &  -36.7$\pm$4.8 &118.9$\pm$3.7 &  [15] \\
HE1318-2111     & 36300$\pm$1000 &5.42$\pm$0.10 & 0.487502$\pm$0.0000001    &   48.9$\pm$0.7 & 48.5$\pm$1.2 & [11,50]  \\
KUV16256+4034   & 23100   &5.38   & 0.4776$\pm$0.00008        &  -90.9$\pm$0.9 & 38.7$\pm$1.2 & [3]  \\
GALEXJ2349+3844 & 23800$\pm$350  &5.38$\pm$0.06 & 0.462516$\pm$0.000005     &    2.0$\pm$1.0 & 87.9$\pm$2.2 &  [20,56] \\
HE0230-4323     & 31100$\pm$500  &5.60$\pm$0.07 & 0.45152$\pm$0.00002       &   16.6$\pm$1.0 & 62.4$\pm$1.6 & [12,55]  \\
HE0929-0424     & 29500$\pm$500  &5.71$\pm$0.05 & 0.4400$\pm$0.0002         &   41.4$\pm$1.0 &114.3$\pm$1.4 &  [10] \\
UVO1419-09      &   -            &   -          & 0.4178$\pm$0.00002        &   42.3$\pm$0.3 &109.6$\pm$0.4 & [12]  \\
{\bf J095101+034757}  & 29800$\pm$300  &5.48$\pm$0.04 & 0.4159$\pm$0.0007         &  111.1$\pm$2.5 & 84.4$\pm$4.2 &  This work \\
KPD1946+4340    & 34200$\pm$500  &5.43$\pm$0.10 & 0.403739$\pm$0.0000008    &   -5.5$\pm$1.0 &156.0$\pm$2.0 &  [21,53] \\
V1405Ori        & 35100$\pm$800  &5.66$\pm$0.11 & 0.398            &  -33.6$\pm$5.5 & 85.1$\pm$8.6 &  [6] \\
Feige48         & 29500$\pm$500  &5.54$\pm$0.05 & 0.376$\pm$0.003           &  -47.9$\pm$0.1 & 28.0$\pm$0.2 &  [22,51] \\
GD687           & 24300$\pm$500  &5.32$\pm$0.07 & 0.37765$\pm$0.00002       &   32.3$\pm$3.0 &118.3$\pm$3.4 &  [23,55] \\
PG1232-136      & 26900$\pm$500  &5.71$\pm$0.05 & 0.3630$\pm$0.0003         &    4.1$\pm$0.3 &129.6$\pm$0.04&  [12,53] \\
PG1101+249      & 29700$\pm$500  &5.90$\pm$0.07 & 0.35386$\pm$0.00006       &   -0.8$\pm$0.9 &134.6$\pm$1.3 &  [24,57] \\
PG1438-029      & 27700$\pm$1000 &5.50$\pm$0.15 & 0.336            &    - & 32.1&  [25,49]  \\
PG1528+104      & 27200   &5.46   & 0.331$\pm$0.0001          &  -49.3$\pm$1.0 & 53.3$\pm$1.6 & [3]  \\
PHL457          & 26500$\pm$1100 &5.38$\pm$0.12 & 0.3128$\pm$0.0007         &    - & 12.8$\pm$0.08&  [26,54] \\
PG0941+280      & 29400$\pm$500  &5.43$\pm$0.05 & 0.311            &   73.7$\pm$4.3 &141.7$\pm$6.3 & [6]  \\
HS2043+0615     & 26200$\pm$500  &5.28$\pm$0.07 & 0.3015$\pm$0.0003         &  -43.5$\pm$3.4 & 73.7$\pm$4.3 & [6,55]  \\
{\bf J102151+301011}  & 30400$\pm$600  &5.67$\pm$0.06 & 0.2966$\pm$0.0001         &  -28.4$\pm$4.8 &114.5$\pm$5.2 &  This work \\
KBS13           & 29700$\pm$500  &5.70$\pm$0.05 & 0.2923$\pm$0.0004         &   7.53$\pm$0.08&22.82$\pm$0.23&  [27,58] \\
CPD-64481       & 27500$\pm$500  &5.60$\pm$0.05 & 0.277263$\pm$0.000005     &   94.1$\pm$0.3 & 23.9$\pm$0.05&  [26,51] \\
GALEXJ0321+4727 & 28000$\pm$400  &5.34$\pm$0.07 & 0.265856$\pm$0.000003     &   69.6$\pm$2.2 & 60.8$\pm$4.5 & [20,56]  \\
HE0532-4503     & 25400$\pm$500  &5.32$\pm$0.05 & 0.2656$\pm$0.0001         &    8.5$\pm$0.1 &101.5$\pm$0.2 & [10]  \\
AADor           & 42000$\pm$1000 &5.46$\pm$0.05 & 0.2614$\pm$0.0002         &   1.57$\pm$0.09&40.15$\pm$0.11&  [28,59] \\
{\bf J165404+303701}  & 24900$\pm$800  &5.39$\pm$0.12 & 0.25357$\pm$0.00001       &   40.5$\pm$2.2 &126.1$\pm$2.6 & [15]  \\
{\bf J012022+395059}  & 29400$\pm$500  &5.48$\pm$0.05 & 0.252013$\pm$0.000013     &  -47.3$\pm$1.3 & 37.3$\pm$2.8 & [44]  \\
PG1329+159      & 29100$\pm$900  &5.62$\pm$0.10 & 0.249699$\pm$0.0000002    &  -22.0$\pm$1.2 & 40.2$\pm$1.1 & [1,47]  \\
{\bf J204613-045418}  & 31600$\pm$500  &5.54$\pm$0.08 & 0.24311$\pm$0.00001       &   87.6$\pm$5.7 &134.3$\pm$7.8 & [15]  \\
PG2345+318      & 27500$\pm$1000 &5.70$\pm$0.15 & 0.2409458$\pm$0.000008    &  -10.6$\pm$1.4 &141.2$\pm$1.1 & [24,49]  \\
PG1432+159      & 26900$\pm$1000 &5.75$\pm$0.15 & 0.22489$\pm$0.00032       &  -16.0$\pm$1.1 &120.0$\pm$1.4 & [24,49]  \\
BPSCS22169-0001 & 39300$\pm$500  &5.60$\pm$0.05 & 0.214            &    - & 16.2& [26,53]  \\
{\bf J113840-003531}  & 31200$\pm$600  &5.54$\pm$0.09 & 0.207536$\pm$0.000002     &   23.3$\pm$3.7 &162.0$\pm$3.8 &  [15] \\
{\bf J082332+113641}  & 31200$\pm$600  &5.79$\pm$0.06 & 0.20707$\pm$0.00002       &  135.1$\pm$2.0 &169.4$\pm$2.5 &  This work \\
HS1741+2133     &  -             &  -           & 0.20$\pm$0.01             & -112.8$\pm$2.7 &157.0$\pm$3.4 & [29]  \\
HE1414-0309     & 29500$\pm$500  &5.56$\pm$0.07 & 0.192$\pm$0.004           &  104.7$\pm$9.5 &152.4$\pm$1.2 & [6,55]  \\
HS2333+3927     & 36500$\pm$1000 &5.70$\pm$0.10 & 0.1718023$\pm$0.0000009   &  -31.4$\pm$2.1 & 89.6$\pm$3.2 & [30]  \\
{\bf J192059+372220}  & 27500$\pm$1000  &5.4$\pm$0.1 & 0.168876$\pm$0.00035 &    16.8$\pm$2.0 & 59.7$\pm$2.5 &  [64] \\
2M1533+3759     & 29200$\pm$500  &5.58$\pm$0.05 & 0.16177042$\pm$0.00000001 &   -3.4$\pm$5.2 & 71.1$\pm$1.0 & [31]  \\
{\bf J083006+475150}  & 25300$\pm$600  &5.38$\pm$0.06 & 0.14780$\pm$0.00007       &   49.9$\pm$0.9 & 77.0$\pm$1.7 &  This work \\
ASAS102322-3737 & 28400$\pm$500  &5.60$\pm$0.05 & 0.13926940$\pm$0.00000004 &    - & 81.0$\pm$3.0 &  [32]  \\
EC00404-4429    &  -             &  -           & 0.12834$\pm$0.000004      &   33.0$\pm$2.9 &152.8$\pm$3.4 & [3]  \\
2M1938+4603     & 29600$\pm$500  &5.43$\pm$0.05 & 0.1257653$\pm$0.000000021 &   20.1$\pm$0.3 & 65.7$\pm$0.6 &  [33] \\
BULSC16335      & 31500$\pm$1800 &5.70$\pm$0.2  & 0.122            &   36.4$\pm$19.6& 92.5$\pm$6.2 & [6]  \\
PG1043+760      & 27600$\pm$800  &5.39$\pm$0.1  & 0.1201506$\pm$0.00000003  &   24.8$\pm$1.4 & 63.6$\pm$1.4 &  [1,47] \\
EC10246-2707    & 28900$\pm$500  &5.64$\pm$0.06 & 0.1185079935$\pm$0.00000000091 & - & 71.6$\pm$1.7 &  [34] \\
HWVir           & 28500$\pm$500  &5.63$\pm$0.05 & 0.115$\pm$0.0008          &  -13.0$\pm$0.8 & 84.6$\pm$1.1 & [35,60]  \\
HS2231+2441     & 28400$\pm$500  &5.39$\pm$0.05 & 0.1105880$\pm$0.0000005   &    - & 49.1$\pm$3.2 &  [36] \\
NSVS14256825    & 40000$\pm$500  &5.50$\pm$0.05 & 0.110374230$\pm$0.000000002 & 12.1$\pm$1.5 & 73.4$\pm$2.0 &  [37] \\
UVEX0328+5035   & 28500   &5.50   & 0.11017$\pm$0.00011       &   44.9$\pm$0.7 & 64.0$\pm$1.5 &  [29,61] \\
PG1336-018      & 32800$\pm$500  &5.76$\pm$0.05 & 0.101015999$\pm$0.00000001&  -25.0& 78.7$\pm$0.6 &  [38,62] \\
{\bf J082053+000843}  & 26000$\pm$1000 &5.37$\pm$0.14 & 0.096$\pm$0.001           &    9.5$\pm$1.3 & 47.4$\pm$1.9 &  [45] \\
HS0705+6700     & 28800$\pm$900  &5.40$\pm$0.10 & 0.09564665$\pm$0.00000039 &  -36.4$\pm$2.9 & 85.8$\pm$3.6 &  [39] \\
KPD1930+2752    & 35200$\pm$500  &5.61$\pm$0.06 & 0.0950933$\pm$0.0000015   &    5.0$\pm$1.0 &341.0$\pm$1.0 &  [40] \\
KPD0422+5421    & 25000$\pm$1500 &5.40$\pm$0.10 & 0.09017945$\pm$0.00000012 &  -57.0$\pm$12.0&237.0$\pm$8.0 &  [41,63] \\
PG1017-086      & 30300$\pm$500  &5.61$\pm$0.10 & 0.0729938$\pm$0.0000003   &   -9.1$\pm$1.3 & 51.0$\pm$1.7 & [42]  \\
{\bf J162256+473051}  & 29000$\pm$600  &5.65$\pm$0.06 & 0.069789         &   -54.7$\pm$1.5   & 47.0$\pm$2.0 & [46]  \\
CD-3011223      & 29200$\pm$400  &5.66$\pm$0.05 & 0.0489790717$\pm$0.0000000038&16.5$\pm$0.3 &377.0$\pm$0.4 & [43]  \\
\hline
\label{tab:allbinaries}
\end{longtable}
\tablebib{
1:  \citet{mor03};
2:   \citet{gei11c}; 
3:   \citet{cop11};
4:   Telting et al. in press\nocite{tel14};
5:    \citet{bar11};
6:   \citet{gei14}
7:   \citet{tel12};
8:   \citet{ede04};
9:   \citet{max00a};  
10:   \citet{kar06};
11:   \citet{nap04a};
12:   \citet{ede05};
13:   \citet{nas12};
14:   \citet{gre04};
15:   \citet{gei11b};
16:   \citet{nap01a};
17:   \citet{gei06};
18:   \citet{for06};
19:   \citet{gei08};
20:   \citet{kaw12};
21:   \citet{blo11};
22:   \citet{oto04};
23:   \citet{gei10a};
24:   \citet{mor99};
25:   \citet{gre05};
26:   \citet{gei12b};
27:   \citet{for08};
28:   \citet{mue10};
29:   \citet{kup14};
30:   \citet{heb04};
31:   \citet{for10};
32:   \citet{sch13};
33:   \citet{ost10};
34:   \citet{bar13};
35:   \citet{ede08};
36:   \citet{ost07};
37:   \citet{alm12};
38:   \citet{vuc07};
39:   \citet{dre01};
40:   \citet{gei07};
41:   \citet{oro99};
42:   \citet{max02};
43:   \citet{gei13};
44:   \citet{ost13};
45:   \citet{gei12};
46:   \citet{sch14};
47:   \citet{max01};
48:   \citet{heb02};
49:   \citet{saf94};
50:   \citet{str07};
51:   \citet{oto06};
52:   \citet{heb86};
53:   \citet{gei10};
54:   \citet{gei13a};
55:   \citet{lis05};
56:   \citet{nem12};
57:   \citet{ede99};
58:   \citet{ost10a};
59:   \citet{kle11};
60:   \citet{woo99};
61:   \citet{ver12};
62:   \citet{cha08};
63:   \citet{koe98};
64:   \citet{sch14a};
65:   \citet{ost14}
}

\begin{longtable}{lllllll}
\caption{Photometry, spectroscopic distances and companion types}\tabularnewline
\hline\hline
\noalign{\smallskip}
Object  &  V [mag] &  J [mag]   &    H [mag] &   Distances [kPc]   &  Comp. Type  &   References \\
\hline
\noalign{\smallskip}
\endfirsthead
\caption{continued.}\tabularnewline
\hline\hline
\noalign{\smallskip}
Object  &  V[mag]   & J [mag]  &  H [mag]   &    Distances [kPc] &   Comp. Type   &   References \\
\hline
\noalign{\smallskip}
\endhead
\hline
\endfoot
\smallskip
PG0850+170       & 13.977$\pm$0.000& 14.531$\pm$0.043 & 14.624$\pm$0.066 & $1.04^{+0.16}_{-0.15}$ & MS/WD & [1,13] \\   
\smallskip
EGB5             & 13.808$\pm$0.04 & 14.482$\pm$0.036 & 14.530$\pm$0.055 & $0.68^{+0.07}_{-0.06}$ & MS/WD & [2,13]\\ 
\smallskip
PG0919+273       & 12.765$\pm$0.009& 13.303$\pm$0.021 & 13.420$\pm$0.030 & $0.39^{+0.00}_{-0.00}$ & WD  & [1,13] \\        
\smallskip
PG1619+522       & 13.297$\pm$0.006& 13.883$\pm$0.027 & 13.969$\pm$0.040 & $0.44^{+0.08}_{-0.06}$ & WD  & [1,13] \\
\smallskip  
KIC7668647       & 15.218$\pm$0.07 & 15.815$\pm$0.066 & 16.056$\pm$0.201$^C$ & $1.54^{+0.12}_{-0.12}$ & WD$^{\rm lc}$ & [2,13] \\
\smallskip  
CS1246           & 14.371$\pm$0.03 & 14.013$\pm$0.039 & 14.032$\pm$0.058 & - & MS/WD$^{\rm lc}$  & [2,13]  \\ 
\smallskip
LB1516           & 12.967$\pm$0.003& 13.520$\pm$0.035 & 13.663$\pm$0.054 & $0.59^{+0.11}_{-0.10}$ & WD  & [3,13] \\  
\smallskip
PG1558-007       & 13.528$\pm$0.006& -                & -                & $0.84^{+0.15}_{-0.12}$ & MS/WD & [1] \\
\smallskip 
KIC11558725      & 14.859$\pm$0.04 & 15.379$\pm$0.046 & 15.352$\pm$0.088 & $1.45^{+0.14}_{-0.13}$ & WD$^{\rm lc}$ & [2,13] \\ 
\smallskip  
PG1110+294       & 14.086$\pm$0.006& 14.626$\pm$0.037 & 14.674$\pm$0.063 & $0.79^{+0.13}_{-0.11}$ & WD & [1,13]  \\ 
\smallskip  
EC20260-4757     & 13.735$\pm$0.01 & 14.424$\pm$0.023 & 14.465$\pm$0.046 & - & MS/WD & [2,13] \\   
\smallskip
Feige108         & 12.973$\pm$0.000& 13.529$\pm$0.024 & 13.704$\pm$0.032 & $0.39^{+0.08}_{-0.07}$ & WD  & [4,13]  \\  
\smallskip 
PG0940+068       &   -             & 14.151$\pm$0.027 & 14.147$\pm$0.049 & - & MS/WD & [13] \\
\smallskip   
PHL861           & 14.826$\pm$0.04 & 15.375$\pm$0.051 & 15.496$\pm$0.103 & $1.45^{+0.14}_{-0.13}$ & MS/WD & [2,13] \\
\smallskip 
{\bf J032138+053840}   & 15.048$\pm$0.04 & 15.148$\pm$0.050 & 15.402$\pm$0.135$^B$ & $1.02^{+0.11}_{-0.10}$ & MS/WD  & [2,13] \\ 
\smallskip
HE1448-0510      & 14.611$\pm$0.04 & 15.199$\pm$0.056 & 15.234$\pm$0.096 & $1.30^{+0.12}_{-0.11}$ & WD & [2,13] \\ 
\smallskip  
PG1439-013       & 13.943$\pm$0.028& 14.506$\pm$0.035 & 14.692$\pm$0.041 & - & MS/WD & [1,13] \\ 
\smallskip
{\bf J095238+625818}   & 14.693$\pm$0.09 & 15.420$\pm$0.067 & 15.609$\pm$0.147$^B$ & $1.13^{+0.16}_{-0.14}$ & WD & [2,13]  \\   
\smallskip
PG1032+406       & 11.519$\pm$0.009& 12.166$\pm$0.022 & 12.275$\pm$0.018 & $0.25^{+0.04}_{-0.04}$ & MS/WD  & [1,13] \\ 
\smallskip
PG0907+123       & 13.970$\pm$0.003& 14.474$\pm$0.030 & 14.666$\pm$0.066 & $1.08^{+0.17}_{-0.15}$ & MS/WD & [1,13] \\   
\smallskip
HE1115-0631      &   -             & 15.623$\pm$0.079 & 15.580$\pm$0.129$^{\rm B}$ & - & MS/WD  & [13] \\
\smallskip   
CD-24731         & 11.748$\pm$0.024& 12.404$\pm$0.027 & 12.583$\pm$0.027 & $0.26^{+0.02}_{-0.02}$ & WD & [5,13] \\  
\smallskip  
PG1244+113       & 14.197$\pm$0.018& 14.821$\pm$0.004 & 14.939$\pm$0.007 & $1.21^{+0.01}_{-0.01}$ & WD  & [1,14] \\ 
\smallskip  
PG0839+399       & 14.389$\pm$0.000& 14.885$\pm$0.035 & 15.080$\pm$0.064 & $1.36^{+0.19}_{-0.16}$ & MS/WD & [1,13] \\ 
\smallskip
{\bf J183249+630910}   & 15.695$\pm$0.01 & 16.236$\pm$0.096 & 16.068$\pm$0.176$^C$ & $2.41^{+0.34}_{-0.30}$ & MS/WD & [2,13]  \\ 
\smallskip  
EC20369-1804     & 13.29 $\pm$0.00 & 13.937$\pm$0.022 & 14.061$\pm$0.044 & - & MS/WD &  [6,13] \\
\smallskip 
TONS135          & 13.302$\pm$0.042& 13.868$\pm$0.029 & 14.017$\pm$0.047 & $0.54^{+0.19}_{-0.14}$ & MS/WD  & [5,13] \\ 
\smallskip
PG0934+186       & 13.138$\pm$0.001& 13.759$\pm$0.025 & 13.972$\pm$0.038 & $0.66^{+0.00}_{-0.00}$ & WD  &  [1,13] \\   
\smallskip
PB7352           & 12.261$\pm$0.01 & 12.819$\pm$0.026 & 12.915$\pm$0.025 & $0.44^{+0.06}_{-0.06}$ & MS/WD  &  [2,13]  \\ 
\smallskip
KPD0025+5402     & 13.919$\pm$0.015& 14.259$\pm$0.031 & 14.361$\pm$0.047 & - & MS/WD  &  [1,13] \\
\smallskip 
KIC10553698      & 14.902$\pm$0.08 & 15.446$\pm$0.047 & 15.538$\pm$0.092 & - & WD$^{\rm lc}$ &  [2,13] \\     
\smallskip 
PG0958-073       & 13.563$\pm$0.002& 14.098$\pm$0.030 & 14.139$\pm$0.039 & $0.63^{+0.05}_{-0.05}$ & MS/WD  &  [4,13] \\  
\smallskip
PG1253+284       & 12.769$\pm$0.000& 12.182$\pm$0.001 & 12.003$\pm$0.001 & - & MS/WD &  [1,13]  \\ 
\smallskip
TON245           & 13.859$\pm$0.04 & 14.312$\pm$0.002 & 14.417$\pm$0.005 & $0.97^{+0.24}_{-0.20}$ & WD & [2,14]  \\
\smallskip  
PG1300+279       & 14.266$\pm$0.023& 14.894$\pm$0.004 & 15.006$\pm$0.007 & $0.94^{+0.16}_{-0.13}$ & WD & [1,14]  \\
\smallskip   
CPD-201123       & 12.173$\pm$0.12 & 12.565$\pm$0.024 & 12.658$\pm$0.027 & $0.64^{+0.13}_{-0.11}$ & MS/WD &  [2,13] \\
\smallskip 
NGC188/II-91     & -               & -                &  -               & -  & MS/WD  & [] \\ 
\smallskip
{\bf J134632+281722}   & 14.908$\pm$0.07 & 15.513$\pm$0.007 & 15.595$\pm$0.010 & $1.53^{+0.21}_{-0.19}$ & WD &  [2,14]   \\  
\smallskip
V1093Her         & 13.967$\pm$0.008& 14.518$\pm$0.034 & 14.677$\pm$0.074 & $0.93^{+0.14}_{-0.13}$ & MS/WD$^{\rm lc}$  &  [1,13]  \\
\smallskip 
PG1403+316       & 13.532$\pm$0.010& 14.179$\pm$0.023 & 14.376$\pm$0.041 & $0.63^{+0.00}_{-0.00}$ & WD &  [1,13] \\
\smallskip   
HD171858         &  9.85$\pm$0.04  & 10.321$\pm$0.023 & 10.432$\pm$0.022 & - & MS/WD  & [7,13] \\
\smallskip 
{\bf J002323-002953}   & 15.577$\pm$0.02 & 16.153$\pm$0.013 & 16.275$\pm$0.026 & $1.60^{+0.14}_{-0.14}$ & WD & [2,14]  \\   
\smallskip
KPD2040+3955     & 14.472$\pm$0.049& 14.560$\pm$0.036 & 14.560$\pm$0.065 & - & MS/WD &  [1,13]  \\ 
\smallskip  
HE2150-0238      &  -              & -                & -                & - & MS/WD & [] \\
\smallskip
{\bf J011857-002546}   & 14.804$\pm$0.05 & 15.184$\pm$0.047 & 15.262$\pm$0.098 & $1.25^{+0.16}_{-0.15}$ & MS/WD & [2,13] \\ 
\smallskip
UVO1735+22       & 11.861$\pm$0.01 & 12.509$\pm$0.021 & 12.650$\pm$0.022 & $0.37^{+0.04}_{-0.04}$ & WD  &  [2,13]  \\   
\smallskip
PG1512+244       & 13.185$\pm$0.04 & 13.957$\pm$0.028 & 13.957$\pm$0.039 & $0.50^{+0.09}_{-0.07}$ & WD  & [2,13]  \\ 
\smallskip 
PG0133+114       & 12.345$\pm$0.000& 12.801$\pm$0.001 & 12.918$\pm$0.002 & $0.37^{+0.05}_{-0.05}$ & WD &  [1,14] \\  
\smallskip 
HE1047-0436      & 14.796$\pm$0.000& 15.527$\pm$0.060 & 15.637$\pm$0.123$^B$ & $1.19^{+0.09}_{-0.09}$ & WD  &  [1,13]  \\
\smallskip   
PG2331+038       & 14.974$\pm$0.025& 15.361$\pm$0.005 & 15.404$\pm$0.012 & $1.20^{+0.02}_{-0.01}$ & WD &  [1,14]  \\   
\smallskip
HE1421-1206      & 15.510$\pm$0.0  & 15.891$\pm$0.073 & 15.872$\pm$0.165$^C$ & $1.72^{+0.18}_{-0.16}$ & MS &  [8,13.15]  \\   
\smallskip
{\bf J113241-063652}   & 16.273$\pm$0.005& 16.684$\pm$0.142$^B$ & -  &  $2.39^{+0.24}_{-0.21}$  & MS/WD  &  [9,13] \\ 
\smallskip
PG1000+408       & 13.327$\pm$0.023& 13.978$\pm$0.027 & 14.244$\pm$0.043 & $0.83^{+0.12}_{-0.10}$ & MS/WD  &  [1,13] \\ 
\smallskip
{\bf J150829+494050}   & 17.516$\pm$0.005& -                & -          & $3.83^{+0.41}_{-0.37}$ & MS/WD  &  [9]  \\ 
\smallskip
PG1452+198       & 12.476$\pm$0.002& 13.055$\pm$0.023 & 13.179$\pm$0.025 & $0.90^{+0.00}_{-0.00}$ & WD  &  [1,13] \\
\smallskip   
HS2359+1942      & 15.639$\pm$0.06 & 16.234$\pm$0.092 & 16.227$\pm$0.212$^C$ & $2.02^{+0.26}_{-0.24}$ & WD & [2,13]  \\
\smallskip     
PB5333           & 12.874$\pm$0.020& 12.879$\pm$0.001 & 12.622$\pm$0.001 & $0.42^{+0.05}_{-0.06}$ & MS/WD &  [1,14]  \\
\smallskip   
HE2135-3749      & 13.896$\pm$0.01 & 14.598$\pm$0.035 & 14.650$\pm$0.051 & $0.63^{+0.05}_{-0.05}$ & WD & [2,13]  \\
\smallskip   
EC12408-1427     & 12.823$\pm$0.02 & 13.390$\pm$0.029 & 13.465$\pm$0.035 & - & MS/WD  & [2,13] \\
\smallskip 
PG0918+029       & 13.415$\pm$0.080& 13.995$\pm$0.002 & 14.092$\pm$0.004 & $0.43^{+0.08}_{-0.07}$ & WD &  [1,14] \\ 
\smallskip  
PG1116+301       & 14.337$\pm$0.019&    -             & -                & $0.84^{+0.14}_{-0.12}$  & MS/WD  &  [1]  \\ 
\smallskip
PG1230+052       & 13.287$\pm$0.018& 13.835$\pm$0.002 & 13.965$\pm$0.003 & $0.67^{+0.01}_{-0.01}$ & MS/WD  &  [1,14]  \\ 
\smallskip
EC21556-5552     & 13.090$\pm$0.04 & 13.718$\pm$0.029 & 13.862$\pm$0.047 & - & MS/WD  &  [2,13]  \\
\smallskip 
V2579Oph         & 12.930$\pm$0.025& 13.369$\pm$0.026 & 13.476$\pm$0.030 & $0.52^{+0.08}_{-0.07}$ & MS/WD$^{\rm lc}$  &  [1,13] \\
\smallskip 
EC13332-1424     & 13.380$\pm$0.03 & 13.895$\pm$0.030 & 13.990$\pm$0.035 & - & MS/WD & [2,13] \\
\smallskip 
TONS183          & 12.598$\pm$0.02 & 13.232$\pm$0.026 & 13.361$\pm$0.028 & $0.52^{+0.05}_{-0.04}$ & MS/WD  & [2,13] \\ 
\smallskip
KPD2215+5037     & 13.739$\pm$0.022& 14.218$\pm$0.040 & 14.313$\pm$0.042 & - & MS/WD & [1,13] \\ 
\smallskip
EC02200-2338     & 12.014$\pm$0.01 & 12.616$\pm$0.026 & 12.748$\pm$0.021 & - & MS/WD & [2,13] \\ 
\smallskip
{\bf J150513+110836}   & 15.378$\pm$0.09 & 16.043$\pm$0.006 & 16.151$\pm$0.015 & $1.44^{+0.26}_{-0.23}$ & WD  & [2,14]  \\
\smallskip  
PG0849+319       & 14.606$\pm$0.000& 15.177$\pm$0.044 & 15.318$\pm$0.095 & $1.46^{+0.23}_{-0.19}$ & MS/WD &  [1,13] \\ 
\smallskip
JL82             & 12.389$\pm$0.003& 12.857$\pm$0.024 & 12.960$\pm$0.025 & $0.57^{+0.08}_{-0.07}$ & MS & [3,13] \\   
\smallskip
PG1248+164       & 14.460$\pm$0.03 & 15.037$\pm$0.037 & 15.013$\pm$0.080 & $0.89^{+0.15}_{-0.13}$ & MS/WD &  [2,13] \\
\smallskip 
EC22202-1834     & 13.802$\pm$0.03 & 14.389$\pm$0.033 & 14.431$\pm$0.049 & - & MS/WD  &  [2,13]  \\
\smallskip   
{\bf J225638+065651}   & 15.314$\pm$0.01 & 15.744$\pm$0.006 & 15.789$\pm$0.012 & $1.39^{+0.12}_{-0.10}$ & WD & [2,14]  \\ 
\smallskip  
{\bf J152222-013018}   & 17.813$\pm$0.02 & 18.424$\pm$0.098 & 18.202$\pm$0.131 & - & MS/WD &  [9,14]  \\ 
\smallskip
PG1648+536       & 14.055$\pm$0.017& 14.553$\pm$0.029 & 14.587$\pm$0.051 & $0.88^{+0.01}_{-0.01}$ & WD  &  [1,13]  \\
\smallskip   
PG1247+554       & 12.259$\pm$0.01 & -                & 11.087$\pm$0.017 & - & MS/WD &  [2,13] \\
\smallskip 
PG1725+252       & 13.008$\pm$0.018& 13.496$\pm$0.026 & 13.641$\pm$0.037 & $0.54^{+0.09}_{-0.07}$ & MS/WD & [1,13]  \\
\smallskip 
EC20182-6534     & 13.29 $\pm$0.0  & 13.782$\pm$0.029 & 13.877$\pm$0.021 & - & MS/WD  & [6,13]  \\
\smallskip
PG0101+039       & 12.065$\pm$0.000& 12.609$\pm$0.001 & 12.724$\pm$0.002 & $0.36^{+0.04}_{-0.03}$ & WD$^{\rm lc}$ & [4,14] \\
\smallskip   
HE1059-2735      & 15.500$\pm$0.06 & 16.051$\pm$0.089 & 16.329$\pm$0.206$^C$ & $2.73^{+0.46}_{-0.43}$ & MS/WD & [2,13] \\
\smallskip 
PG1519+640       & 12.458$\pm$0.001& 13.007$\pm$0.023 & 13.185$\pm$0.026 & $0.39^{+0.00}_{-0.00}$ & MS/WD  &  [1,13] \\ 
\smallskip
PG0001+275       &   -             & 13.833$\pm$0.024 & 13.971$\pm$0.041 &-  & MS/WD  &  [13] \\
\smallskip 
PG1743+477       & 13.787$\pm$0.009& 14.313$\pm$0.024 & 14.526$\pm$0.060 & $0.76^{+0.12}_{-0.10}$ & WD  &  [1,13]  \\
\smallskip 
{\bf J172624+274419}   & 15.994$\pm$0.01 & 16.467$\pm$0.101 & - &  $1.76^{+0.15}_{-0.13}$ & WD  &  [2,13]  \\ 
\smallskip
HE1318-2111      & 14.718$\pm$0.09 & 15.218$\pm$0.049 & 15.288$\pm$0.089 & $1.64^{+0.30}_{-0.26}$ & MS/WD  &  [2,13] \\
\smallskip 
KUV16256+4034    & 12.64 $\pm$0.21 & 13.067$\pm$0.035 & 13.224$\pm$0.035 & $0.49^{+0.04}_{-0.05}$ & MS/WD  &  [7,13] \\
\smallskip 
GALEXJ2349+3844  & 11.73 $\pm$0.13 & 12.040$\pm$0.024 & 12.156$\pm$0.031 & - & MS/WD  &  [7,13] \\      
\smallskip 
HE0230-4323      & 13.768$\pm$0.02 & 13.948$\pm$0.032 & 13.804$\pm$0.044 & $0.82^{+0.09}_{-0.08}$ & MS &  [2,13] \\ 
\smallskip        
HE0929-0424      & 16.165$\pm$0.15 & 16.646$\pm$0.112$^B$ & -                & $2.06^{+0.31}_{-0.28}$ & MS/WD &  [2,13] \\
\smallskip 
UVO1419-09       & 12.115$\pm$0.09 & 12.692$\pm$0.023 & 12.835$\pm$0.025 & - & WD &  [2,13]  \\  
\smallskip 
{\bf J095101+034757}   & 15.895$\pm$0.02 & 15.972$\pm$0.007 & 15.705$\pm$0.013 & $2.37^{+0.17}_{-0.15}$ & MS/WD  &  [2,14]  \\ 
\smallskip
KPD1946+4340     & 14.299$\pm$0.002& 14.683$\pm$0.031 & 14.836$\pm$0.055 & - & WD & [1,13]  \\ 
\smallskip  
V1405Ori         & 15.142$\pm$0.09 & 14.574$\pm$0.031 & 14.677$\pm$0.045 & - & MS$^{\rm lc}$ &  [2,13]  \\ 
\smallskip  
Feige48          & 13.456$\pm$0.000& 13.983$\pm$0.027 & 14.137$\pm$0.043 &  $0.73^{+0.05}_{-0.06}$ & WD$^{\rm lc}$  & [1,13]  \\
\smallskip         
GD687            & 14.077$\pm$0.03 & 14.618$\pm$0.033 & 14.874$\pm$0.077 & $1.04^{+0.12}_{-0.11}$ & WD &  [2,13]  \\   
\smallskip
PG1232-136       & 13.336$\pm$0.033& 13.758$\pm$0.028 & 13.897$\pm$0.040 & $0.50^{+0.05}_{-0.04}$ & WD &  [1,13]  \\   
\smallskip
PG1101+249       & 12.775$\pm$0.03 & 13.187$\pm$0.027 & 13.257$\pm$0.039 & $0.36^{+0.03}_{-0.04}$ & WD  &   [2,13]  \\
\smallskip   
PG1438-029       &    -            & 14.168$\pm$0.029 & 14.240$\pm$0.053 & - & MS  &  [13]  \\ 
\smallskip  
PG1528+104       & 13.569$\pm$0.010& 14.082$\pm$0.002 & 14.156$\pm$0.004 & $0.77^{+0.01}_{-0.01}$ & WD & [1,14]  \\   
\smallskip
PHL457           & 12.947$\pm$0.010& 13.499$\pm$0.026 & 13.595$\pm$0.021 & $0.62^{+0.12}_{-0.11}$ & MS  &  [2,13]  \\   
\smallskip
PG0941+280       & 13.265$\pm$0.07 & 13.799$\pm$0.029 & 13.899$\pm$0.042 & $0.75^{+0.09}_{-0.07}$ & WD$^{\rm lc}$ & [2,13]  \\   
\smallskip
HS2043+0615      & 15.420$\pm$ 0   & 16.098$\pm$0.093 & -                & $2.02^{+0.20}_{-0.19}$ & MS & [8,13]  \\   
\smallskip
{\bf J102151+301011}   & 18.218$\pm$0.007& -                & -          & $5.74^{+0.55}_{-0.51}$  & MS/WD  & [9] \\ 
\smallskip
KBS13            & 13.633$\pm$0.01 & 14.018$\pm$0.024 & 14.063$\pm$0.032 & - & MS$^{\rm lc}$ &  [2,13] \\  
\smallskip 
CPD-64481        & 11.291$\pm$0.01 & 11.878$\pm$0.022 & 11.994$\pm$0.028 & $0.23^{+0.02}_{-0.02}$ & MS  &  [2,13]  \\   
\smallskip
GALEXJ0321+4727  & 11.73 $\pm$0.15 & 11.795$\pm$0.001 & 11.923$\pm$0.001 & - & MS & [7,14]  \\   
\smallskip
HE0532-4503      & 15.98 $\pm$0.0  & 16.563$\pm$0.146$^C$ & -                & $2.55^{+0.20}_{-0.18}$ & MS/WD &  [10,13]  \\ 
\smallskip
AADor            & 11.90 $\pm$0.0  & 11.795$\pm$0.028 & 11.965$\pm$0.029 & - & MS$^{\rm lc}$ & [2,13] \\  
\smallskip 
{\bf J165404+303701}   & 15.409$\pm$0.04 & 15.938$\pm$0.008 & 16.030$\pm$0.016 & $1.78^{+0.37}_{-0.30}$ & MS/WD  &  [2,14]  \\ 
\smallskip
{\bf J012022+395059}   & 15.341$\pm$0.07 & 16.016$\pm$0.078 & 15.925$\pm$0.175$^C$ & $1.79^{+0.19}_{-0.18}$ & MS &  [2,13]  \\  
\smallskip 
PG1329+159       & 13.507$\pm$0.03 & 14.047$\pm$0.002 & 14.221$\pm$0.004 & $0.67^{+0.12}_{-0.10}$ & MS &  [2,13]  \\   
\smallskip
{\bf J204613-045418}   & 16.324$\pm$0.01 & 16.677$\pm$0.143$^B$ & 16.308$\pm$0.217$^D$ & $2.81^{+0.33}_{-0.30}$  & MS/WD & [2,13] \\  
\smallskip
PG2345+318       & 14.178$\pm$0.005& 14.690$\pm$0.039 & 14.833$\pm$0.071 & $0.19^{+0.05}_{-0.03}$ & WD & [1,13]  \\ 
\smallskip  
PG1432+159       & 13.896$\pm$0.013& 14.445$\pm$0.028 & 14.530$\pm$0.050 & $0.64^{+0.15}_{-0.12}$  & WD &  [1,13]  \\  
\smallskip 
BPSCS22169-0001  & 12.848$\pm$0.01 & 13.456$\pm$0.023 & 13.551$\pm$0.024 & - & MS  &  [2,13]  \\
\smallskip   
{\bf J113840-003531}   & 14.467$\pm$0.03 & 15.161$\pm$0.043 & 15.137$\pm$0.085 & $1.23^{+0.17}_{-0.16}$  & WD & [2,13] \\ 
\smallskip  
{\bf J082332+113641}   & 16.658$\pm$0.005& -                & -              & $2.48^{+0.22}_{-0.21}$ & MS/WD & [2]  \\   
\smallskip
HS1741+2133      & 13.990$\pm$0.01 & 14.386$\pm$0.036 & 14.616$\pm$0.060 & - & WD   &  [2,13]  \\
\smallskip   
HE1415-0309      & 16.487$\pm$0.0  & -                & -                & $2.76^{+0.27}_{-0.26}$ & WD &  [9]  \\   
\smallskip
HS2333+3927      & 14.794$\pm$0.01 & 14.986$\pm$0.048 & 15.018$\pm$0.084 & $1.25^{+0.17}_{-0.15}$  & MS$^{\rm lc}$ & [2,13]  \\
\smallskip   
{\bf J192059+372220}   & 15.745$\pm$0.01 & 16.186$\pm$0.083 & 16.272$\pm$0.224$^D$ & - & MS & [2,13]  \\   
\smallskip
2M1533+3759      & 12.964$\pm$0.17 & 13.652$\pm$0.026 & 13.736$\pm$0.031 & $0.55^{+0.09}_{-0.08}$  & MS$^{\rm lc}$ & [2,13]  \\  
\smallskip 
{\bf J083006+475150}   & 16.043$\pm$0.03 & 16.737$\pm$0.143$^B$ & 16.477$\pm$0.220$^D$ & $2.44^{+0.27}_{-0.24}$ & WD & [2,13] \\
\smallskip   
ASAS102322-3737  & 11.707$\pm$0.07 & 12.028$\pm$0.021 & 12.112$\pm$0.027 & $0.28^{+0.03}_{-0.03}$  & MS$^{\rm lc}$ & [2,13]  \\   
\smallskip
EC00404-4429     & 13.674$\pm$0.02 & 14.220$\pm$0.030 & 14.424$\pm$0.052 & - & WD  &  [2,13] \\  
\smallskip 
2M1938+4603      & 12.063$\pm$0.01 & 12.757$\pm$0.022 & 12.889$\pm$0.020 & - & MS$^{\rm lc}$  &  [2,13]  \\
\smallskip   
BULSC16335       & 16.395$\pm$0.028& 12.868$\pm$0.022 & 12.072$\pm$0.021 & - & MS$^{\rm lc}$ &  [11,13]  \\   
\smallskip
PG1043+760       & 13.768$\pm$0.016& 14.278$\pm$0.030 & 14.359$\pm$0.049 & $0.89^{+0.14}_{-0.12}$  & WD & [1,13]  \\
\smallskip   
EC10246-2707     & 14.38 $\pm$0.0  & 14.830$\pm$0.036 & 14.842$\pm$0.052 &  $0.92^{+0.08}_{-0.07}$& MS$^{\rm lc}$ & [12,13]  \\ 
\smallskip 
HWVir            & 10.577$\pm$0.069& 10.974$\pm$0.027 & 11.093$\pm$0.022 & $0.17^{+0.01}_{-0.02}$ & MS$^{\rm lc}$ & [1,13]  \\   
\smallskip
HS2231+2441      & 14.153$\pm$0.05 & 14.669$\pm$0.035 & 14.732$\pm$0.054 & $1.11^{+0.11}_{-0.10}$ & MS$^{\rm lc}$ & [2,13]  \\   
\smallskip
NSVS14256825     & 13.389$\pm$0.25 & 13.658$\pm$0.026 & 13.797$\pm$0.026 & - & MS$^{\rm lc}$ & [2,13]  \\  
\smallskip 
UVEX0328+5035    & 14.263$\pm$0.02 & 14.121$\pm$0.004 & 13.915$\pm$0.004 & - & MS  &  [2,14] \\  
\smallskip 
PG1336-018       & 13.661$\pm$0.27 & 14.594$\pm$0.037 & 14.646$\pm$0.050 & $0.61^{+0.07}_{-0.07}$ & MS$^{\rm lc}$ &  [2,13] \\  
\smallskip 
{\bf J082053+000843}   & 15.168$\pm$0.05 & 15.712$\pm$0.008 & 15.818$\pm$0.014 & $1.69^{+0.42}_{-0.33}$ & BD & [2,14]  \\   
\smallskip
HS0705+6700      & 14.923$\pm$0.52 & 15.103$\pm$0.039 & 15.233$\pm$0.086 & $1.58^{+0.74}_{-0.50}$ & MS$^{\rm lc}$  &  [2,13] \\   
\smallskip
KPD1930+2752     & 13.833$\pm$0.035& 13.983$\pm$0.029 & 13.968$\pm$0.045 & - & WD$^{\rm lc}$ & [1,13]  \\   
\smallskip
KPD0422+5421     & 14.682$\pm$0.018& 14.425$\pm$0.031 & 14.421$\pm$0.046 & - & WD$^{\rm lc}$ & [1,13]  \\  
\smallskip 
PG1017-086       & 14.426$\pm$0.025& 14.866$\pm$0.042 & 15.036$\pm$0.074 & $1.05^{+0.16}_{-0.14}$ & MS  & [1,13]  \\   
\smallskip
{\bf J162256+473051}   & 16.188$\pm$0.02 & 16.732$\pm$0.136$^B$ & -                & $2.25^{+0.23}_{-0.21}$ & BD & [2,13]  \\     
\smallskip       
CD-3011223       & 12.296$\pm$0.03 & 12.886$\pm$0.029 & 12.932$\pm$0.023 & $0.35^{+0.03}_{-0.03}$ & WD  & [2,13]  \\   
\noalign{\smallskip}
\hline
\label{tab:allbinaries1}
\end{longtable}
\tablefoot{
lc: identified photometrically; B,C,D: 2MASS colours of quality B, C or D which were excluded from the analysis
}
\tablebib{
1:  \citet{wes92};
2:   UCAC4; 
3:   \citet{lan07};
4:   \citet{lan09};
5:    \citet{mer92};
6:   \citet{odo13}
7:   \citet{hog00};
8:   NOMAD;
9:   SDSS, \citet{jes05};
10:   SPM4.0;  
11:   \citet{uds02};
12:   \citet{kil97};
13: 2MASS;
14: UKIDSS;
15: Koen priv. comment
}

\end{longtab}

}
\end{appendix}

\end{document}